\documentclass[acmsmall, screen]{acmart}

\AtBeginDocument{%
  \providecommand\BibTeX{{%
    \normalfont B\kern-0.5em{\scshape i\kern-0.25em b}\kern-0.8em\TeX}}}

\usepackage{booktabs, cancel, comment, pgfplots, starfont, enumitem}   %
\usepackage{pgfplotstable}

\usepgfplotslibrary{fillbetween}

\allowdisplaybreaks

\usepackage{subcaption, dsfont} %
\usepackage{listings}
\usepackage{stmaryrd}
\usepackage{mathtools}
\usepackage{amsmath}
\usepackage{accents}
\usepackage{pifont}
\usepackage[inkscapeformat=png]{svg}
\usepackage{svg}
\newcommand{\HRule}[2][\medskipamount]{\par
	\vspace*{\dimexpr-\parskip-\baselineskip+#1}
	\noindent\rule{\linewidth}{0.5pt}\par
	\vspace*{\dimexpr-\parskip-.5\baselineskip+#2}}

\newcommand{\cmark}{\ding{51}}%
\newcommand{\xmark}{\ding{55}}

\newcommand{\true}[0]{ \texttt{T} }
\newcommand{\decondition}[1]{ \mathit{DC}({#1})}
\newcommand{\obdd}[1]{ \mathit{OBDD}({#1})}
\newcommand{\false}[0]{ \texttt{F} }

\newcommand{\unifobs}[2]{ \texttt{unifObs(}#1\texttt{, #2)} }
\newcommand{\pointgamma}[4]{ \texttt{pointGamma(\(#1, #2, #3, #4\))} }
\newcommand{\lessthan}[3]{ \texttt{less\_than(#1, #2, #3)} }
\newcommand{\flipcount}[1]{\texttt{flip\_count}\left( { #1 } \right) }
\newcommand{\flip}[1]{\texttt{f}_{\texttt{#1}}}
\usepackage{algpseudocode}
\usepackage{float}
\usepackage[linewidth=0pt]{mdframed}
\usepackage{blindtext}
\usepackage{hyperref}
\usepackage{fancyvrb}
\usepackage{bbm}
\usepackage[inference]{semantic}
\algdef{SE}[SUBALG]{Indent}{EndIndent}{}{\algorithmicend\ }%
\algtext*{Indent}
\algtext*{EndIndent}
\hypersetup{
	colorlinks=true,
	linkcolor=blue,
	filecolor=magenta,      
	urlcolor=cyan,
	pdftitle={Overleaf Example},
	pdfpagemode=FullScreen,
}

\graphicspath{ {./} }
\usepackage{tikz, tikz-cd}

\usetikzlibrary{patterns,calc,backgrounds, shapes.geometric}

\tikzstyle{nnf}=[
>=stealth,font=\small,auto,scale=0.7,every node/.style={scale=0.7}
]
\tikzstyle{extnode}=[
draw,circle,inner sep=2pt,fill=white
]

\tikzstyle{bddroot}=[
draw,fill=white,inner sep=2.5pt, font=\small
]

\tikzstyle{leafnode}=[
draw,fill=gray!20,inner sep=3.5pt
]
\tikzstyle{constnode}=[
draw,fill=white,inner sep=3.5pt
]
\tikzstyle{label}=[
fill=white,inner sep=2.5pt
]

\tikzstyle{acarrow}=[
decoration={markings,mark=at position 1 with {\arrow[scale=0.6]{>}}},
postaction={decorate},
shorten >=0.4pt,
>=latex,
line width=0.1
]

\tikzstyle{bnarrow}=[
decoration={markings,mark=at position 1 with {\arrow[scale=1.5]{>}}},
postaction={decorate},
shorten >=0.7pt,
>=latex,
line width=0.3
]
\tikzstyle{bayesnet}=[
>=latex, thick, auto
]
\tikzstyle{bnnode}=[
draw,ellipse,minimum size=7mm,inner sep=1pt,font=\small
]
\tikzstyle{cpt}=[
font=\footnotesize
]

\tikzstyle{graph}=[
>=stealth,font=\small,auto,scale=1,every node/.style={scale=1}
]
\tikzstyle{node}=[
draw,circle,inner sep=3pt,fill=white
]

\tikzstyle{bdd}=[
>=latex, thick, >=stealth, font=\small,auto,scale=0.9,every node/.style={scale=0.9}
]
\tikzstyle{bddnode}=[
draw,circle,inner sep=0pt,fill=white,minimum size=5.5mm
]

\tikzstyle{highedge}=[
line width=0.9
]
\tikzstyle{lowedge}=[
line width=0.9,dotted
]
\tikzstyle{bddterminal}=[
draw,fill=gray!20,inner sep=2.5pt, font=\small
]

\pdfoutput=1
\usepackage[titlenumbered,ruled,resetcount]{algorithm2e}

\usepackage{amsthm}
\usepackage{graphicx}
\usepackage{wrapfig}
\graphicspath{ {./} }
\newtheorem{theorem}{Theorem}
\newtheorem{lemma}[theorem]{Lemma}

\newtheoremstyle{Definition}%
{}%
{}%
{}%
{}%
{\upshape}%
{:}%
{}%
{}%
\newtheorem{definition}{Definition}
\newtheorem{example}{Example}

\newcommand{\pdf}[2]{\pi_{#1, #2}}

\newcommand{\Llet}[2]{ {\texttt{let}~#1~\texttt{in}~#2} } %
\newcommand{\Lobs}[1]{ {\texttt{observe}~#1}}
\newcommand{\Lflip}[1]{ {\texttt{flip}~#1}}

\newcommand{\Lfst}[1]{ {\texttt{fst(}#1\texttt{)}}}
\newcommand{\Lsnd}[1]{ {\texttt{snd(}#1\texttt{)}}}
\newcommand{\Lite}[3]{ {\texttt{if}~#1~\texttt{then}~#2~\texttt{else}~#3}}

\newcommand{\dice}[0]{\texttt{Dice}}
\newcommand{\hybit}[0]{\texttt{HyBit}}

\newcommand{\comp}[0] {\rightsquigarrow}

\newcommand{\xbroadand}[1]{{\underset{#1}{\land}}}
\newcommand{\xpointor}[1]{{\overset{{.}}{\underset{#1}{\lor}}}}

\newcommand{\defeq}[0] {\triangleq}

\newcommand{\bool}[0]{ \mathrm{B} }
\newcommand{\tbool}[0]{ \texttt{bool} }

\newcommand{\prog}[1]{ \texttt{p}_{\texttt{#1}}}
\newcommand{\var}[1]{ \texttt{var}_{\texttt{#1}}}
\newcommand{\ord}[1]{ \texttt{ord}({#1})}
\newcommand{\yvar}[1]{ \texttt{y}_{\texttt{#1}}}

\newcommand{\te}[0]{ \texttt{exp}}  %

\newcommand{\dbracket}[1]{ \left\llbracket{} { #1 } \right\rrbracket}

\newcommand{\bin}[2]{\llparenthesis{} { #1 } \rrparenthesis_{ #2 }}

\makeatletter
\renewcommand{\fnum@figure}{Figure \thefigure}
\makeatother

\setcopyright{rightsretained}
\acmDOI{10.1145/3656412}
\acmYear{2024}
\copyrightyear{2024}
\acmSubmissionID{pldi24main-p188-p}
\acmJournal{PACMPL}
\acmVolume{8}
\acmNumber{PLDI}
\acmArticle{182}
\acmMonth{6}
\received{2023-11-16}
\received[accepted]{2024-03-31}

\begin{document}

	\title{Bit Blasting Probabilistic Programs}
	
	\author{Poorva Garg}
	\orcid{0000-0003-4753-3974}
	\affiliation{%
		\institution{University of California, Los Angeles}
		\country{USA}
	}
	\email{poorvagarg@cs.ucla.edu}
	
	\author{Steven Holtzen}
	\orcid{0000-0002-8190-5412}
	\affiliation{%
		\institution{Northeastern University}
		\city{Boston}
		\country{USA}
	}
	\email{s.holtzen@northeastern.edu}
	
	\author{Guy Van den Broeck}
	\orcid{0000-0003-3434-2503}
	\affiliation{%
		\institution{University of California, Los Angeles}
		\country{USA}
	}
	\email{guyvdb@cs.ucla.edu}
	
	\author{Todd Millstein}
	\orcid{0000-0002-2031-1514}
	\affiliation{%
		\institution{University of California, Los Angeles}
		\country{USA}
	}
	\email{todd@cs.ucla.edu}

	\begin{abstract}
	Probabilistic programming languages (PPLs) are an expressive means for creating and reasoning about probabilistic models.  Unfortunately {\em hybrid} probabilistic programs that involve both continuous and discrete structures are not well supported by today's PPLs.  In this paper we develop a new approximate inference algorithm for hybrid probabilistic programs that first discretizes the continuous distributions and then performs discrete inference on the resulting program.  The key novelty is a form of discretization that we call \emph{bit blasting}, which uses a binary representation of numbers such that a domain of $2^b$ discretized points can be succinctly represented as a discrete probabilistic program over  \(\mathit{poly}(b)\) Boolean random variables.  Surprisingly, we prove that many common continuous distributions can be bit blasted in a manner that incurs no loss of accuracy over an explicit discretization and supports efficient probabilistic inference. We have built a probabilistic programming system for hybrid programs called \hybit{}, which employs bit blasting followed by discrete probabilistic inference.  We empirically demonstrate the benefits of our approach over existing sampling-based and symbolic inference approaches%
	\end{abstract}


	\maketitle	
	\section{Introduction}
	
	Probabilistic programming languages (PPLs) are an expressive means for creating and reasoning about probabilistic models.  Many such models are naturally {\em hybrid}, involving both continuous (e.g., Gaussian distributions) and discrete structures (e.g., Bernoulli random variables, {\tt if} statements and other control flow). For example, hybrid models arise in applications such as medical diagnosis, gene expression and cyber-physical systems~\cite{chen2020probabilistic, cyberphysical}.
	
	Unfortunately, hybrid programs are not well supported by today's PPLs.  The primary analysis task in probabilistic programming languages is {\em probabilistic inference}, computing
	the probability that an event occurs according to the distribution defined by the program. Existing inference algorithms employ forms of sampling to perform approximate inference. Some approaches, notably Hamiltonian Monte Carlo, used in the PPLs Pyro and Stan~\cite{bingham2019pyro, Gorinova_2021}, do not support discrete random variables, instead requiring them to be (manually or automatically) marginalized out.  However, this approach has numerous fatal cases that explode exponentially in the number of discrete variables.\footnote{Indeed, the Pyro documentation states that it cannot support more than 25 discrete variables in CUDA and 64 discrete variables on a CPU~\cite{pyro_documentation}.} Other sampling-based approaches are universal and so can handle discreteness, such as importance sampling, Markov Chain Monte Carlo and Sequential Monte Carlo, etc.~\cite{koller2009probabilistic}.  However, these algorithms are known to struggle with multimodal distributions~\cite{yao2021stacking}, which arise through discreteness, as well as with programs that condition on low-probability events.
	\begin{table}[t]
		\caption{Distributions with sound and efficient bit blasting subsumed by mixed-gamma densities.}
		\begin{tabular}{r l c r l}
			Distribution & Density &\qquad\quad& Distribution & Density\\
			\toprule
			Uniform & 1 &&
			Gamma & $\frac{\beta^{\alpha}x^{\alpha-1}e^{-\beta x}}{\Gamma(\alpha)}$
			\\
			Linear & $x$ &&
			Laplace & $\frac{1}{2b} e^{\frac{-|x - \mu|}{b}}$
			\\  
			Polynomial & $x^n$ &&
			Chi-squared & $\frac{1}{2^{\frac{k}{2}} \Gamma(\frac{k}{2})} x^{\frac{k}{2} - 1} e^{\frac{-x}{2}}$
			\\
			Exponential & $\lambda e^{-\lambda x}$ &&
			Student-T & $c {(1 + \frac{x^2}{\nu})}^{- \frac{\nu + 1}{2}}$
		\end{tabular}
		\label{table:distributions}
		\vspace{-10pt}
	\end{table}
	
	In this paper we develop a new inference algorithm for hybrid probabilistic programs via {\em discretization}: we convert the continuous distributions in a hybrid program to discrete distributions. This yields a fully discrete probabilistic program on which existing algorithms for discrete inference can be used.  Discretization approximates a continuous distribution as a sequence of {\em intervals}, with each interval associated with the probability of the value falling in that interval.  Forms of discretization have been used in prior work~\cite{https://doi.org/10.48550/arxiv.2204.02948, huang2021aqua, Albarghouthi2017_1, 10.1145/2491411.2491423} but they all scale linearly in the number of intervals.  This imposes a clear tradeoff: one needs many small intervals in order to avoid losing too much precision, but the cost of inference quickly becomes prohibitive as the number of intervals grows.

We introduce a new approach to discretization that we call \emph{bit blasting}, by analogy with the technique of the same name in verification~\cite{5361322}.	The key property of a bit blasted discretization is that it uses only \(\mathit{poly}(\log{n})\) Boolean random variables to represent a discretization on $n$ intervals. This is achieved by employing a binary representation of numbers and representing discretizations as discrete probabilistic programs over this binary representation.  At first blush, this succinct representation would appear to lose too much accuracy to be a viable strategy, but we present both theoretical and empirical results to the contrary.

First, we prove that a large class of common continuous densities can be bit blasted {\em soundly}, that is with no loss of accuracy versus a na\"ive discretization.  Table~\ref{table:distributions} lists example distributions that are in this class; we refer to the entire class as {\em mixed-gamma} distributions.  

For instance, consider discretizing a continuous uniform distribution between 0 and 1. Na\"ive discretization to \(2^{32}\) intervals requires enumeration of $2^{32}$ values. Instead, this distribution can be represented in binary as a tuple of 32 Bernoulli random variables of the form {\tt flip$(0.5)$}, i.e., coin tosses that are equally likely to have the value 0 or 1. This observation is not new, and a similar result holds for exponential distributions as well~\cite{10.1214/aoms/1177693058}. However, the bit blasted discretizations of other mixed-gamma densities are novel.  Further, unlike the case for uniforms and exponentials, these discretizations are not defined as simply a tuple of independent Bernoulli random variables but rather require full-fledged discrete probabilistic programs over such variables.

A succinct representation does not necessarily imply efficient inference, which is hard in general.  As our second contribution, however, we prove that bit blasted mixed-gamma distributions are not only sound and succinct, but they also support polynomial-time inference in the number of bits of precision. Specifically, we prove that the \emph{knowledge compilation} approach to discrete probabilistic inference~\cite{holtzen2020dice, Raedt2007, Fierens2015, Chavira2006, Chavira2008}, which reduces inference to weighted model counting on a boolean formula, has this property for bit blasted mixed-gamma distributions. 
Therefore, the process of bit blasting a mixed-gamma distribution followed by discrete inference via knowledge compilation is guaranteed to take time that is polynomial in the bitwidth used for discretization. 

	Third, we have used the above theoretical results to design a new probabilistic programming system called \hybit{} that performs non-stochastic approximate inference for hybrid probabilistic programs via bit blasting. \hybit{}%
	is a discrete probabilistic language that includes support for fixed-point binary numbers and associated arithmetic operations, and it is implemented as an embedded domain specific language in Julia~\cite{Julia-2017}. The \hybit{} API allows users to produce bit blasted fixed-point approximations of arbitrary continuous distributions. Mixed-gamma distributions can be represented by their sound bit blasted discrete distributions. For other distributions the API enables users to employ piece-wise discrete approximations, where each piece is itself a bit blasted mixed-gamma distribution.

	\hybit{} leverages knowledge compilation to perform exact inference on the given discrete probabilistic program. In the worst case, the inference for an arbitrary hybrid probabilistic program via bit blasting can be exponential in the number of bits, but we empirically demonstrate the benefits of our approach. We show that \hybit{} performs better than existing sampling-based and symbolic inference approaches on a comprehensive benchmark suite of hybrid probabilistic programs.
	
	Overall, we present the following contributions in this paper:
	\begin{enumerate}
		\item We motivate the challenges for inference on hybrid probabilistic programs in Section~\ref{sec:motivation}.
		\item We present a new form of discretization called bit blasting that is characterized by its succinctness in the number of discrete intervals in Section~\ref{sec:lexbit}.
		\item We present a class of continuous distributions, namely mixed-gamma distributions, for which a sound bit blasted representation exists. We formalize this construction and prove its properties in Section~\ref{sec:lexbit}. We further prove that knowledge compilation based inference scales polynomially in the bitwidth for these distributions.
		\item We describe the \hybit{} PPL and its new inference algorithm via bit blasting in Section~\ref{sec:hybit}.
		\item In Section~\ref{sec:experiments}, we empirically compare \hybit{} with other PPLs on benchmarks obtained from the existing literature. We also characterize the behavior of \hybit{} with respect to its hyperparameters, i.e.\ number of bits and pieces. 
	\end{enumerate}
	
	\noindent HyBit is available at \href{https://github.com/Tractables/Dice.jl/tree/hybit}{https://github.com/Tractables/Dice.jl/tree/hybit}.

	\section{Motivating Examples} \label{sec:motivation}
	This section motivates the challenge of inference for hybrid probabilistic programs using three examples. First, we present an example from computational biology with inherent logical structure. Next, we show an example from the literature with a multimodal posterior arising due to discrete control flow. Finally, we show an example of low probability observations through conjugate Gaussians. We then investigate the performance of various inference algorithms, including \hybit{}. The examples demonstrate the advantages of \hybit{} over other approximate inference algorithms. We provide a detailed comparison with several approximate and exact baselines in Section~\ref{sec:experiments}.

	\subsection{Logical Structure}\label{sec:gene}
	We present a simplified example from computational biology which relates genetic expression with blood sugar levels. Figure~\ref{fig:gene} shows the probabilistic program, where the task is to get the updated belief of a gene's occurrence in a patient given their blood sugar levels. 

	The first four lines of the probabilistic program in Figure~\ref{fig:gene} use a beta distribution as the prior probability of each of $T$ genes occurring in the general population.  The syntax \texttt{flip}$(\theta)$ denotes a Bernoulli random variable with success probability $\theta$. On line~\ref{line:collectgene}, the program uses the syntax \texttt{reduce(|, gene)} to denote the expression \(\bigvee_{i = 1}^{T} \texttt{gene[i]}\).  In other words, the patient is considered to have diabetes if at least one of the genes is expressed.  What follows is multiple readings of the patient's blood sugar level. For each reading a random variable first defines the blood sugar depending on whether they have diabetes.   Then we use the syntax \texttt{observe} (\(y, v\)) to condition on the random variable \(y\) having the value \(v\) --- in the program this is used to condition on actual blood-sugar readings from the patient.  Finally, on line~\ref{line:expectation}, the program queries for the expectation of the posterior distribution of the occurrence of the first gene.

	\begin{wrapfigure}{r}{0.53\textwidth}
		\begin{lstlisting}[mathescape=true, escapechar=&]
&\label{line:loop}&for i in 1:T
&\label{line:belief}&    gene_occurrence[i] = beta(1, 1)
&\label{line:othergene}&    gene[i] = flip(gene_occurrence[i])
&\label{line:loopend}&end
&\label{line:collectgene}&diabetes = reduce(|, gene)
&\label{line:person1}&blood_sugar1 = if diabetes normal(80, 2) 
						else normal(135, 2) end
&\label{line:data1}&observe(blood_sugar1, 79)
&\label{line:person2}&blood_sugar2 = if diabetes normal(80, 2) 
						else normal(135, 2) end
&\label{line:data2}&observe(blood_sugar2, 136)
&\label{line:expectation}&return Expectation(gene_occurrence[1])
		\end{lstlisting}
		\caption{Example 1: Gene Expression}\label{fig:gene}
	\end{wrapfigure}

\begin{wrapfigure}[16]{r}{0.45\textwidth}
		\centering
		\includegraphics[width=0.45\textwidth]{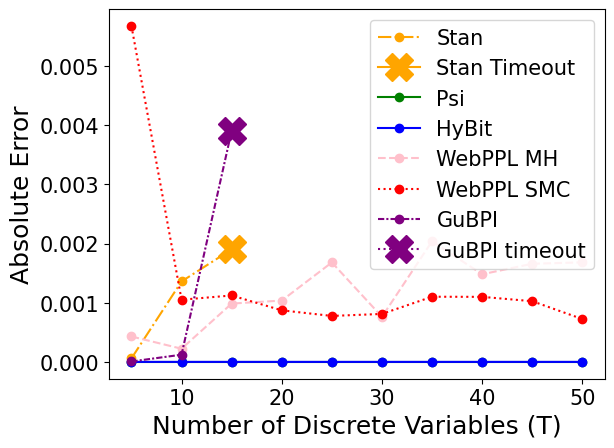}
		\caption{Scaling on Logical Constraints. \hybit{} scales to 50 genes with the least absolute error. The graph for Psi and HyBit overlap closely.}\label{fig:logical_plot}
	\end{wrapfigure}

	Figure~\ref{fig:logical_plot} shows the results of inference with a timeout of 20 minutes on this program using different inference algorithms, as the number of genes (T) increases. Stan uses Hamiltonian Monte Carlo, which does not directly support discrete random variables.  Instead, they are marginalized out, either manually or automatically using variable elimination~\cite{gorinova2019automatic}. As shown in the figure, Stan times out when there are more than 15 genes --- as its strategy scales exponentially in T. The same issue of exponential blowup plagues GuBPI~\cite{https://doi.org/10.48550/arxiv.2204.02948}, which employs a combination of symbolic evaluation and discretization to find upper and lower bounds.  As the figure shows, the universal sampling methods MCMC with a Metropolis Hastings kernel (WebPPL MH) and Sequential Monte Carlo (WebPPL SMC) can scale and provide reasonable accuracy. Exact inference strategy Psi~\cite{gehr2016psi} also scales well with the number of discrete random variables when the program is written to avoid a large discrete state space. AQUA discretizes hybrid programs~\cite{huang2021aqua} but does not support this program.
	
	Our system and approach, \hybit{}, scales to 50 genes and has the least absolute error among approximate inference algorithms. 	The program, when written in \hybit{}, is represented by its discrete bit-level abstraction. The user, while writing the program can employ the \hybit{} API to replace all continuous distributions (specifically on Lines~\ref{line:belief},~\ref{line:person1},~\ref{line:person2} in Figure~\ref{fig:gene}) with their bit blasted discrete approximations. As a result, we now have a discrete program with distributions over Boolean and fixed-point numbers. Note that now \texttt{observe} on Line~\ref{line:data1} conditions on the fact that the value of the discrete distribution for \texttt{blood\_sugar1} lies in the interval corresponding to the value \texttt{79}. 
	
	In the experiment shown we use a bitwidth of 25 --- each continuous distribution is discretized as a program over a tuple of 25 bits interpreted as a fixed-point number. Further, we approximate these continuous distribution using 4096 pieces, each of which is a bit blasted exponential distribution. Naive discretization with 25 bits would be prohibitively slow, as it yields $2^{25}$ intervals i.e.\ \(134\)M intervals. However, our bit blasted program only uses \(53\)K coin flips (Boolean random variables) to represent them. Moreover, knowledge compilation based inference~\cite{holtzen2020dice, Fierens2015}  automatically identifies and exploits conditional independences in the program's logical structure and helps to scale inference. More details of this experiment can be found in the appendix.

	\subsection{Handling Multimodality}
		
	This section presents an example of a multimodal distribution to highlight another challenge for inference on hybrid probabilistic programs. Multimodal distributions have multiple peaks separated by low probability regions. These distributions commonly emerge in various applications such as sensor network localization, cosmology and many more~\cite{Tak_2018, Shaw_2007}. We adapt an example from the existing literature~\cite{yao2021stacking}, as shown in Figure~\ref{fig:multimodal}.

\begin{wrapfigure}{r}{0.53\textwidth}
	\begin{lstlisting}[mathescape=true, escapechar = &]
mu1 = uniform(-16, 16)
mu2 = uniform(-16, 16)
&\label{line:mm:datapts}&datapts = [5, 5, 5, 5, 5, 5, -5, -5, -5] 
for data in datapts
	y = if flip($\frac{2}{3}$) normal(mu1, 1) 
		  else normal(mu2, 1) end
	observe(y, data)
end
return mu1
	\end{lstlisting}
	\caption{Example 2: Yao-Vehtari-Gelman model}\label{fig:multimodal}
\end{wrapfigure}

	The probabilistic program shown in Figure~\ref{fig:multimodal} is hard for existing probabilistic inference approaches. The \texttt{datapts} on Line~\ref{line:mm:datapts} has nine entries, with two-thirds being \(5\) and the other one-third being \(-5\). This leads to the posterior for \((\mu_1, \mu_2)\) being bimodal around \((5, -5)\) and \((-5, 5)\). As the number of data points increases with the same proportion, the posterior of \((\mu_1, \mu_2)\) converges to \((5, -5)\). However, in the presence of limited data points, the posterior for \(\mu_1\) is bimodal around \(5\) and \(-5\).

	\begin{figure}[t]
		\centering
		\begin{tabular}[t]{cccc}
			\begin{subfigure}[t]{0.22\textwidth}
				\includegraphics[width=\textwidth]{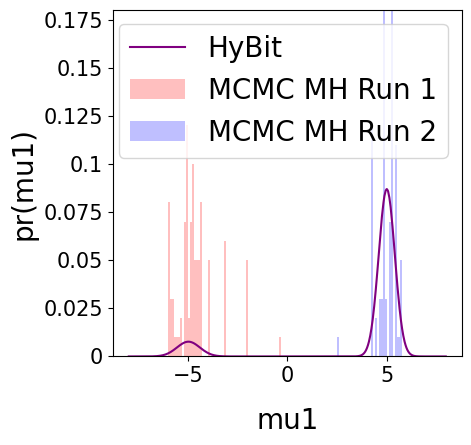}
				\caption{\label{fig:mm_mh}}
			\end{subfigure}\quad
			\begin{subfigure}[t]{0.22\textwidth}
				\includegraphics[width=\textwidth]{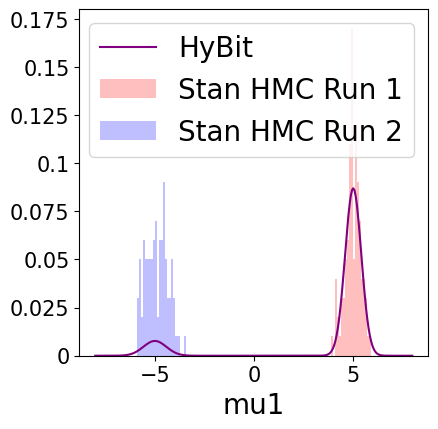}\caption{\label{fig:mm_hmc}}
			\end{subfigure}&
			\begin{subfigure}[t]{0.22\textwidth}
				\includegraphics[width=\textwidth]{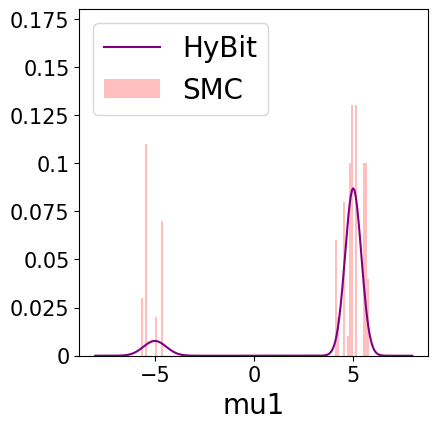}\caption{\label{fig:mm_smc}}
			\end{subfigure}&
			\begin{subfigure}[t]{0.22\textwidth}
				\includegraphics[width=\textwidth]{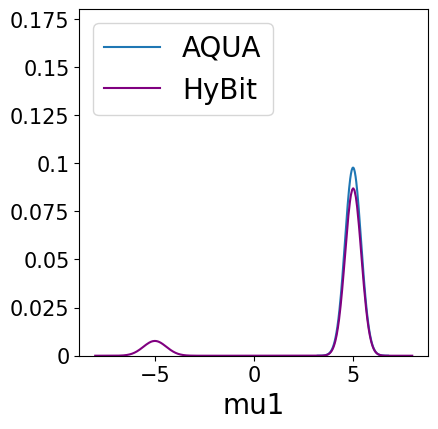}
				\caption{\label{fig:mm_aqua}}
			\end{subfigure}
		\end{tabular}    
		\caption{Posterior from different baselines compared with that of \hybit{} for Figure~\ref{fig:multimodal}. \hybit{} and WebPPL SMC are able to identify both the modes in the posterior distribution. (a) Different runs of WebPPL MCMC with a Metropolis Hastings kernel converge to different modes. (b) Different runs of Stan HMC converge to different modes. (c) Different runs of WebPPL SMC are able to find both the modes. (d) AQUA with its adaptive interval strategy only finds the more probable mode}\label{figure8}
	\end{figure}
	
	The existence of multiple modes challenges sampling-based algorithms as they tend to get stuck in one of the modes. Specifically, WebPPL using MCMC with a Metropolis Hastings kernel and both Stan and WebPPL using HMC end up arbitrarily in one of the modes and fail to explore the other mode. Figures~\ref{fig:mm_mh} and ~\ref{fig:mm_hmc} show the results obtained using WebPPL MCMC and Stan HMC respectively, where two different runs get stuck in two different modes. On the other hand, \hybit{} performs exact inference on its discrete abstraction and so explores the distribution globally, allowing it to identify both modes.  Sequential Monte Carlo (SMC) (Figure~\ref{fig:mm_smc}), Psi and GuBPI also explore the distribution globally and thus, are able to find both modes. Finally, to address the computational challenge of direct discretization, AQUA adapts its discretizing intervals to focus on high probability regions, and ends up only identifying the higher probability mode (Figure~\ref{fig:mm_aqua}).
	
	\subsection{Handling Low Probability Observations}
		\begin{wrapfigure}{r}{0.4\textwidth}
		\begin{lstlisting}[mathescape=true, escapechar=|]
|\label{line:mu}|mu = normal(0, 1)
|\label{line:observe1}|observe(normal(mu, 1), 8)
|\label{line:observe2}|observe(normal(mu, 1), 9)
|\label{line:posterior}|return mu
		\end{lstlisting}
		\caption{Example 3: Conjugate Gaussians}\label{fig:conj_gauss}
	\end{wrapfigure}
	This section presents conjugate Gaussians (Figure~\ref{fig:conj_gauss})  with low probability observations. In Figure~\ref{fig:conj_gauss}, the posterior distribution for random variable \texttt{mu} is queried after conditioning on low probability data on Lines~\ref{line:observe1} and~\ref{line:observe2}.

	Why are low probability observations hard? Intuitively, general-purpose sampling algorithms begin sampling from the prior distribution and struggle to find samples with considerable weight. 
	Only after a very large number of samples do these algorithms manage to sample from the true posterior. On the other hand, \hybit{} explores the domain of the posterior distribution globally, via exact inference on a bit blasted abstraction, and so it is insensitive to this issue.
	
	\begin{wrapfigure}{l}{0.58\textwidth}
		\includegraphics[width = 0.58\textwidth]{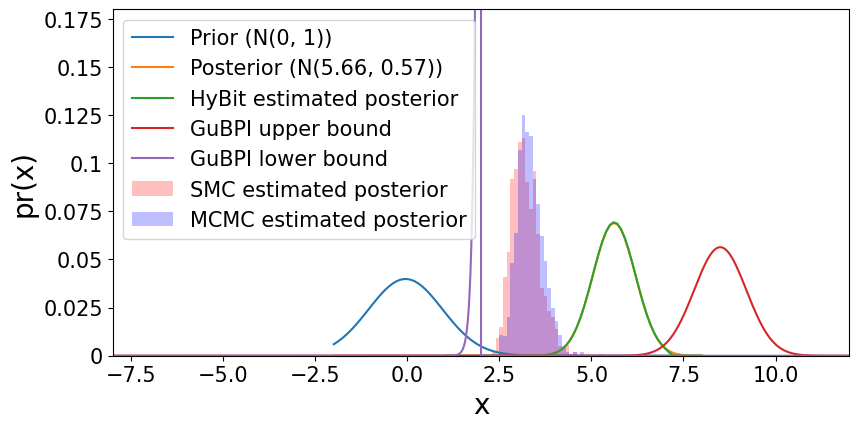}
		\caption{Results for Example 3. The \hybit{} estimated posterior overlaps closely with the true posterior distribution.}\label{fig:conjugate}
	\end{wrapfigure}
		
	Figure~\ref{fig:conjugate} plots the true prior and posterior distributions along with results from different inference algorithms. For the sampling-based algorithms MCMC with a Metropolis Hastings kernel and SMC, we obtained and plotted 1000 samples after running the corresponding WebPPL~\cite{dippl} program. The importance sampling algorithm was not able to obtain any sample with non-negligible weight for this program. The samplers are shifting the posterior towards the true posterior distribution but require many more samples to achieve that. Even after sampling about 16M and 65K samples respectively, the expectation of the samples obtained from MCMC and SMC have absolute error of 0.549798 and 1.520776. GuBPI reports upper and lower bounds on the probability for each interval and incurs an absolute error of 2.33. On the other hand, the posterior distribution from \hybit{} overlaps perfectly with the ground truth. Stan HMC handles low probability observations well and obtains high accuracy. Psi also obtains the exact symbolic expression for the posterior distribution. Finally, the mean of AQUA's reported posterior had an error of 5.66 as it fails to make any update to the prior of \(\mu\).

\section{Bit Blasting: Key Insights}\label{sec:lexbit}

To scale inference on hybrid probabilistic programs with respect to discrete structure, we need an algorithm that treats discreteness as first class, and that discretizes away continuous structure.
This section defines the semantic notion of bit blasting and sets it up as a special case of discretization with desirable properties. Then, we provide bit blasting functions for common classes of continuous distributions. We provide discretization techniques that are sound (accurate up to \(b\) bits), succinct, and amenable to efficient inference.

\subsection{Discretization and Bit Blasting}

In the standard terminology of probability theory~\cite{Rosenthal2006}, a probability space \((\Omega, \Sigma, \mu)\) consists of a sample space \(\Omega\),  a \(\sigma\)-algebra on \(\Omega\) denoted  \(\Sigma\), and a probability measure on \(\Sigma\) denoted~\(\mu\).
In a general sense, a discretization function takes as input such a probability space \((\Omega, \Sigma, \mu)\) and outputs a discrete probability space \((\Omega_D, \Sigma_D, \mu_D)\) where \(\Omega_D\) is a countable set.

We will study a more specific notion of discretization: one that takes as input a continuous distribution over a finite interval and outputs a discrete distribution over \(2^b\) points for some number of bits $b$. 
Formally, let \([l, u)\) be an interval with \(l, u \in \mathrm{R}\). 
For the input probability space, we use \(\mathcal{B}([l, u))\) to denote the Borel \(\sigma\)-algebra of subsets of interval \([l, u)\)~\cite{Rosenthal2006}. 
For the output probability space, we write \(\mathcal{P}(S)\) to refer to the power set (a \(\sigma\)-algebra) of set \(S\). Moreover, we will assume the sample space to be discretized as follows. 
\begin{definition}[$b$-bit interval]\label{def:discretize-unit-interval}
	A $b$-bit interval $[l, u]_b$ is the set of points obtained by dividing \([l, u)\) into \(2^b\) intervals:
	\([l, u]_b = \{r~|~r2^b \in \mathrm{Z}, r \in [l, u)\}\)
\end{definition}

\noindent We are now ready to define the notion of discretization function used in this paper.

\begin{definition}[$b$-bit discretization function] 
A $b$-bit discretization function takes as input a probability space \(([l, u), \mathcal{B}([l, u)), \mu)\), a bit width $b \in \mathrm{Z}^+$ and outputs a discrete probability space \(([l, u]_b, \mathcal{P}([l, u]_b), \mu_D)\) for some measure $\mu_D$.
\end{definition}

\begin{example}\label{example:discretization}
	Let \(\pi\) be a uniform distribution on the unit interval, described by the probability space \(([0, 1), \mathcal{B}([0, 1)), \mu)\) such that the probability density function specified by \(\mu\) is \(1\) on the unit interval.
	Consider a function \(d_1\) that takes \(\pi\) as input and outputs the probability space \((S, \mathcal{P}(S), \mu_D)\) where  \(S = \{0, 0.25, 0.5, 0.75\}\) and \(\mu_D\) is a probability measure on \(\mathcal{P}(S)\). Then \(d_1\) is a \(2\)-bit discretization function. As a concrete example \(\mu_D\) can be defined as \(\{0 \rightarrow 0.1, 0.25 \rightarrow 0.2, 0.5 \rightarrow 0.3, 0.75 \rightarrow 0.4\}\)\footnote{We define discrete probability measures \(\mu_D\) using a mapping from single elements in the sample space \(S\) to their probabilities, which can then be used to compute \(\mu_D\) for any set in \(\mathcal{P}(S)\).}.
\end{example}

As the example shows, one can come up with any arbitrary \(b\)-bit discretization function. To qualify them further, we need a notion of accuracy --- soundness up to \(b\) bits defined as follows.

\begin{definition} [soundness of $b$-bit discretization function]\label{soundness}
	For any integer \(b > 0\), a \(b\)-bit discretization function is \(b\)-sound for a particular input probability space \(([l, u), \mathcal{B}([l, u)), \mu)\) if it outputs a discrete probability space \(([l, u)_b, \mathcal{P}([l, u]_b), \mu_D)\) such that the following holds:
	\[\forall x \in {[l, u]}_b \;\;\;\;\;\; \int_{x}^{x + \frac{1}{2^b}} d \mu(y) = \mu_D(\{x\})\]
\end{definition}

\begin{example}\label{example:sound}
	Let \(d_2\) be a \(b\)-bit discretization function that takes \(\pi\) (as defined in Example~\ref{example:discretization}) as input and outputs the probability space \((S, \mathcal{P}(S), \widetilde{\mu_D})\) where \(S = \{0, 0.25, 0.5, 0.75\}\) and \(\widetilde{\mu_D}\) can be defined as \(\{0 \rightarrow 0.25, 0.25 \rightarrow 0.25, 0.5 \rightarrow 0.25, 0.75 \rightarrow 0.25\}\). Then \(d_2\) is a sound \(2\)-bit discretization function for the probability space \(\pi\).
\end{example}

Before we define a \(b\)-bit blasting function, we need to fix a generic representation of discrete probability distributions. To that purpose, we define the concept of a discrete probabilistic closure, akin to probabilistic Turing machines~\cite{arora2006computational}.

Each probabilistic closure is a deterministic function from a set of biased coin flips to a discrete set. This induces a probability distribution on the output of the function through probabilities associated with coin flips. It also consists of an accepting Boolean formula that handles observations and limits the set of values that input flips can take. We provide a formal definition below:
\begin{definition}[discrete probabilistic closure]\label{def:discrete-prob-closure}
	A discrete probabilistic closure is defined as the tuple $(\varphi, \gamma, w)$ where $w$ is a vector of Boolean random variables (or biased coin flips), $\varphi$ is a deterministic function from \(\{\true, \false\}^{|w|}\) to a finitely countable set S and $\gamma$ is a deterministic Boolean formula over variables in $w$.
	
	The semantics of a discrete probabilistic closure, i.e. $\langle(\varphi, \gamma, w)\rangle$ defines a probability space \((S, \mathcal{P}(S), \mu)\) such that
	\[\forall T \subseteq S, \mu(T) = \frac{\mathrm{E}_{w}((\varphi(w) \in T) \wedge \gamma)}{\mathrm{E}_{w}(\gamma)} \;\;\;\;\;\; \mathit{ if } \; \mathrm{E}_{w}(\gamma) \neq 0 \]
	
\end{definition}

\begin{example}~\label{example:naive}
From Example~\ref{example:sound}, consider \(\widetilde{\mu_D} = \{0 \rightarrow 0.25, 0.25 \rightarrow 0.25, 0.5 \rightarrow 0.25, 0.75 \rightarrow 0.25\}\). \(\widetilde{\mu_D}\) can be represented using a discrete probabilistic closure using the function \texttt{na\"ive\_unif}, the weight function \(w\) and the accepting Boolean formula \(\gamma\) as follows:
\[\gamma = \true \;\;\;\;\;\;\;\;\;\;\;\;\;\;\;\;\;\; w = [f_0 \rightarrow \Lflip{\frac{1}{4}}, f_1 \rightarrow \Lflip{\frac{1}{3}}, f_2 \rightarrow \Lflip{\frac{1}{2}}]\]
Now, one can calculate the probability that \texttt{na\"ive\_unif} returns 0.5 as follows:
\[\mathit{Pr}(\texttt{val} = 0.5) = \mathit{Pr}(\neg f_0) \mathit{Pr}(\neg f_1) \mathit{Pr}(f_2) = (1 - 0.25) (1 - 0.33) 0.5 = 0.25\]
\end{example}

\begin{wrapfigure}[9]{l}{0.38\textwidth}
		\begin{lstlisting}[mathescape=true, escapechar=|]
function |\texttt{na\"ive\_unif}|(|\(f_0, f_1, f_2\)|)
	val = if |\(f_0\)| then 0
	else if |\(f_1\)| then 0.25
	else if |\(f_2\)| then 0.5
	else 0.75
	return val 
		\end{lstlisting}
	\end{wrapfigure}
Note that as explained by Example~\ref{example:naive}, any discrete distribution \(\mu_D\) over \(2^b\) values can be represented as a discrete probabilistic closure \((\varphi, \gamma, w)\) where \(|w| = 2^b - 1\). 

Dice~\cite{holtzen2020dice} and Problog~\cite{Fierens2015} are examples of PPLs that directly fit into the paradigm of a discrete probabilistic closure.

We want a discrete probabilistic closure to be more succinct --- of size polynomial in the number of bits of precision, \(b\). To this purpose, we define a bit blasting function.

\begin{definition}[$b$-bit blasting function]\label{def:bitblasting}
	A $b$-bit blasting function ${[.]}_b$ is a $b$-bit discretization function that outputs a discrete probabilistic closure $(\varphi, \gamma, w)$ that uses a number of Boolean random variables that is polynomial in the number of bits $b$, that is, \(|w| \in O(\text{poly}(b))\)
\end{definition}

It follows from Definitions~\ref{soundness}, \ref{def:discrete-prob-closure} and~\ref{def:bitblasting} that for an integer \(b > 0\), a \(b\)-bit blasting function is sound for a given probability space \(([l, u), \mathcal{B}([l, u)), \mu)\) if
\[\forall x \in {[l, u]}_b \;\;\;\;\;\; \int_{x}^{x + \frac{1}{2^b}}d \mu(y) = \langle [\mu]_b \rangle,\]
\begin{figure}[ht]
	\centering
	\begin{tikzpicture}[every text node part/.style={align=center}, scale=0.8, transform shape]
		\draw (0,0) rectangle (3,1) node[pos=.5] {\(([l, u), \mathcal{B}([l, u]), \mu)\)};
		\draw (8,0) rectangle (11,1) node[pos=.5] {$(\varphi, \gamma, w)$};
		\draw (3,-1.75) rectangle (8,-0.75) node[pos=.5, text width=2.5cm] {\(([l, u]_b, \mathcal{P}([l, u]_b), \mu_D)\)};
		
		\draw [-stealth](3,0.5) -- (8,0.5);
		\draw [-stealth](8,0.5) -- (5.5,-0.75);
		\draw [-stealth](3, 0.5) -- (5.5, -0.75);
		
		\draw(5.5, 1) node{Bit blast($[.]_b$)};
		\draw(7.5, -0.3) node{$\langle . \rangle$};
		\draw(3.5, -0.3) node{\(\int\)};
	\end{tikzpicture}
	\caption{Commutative diagram for a $b$-sound bit blasting function}\label{fig:commute}
\end{figure}
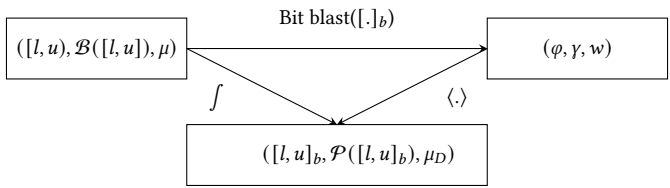

\begin{example}~\label{example:bitblast}
	Again, consider the result of the sound 2-bit discretization function from Example~\ref{example:sound}, that is \(\widetilde{\mu_D} = \{0 \rightarrow 0.25, 0.25 \rightarrow 0.25, 0.5 \rightarrow 0.25, 0.75 \rightarrow 0.25\}\). Measure \(\widetilde{\mu_D}\) can alternatively be represented using a discrete probabilistic closure using the function \texttt{bitblast\_unif}, the weight function \(w\) and the accepting Boolean formula \(\gamma\) as follows:
	\[\gamma = \true \;\;\;\;\;\;\;\;\;\;\;\;\;\;\;\;\;\; w = [f_0 \rightarrow \Lflip{\frac{1}{2}}, f_1 \rightarrow \Lflip{\frac{1}{2}}]\]
	Note that in \texttt{bitblast\_unif}, \(f_0\) and \(f_1\) are being used as bits in the binary representation of the variable \texttt{val}. Now, one can calculate the probability that \texttt{bitblast\_unif} returns 0.5 as follows:
	\[\mathit{Pr}(\texttt{val} = 0.5) = \mathit{Pr}(f_0) \mathit{Pr}(\neg f_1) = (0.5) (1 - 0.5) = 0.25\]
\end{example}

We would like to point out that \texttt{bitblast\_unif} uses only 2 coin flips to represent a distribution \begin{wrapfigure}[7]{l}{0.4\textwidth}
	\begin{lstlisting}[mathescape=true, escapechar=|]
function |\texttt{bitblast\_unif}|(|\(f_0, f_1\)|)
	bit|\(_0\)| = if |\(f_0\)| then 1 else 0
	bit|\(_1\)| = if |\(f_1\)| then 1 else 0
	val = |\(\frac{\texttt{bit}_0}{2} + \frac{\texttt{bit}_1}{4}\)|
	return val 
	\end{lstlisting}
\end{wrapfigure}over 4 values. It can be generalized to use \(b\) coin flips to represent a uniform distribution over \(2^b\) values. On the other hand, \texttt{na\"ive\_unif} uses 3 coin flips and would generalize to use \(2^b - 1\) coin flips. Hence, \texttt{bitblast\_unif} is a viable output for a \(b\)-bit blasting function. In the next section, we provide details for the \(b\)-bit blasting functions for exponential and mixed-gamma distributions.

\subsection{Concrete Bit Blasting Function: Preliminaries}
Next, our goal is to provide a concrete instantiation of a sound bit blasting function for mixed gamma distributions, which have probability density functions as defined below.

\begin{definition}[generalized-gamma density]\label{def:general-gamma}
	Given parameters \(\alpha \in \mathrm{Z}^+\) and \(\beta \in \mathrm{R}\), a generalized-gamma density \(\pdf{\alpha}{\beta}\) is a probability density function over the interval \([0, 1)\) of the form
	\[\pi_{\alpha, \beta}(x) = \frac{x^{\alpha} e^{\beta x}}{\int_{[0, 1)} y^{\alpha} e^{\beta y} \; dy}.\]
\end{definition}

\begin{definition}[mixed-gamma density]\label{def:mixed-gamma}
	Given a collection of \(N \in \mathrm{Z}^+\) generalized-gamma densities \(\pi_{\alpha_i, \beta_i}\) with their associated weights $a_i \in [0, 1]$ such that \(\sum_{i=1}^{N} a_i = 1\),
	a mixed-gamma density \(\Upsilon\) is a probability density function over the interval \([0, 1)\) of the form
	\[\Upsilon(x) = \sum_{i = 1}^{N} a_i \pi_{\alpha_i, \beta_i}(x).\]
\end{definition}

For notational convenience, we confine the continuous distributions to the unit interval to get discrete distributions over a \(b\)-bit unit interval. We generalize our approach to any finite interval for building the probabilistic programming system \hybit{} based on bit blasting.

To describe our construction of the bit blasting function, we make use of Dice~\cite{holtzen2020dice}. Dice already compiles its programs to weighted Boolean formulas (via the \(\comp\) judgement) that fit the definition of a discrete probabilistic closure.\footnote{Dice compiles to weighted Boolean formulas \((\varphi, \gamma, w)\) where \(\varphi\) outputs (tuples of) Boolean values, \(\gamma\) represents observations in a Dice program and \(w\) consists of weights associated with Boolean variables (biased coin flips)} This allows us to only define a \(\comp_b\) judgment from probability density functions to Dice programs to specify a bit blasting function. Dice also defines a distributional semantics function \(\dbracket{.}_D : \texttt{p} \rightarrow V \rightarrow [0, 1]\) that takes as input a Dice program \texttt{p} and outputs a normalized probability distribution (described as a function from the set of values $V$ to probabilities). We use the function \(\dbracket{.}_D\) to argue about soundness of our construction later. More details about syntax and semantics of Dice can be found in the appendix.

The compilation judgement for mixed-gamma densities are of the form \(\Upsilon \comp_b \texttt{p}\) where \texttt{p} are Dice programs. We further provide the following definitions for \(b\)-equivalence between densities and Dice programs and \(b\)-succinctness of Dice programs. 

\begin{definition}[binarizing function]\label{def:binarize}
	The binarizing function for $b$ bits, $\bin{.}{b}$ takes as input a number $r \in {[0, 1]}_b$ and returns a \(b\)-bit tuple \((v_1, (v_2, (\ldots v_b)\ldots))\) such that
	\(r = \sum_{i = 1}^{b} \frac{v_i}{2^i}\)
\end{definition}

\begin{definition}[b-equivalence]\label{def:b-equivalence}
	A mixed-gamma density \(\Upsilon\) and a Dice program \texttt{p} are \(b\)-equivalent for some \(b \in \mathrm{Z}^+\) if for all \(r \in [0, 1]_b\)
	\[\int_{r}^{r + \frac{1}{2^b}} \Upsilon(y) \; dy = \dbracket{\texttt{p}}_D(\bin{r}{b})\]
\end{definition}

\noindent Note that \(b\)-equivalence is analogous to \(b\)-soundness but is specialized for Dice programs.

\begin{definition}\label{def:flipcount}
	The flip count function \(\flipcount{.}\) takes as input a Dice program \(\prog{}\) and outputs the number of Boolean random variables (coin flips). 
\end{definition}

\begin{definition}
	Compilation \(\comp_b\) is \(b\)-succinct for \(\Upsilon\) if \(\exists k > 0, \forall b \in \mathrm{Z}^+\) such that if \(\Upsilon \comp_b \prog{}\), then \(\flipcount{\prog{}} \leq kb\)
\end{definition}

We define \(b\)-succinctness such that the Dice program \(\prog{}\) employs coin flips linear in the number of bits \(b\). Observe that \(b\)-succinctness imposes a stricter condition than that required by a \(b\)-bit blasting function (which requires \(\mathit{poly}(b)\) coin flips). This implies that if we have a \(b\)-succinct judgment for a mixed-gamma density, then we can have a \(b\)-bit blasting function for that distribution. We describe this in more detail later. 

\subsection{Judgment \(\comp_b\) and bit blasting}

This section describes the rules for the judgement \(\comp_b\). We first describe the rules for an exponential distribution and then move on to generalized gamma distributions. Finally, we describe the rule for mixed-gamma densities. For each of the rules \(\Upsilon \comp_b \prog{}\), we prove the \(b\)-equivalence of the density \(\Upsilon\) and the Dice program \(\prog{}\) and the \(b\)-succinctness of \(\prog{}\). Detailed proofs can be found in the appendix. We also describe how \(\comp_b\) allows us to construct a bit blasting function for mixed-gamma densities.

\subsubsection{Exponential Distribution, \(\pdf{0}{\beta}\)}

Let us first consider the uniform distribution (\(\pdf{0}{0}\)), a special case of an exponential distribution. If we bit blast a uniform distribution using $b$ bits into $2^b$ intervals, we end up with a discrete distribution $D_b$ over \([0, 1]_b\) with \(2^b\) discrete points each having probability $\frac{1}{2^b}$. A straightforward discretization strategy enumerates $2^b$ values using $2^b-1$ coin flips. But the same can be achieved using a tuple of $b$ bits, where each bit is an unbiased coin \texttt{flip}$(0.5)$. 

The strategy to bit blast a uniform distribution using its binary representation works because of the independence between binary digits. The same strategy can be extended to general exponential distributions as well. This fact was shown in a classic paper in the statistics literature~\cite{10.1214/aoms/1177693058}. We formalize that idea using the following rule.
\begin{definition}\label{def:flip-parameter}
	The function \texttt{flip\_param}\(:\mathrm{R} \times \mathrm{Z}^+ \rightarrow [0, 1]\) is defined \(\texttt{flip\_param}(\beta, b) = \frac{e^{\frac{\beta}{2^b}}}{1 + e^{\frac{\beta}{2^b}}}\)
\end{definition}
\begin{align*}
	\inference[]
	{
		\texttt{fresh \(\yvar{1}\), \(\yvar{2}\), \ldots \(\yvar{b}\)}
		\\
		\texttt{flip\_param}(\beta, 1) =\theta_1
		\;\;\;\; \texttt{flip\_param}(\beta, 2) = \theta_2 \;\;\;\;
		\ldots
		\;\;\;\; \texttt{flip\_param}(\beta, b) = \theta_b
	}
	{ \pdf{0}{\beta} \comp_b 
		\begin{array}{l}
			\texttt{let \(\yvar{1}\) = flip($\theta_1$) in}
			\\ \texttt{let \(\yvar{2}\) = flip($\theta_2$) in}
			\\ \ldots
			\\ \texttt{let \(\yvar{b}\) = flip($\theta_b$) in}
			\\ \texttt{(\(\yvar{1}\), (\(\yvar{2}\), (\ldots, \(\yvar{b}\)))\ldots))}
		\end{array}
	} \label{Trans-expo0}~\tag{Expo0}
\end{align*}
We prove the \(b\)-equivalence of the density  \(\pdf{0}{\beta}\) and the Dice program in rule~\ref{Trans-expo0} and \(b\)-succinctness of the latter in the following lemmas. It is more straightforward to see the succinctness of the Dice program --- it uses only \(b\) coin flips.
 
\begin{lemma}\label{lemma:alpha-0-density}
	\(\forall b \in \mathrm{Z}^+, \beta \in \mathrm{R}\), \texttt{p}, if \(\pdf{0}{\beta} \comp_b \texttt{p}\), then \(\pdf{0}{\beta}\) and \texttt{p} are \(b\)-equivalent.
\end{lemma}

\begin{lemma}\label{succinct-alpha-0}
	\(\forall \beta \in \mathrm{R}, \comp_b\) is \(b\)-succinct for \(\pdf{0}{\beta}\) 
\end{lemma}

We provide detailed proofs of the above lemmas in the appendix. For rest of the paper, we consider exponential distributions as primitives to build other distributions, since these are the only ones that enjoy the property of independent bits. But, sound bit blasting functions are still possible for other distributions, as we show next.

\subsubsection{Gamma Distribution \(\pdf{1}{\beta}\)} 

To come up with a sound bit blasting function for \(\pdf{1}{\beta}\), we present a key mathematical insight. 
Consider the program in Figure~\ref{fig:gammaprogram}. Continuous random variables \texttt{X} and \texttt{Y} have a uniform (\(\pdf{0}{0}\)) and exponential (\(\pdf{0}{\beta}\)) distribution respectively. It returns the new distribution of \texttt{X} after conditioning on the inequality \texttt{Y < X}. It turns out that the posterior distribution is a specific gamma distribution \(\pdf{1}{\beta}\). We show the resulting calculation below, where \textit{pdf} refers to the probability density function.
\begin{flalign*}
	\mathit{pdf}(X | Y < X) &\propto \int_{y = 0}^{1} \mathit{pdf}(Y) \cdot \mathit{pdf}(X) \cdot \mathds{1}(Y < X) \; dy = \int_{y = 0}^{x} 1 \cdot e^{\beta x} \; dy \propto xe^{\beta x}  \tag{1}
\end{flalign*}
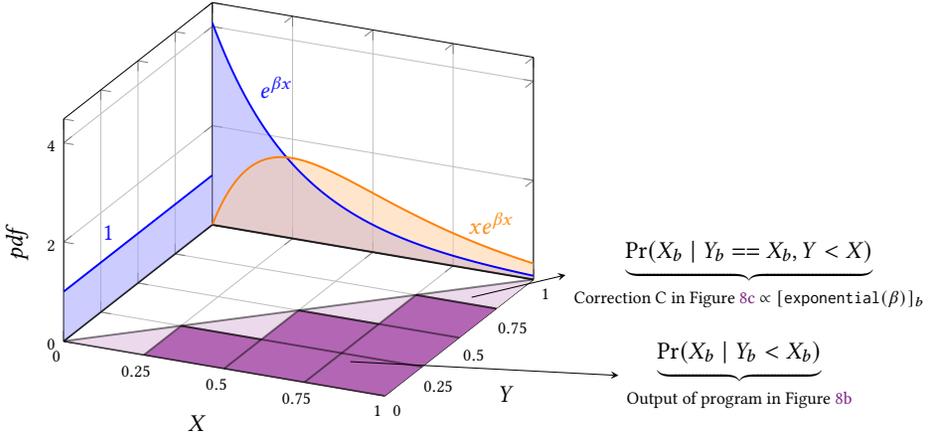
\begin{figure}[t]
	\begin{subfigure}[t]{0.28\linewidth}
		\centering
		\begin{lstlisting}[mathescape=true]
Y ~ uniform(0, 1)
X ~ exponential($\beta$)
observe(Y < X)
return X
		\end{lstlisting}
		\caption{
			Continuous probabilistic program for density~\(xe^{\beta x}\) 
		}\label{fig:gammaprogram}
	\end{subfigure}
	\quad
	\begin{subfigure}[t]{0.32\linewidth}
		\centering
	\begin{lstlisting}[mathescape=true]
Y$_b$ ~ [uniform(0, 1)]$_b$
X$_b$ ~ [exponential($\beta$)]$_b$
observe(Y$_b$ < X$_b$)
return X$_b$
	\end{lstlisting}
	\caption{bit blasted probabilistic program analogous to program~\ref{fig:gammaprogram}
	}\label{fig:gammaprogram2}
	\end{subfigure}
	\quad
	\begin{subfigure}[t]{0.32\linewidth}
		\centering
		\begin{lstlisting}[mathescape=true]
Y$_b$ ~ [uniform(0, 1)]$_b$
X$_b$ ~ [exponential($\beta$)]$_b$
C ~ [exponential($\beta$)]$_b$
observe(Y$_b$ < X$_b$)
Z = if flip($\theta$) then C else X$_b$
return Z
		\end{lstlisting}
		\caption{Sound bit blasted probabilistic program for density \(xe^{\beta x}\)
		}\label{fig:gammaprogram3}
	\end{subfigure}
	\qquad
	\begin{subfigure}[b]{\textwidth}
		\begin{center}
		\resizebox{0.9\textwidth}{!}{
		\begin{tikzpicture}
			\begin{axis}[domain=0:3, xtick = {0, 0.25, ..., 1},
				ytick = {0, 0.25, ..., 1},
				grid,
				zmin=0,
				xlabel = $X$, ylabel = $Y$,
				zlabel = \(\mathit{pdf}\),
				ticklabel style = {font = \scriptsize}, clip=false]
				
				\draw[-stealth] (axis direction cs:0.8, 0.2, 0) -- +(axis direction cs:0.65, 0.4, -0.5) node [anchor=west] {\(\underbrace{\Pr({X}_b \mid {Y}_b < {X}_b)}_{\text{Output of program in Figure~\ref{fig:gammaprogram2}}}\)};

				\draw[-stealth] (axis direction cs:0.9, 0.8, 0) -- +(axis direction cs:0.2, 0.2, 0.2) node [anchor=west] {\(\underbrace{\Pr({X}_b \mid {Y}_b == {X}_b, Y < X)}_{\text{Correction C in Figure~\ref{fig:gammaprogram3}} \; \propto \; [\texttt{exponential}(\beta)]_b}\)};

				\addplot3[domain=0:1,samples=80,samples y=0,mark=none,blue, opacity=1,thick, name path=A]({x},{1},{4*exp(-4*x)/(1 - exp(-4))}); 
				\addplot3[domain=0:1,samples=80,samples y=0,mark=none,black, opacity=0.5,thick, name path=B]({x},{1},{0});
				\addplot3[blue!40, opacity=0.5] fill between[of=A and B];
				\node[blue!90] at (axis cs:0.2,1,3){\(e^{\beta x}\)};
				
				\addplot3[domain=0:1,samples=80,samples y=0,mark=none,blue, opacity=1,thick, name path=Q]({0},{x},{1});
				\addplot3[domain=0:1,samples=80,samples y=0,mark=none,black, opacity=0.5,thick, name path=S]({0},{x},{0});
				\addplot3[blue!40, opacity=0.5] fill between[of=Q and S];
				\node[blue!90] at (axis cs:0,0.3,1.5){\(1\)};
				
				\addplot3[domain=0:1,samples=80,samples y=0,mark=none,orange, opacity=1,thick, name path=R]({x},{1},{x*exp(-4*x) / ((1 - 5/exp(4))/16)});
				\addplot3[domain=0:1,samples=80,samples y=0,mark=none,black, opacity=0.5,thick, name path=T]({x},{1},{0});
				\addplot3[orange!40, opacity=0.5] fill between[of=R and T];
				
				\node[orange!90] at (axis cs:0.87,1,1){\(xe^{\beta x}\)};
				
				\addplot3[domain=0.75:1,samples=80,samples y=0,mark=none,black, opacity=0.5,thick, name path=C]({x},{0.75},{0});
				\addplot3[domain=0.75:1,samples=80,samples y=0,mark=none,black, opacity=0.5,thick, name path=D]({x},{0},{0});
				\addplot3[domain=0.75:1,samples=80,samples y=0,mark=none,black, opacity=0.5,thick]({x},{0.5},{0});
				\addplot3[domain=0.75:1,samples=80,samples y=0,mark=none,black, opacity=0.5,thick]({x},{0.25},{0});
				\addplot3[violet!60] fill between[of=C and D];

				\addplot3[domain=0.5:0.75,samples=80,samples y=0,mark=none,black, opacity=0.5,thick, name path=E]({x},{0.5},{0});
				\addplot3[domain=0.5:0.75,samples=80,samples y=0,mark=none,black, opacity=0.5,thick, name path=F]({x},{0},{0});
				\addplot3[domain=0.5:0.75,samples=80,samples y=0,mark=none,black, opacity=0.5,thick]({x},{0.25},{0});
				\addplot3[violet!60] fill between[of=E and F];
				
				\addplot3[domain=0.25:0.5,samples=80,samples y=0,mark=none,black, opacity=0.5,thick, name path=G]({x},{0.25},{0});
				\addplot3[domain=0.25:0.5,samples=80,samples y=0,mark=none,black, opacity=0.5,thick, name path=H]({x},{0},{0});
				\addplot3[violet!60] fill between[of=G and H];
				
				\addplot3[domain=0:0.25,samples=80,samples y=0,mark=none,black, opacity=0.5,thick]({0.25},{x},{0});
				\addplot3[domain=0:0.5,samples=80,samples y=0,mark=none,black, opacity=0.5,thick]({0.5},{x},{0});
				\addplot3[domain=0:0.75,samples=80,samples y=0,mark=none,black, opacity=0.5,thick]({0.75},{x},{0});
				
				\addplot3[domain=0:0.25,samples=80,samples y=0,mark=none,black, opacity=0.5,thick, name path=I]({x},{x},{0});
				\addplot3[domain=0:0.25,samples=80,samples y=0,mark=none,black, opacity=0.5,thick, name path=J]({x},{0},{0});
				\addplot3[violet!15] fill between[of=I and J];
				
				\addplot3[domain=0.25:0.5,samples=80,samples y=0,mark=none,black, opacity=0.5,thick, name path=K]({x},{x},{0});
				\addplot3[domain=0.25:0.5,samples=80,samples y=0,mark=none,black, opacity=0.5,thick, name path=L]({x},{0.25},{0});
				\addplot3[violet!15] fill between[of=K and L];
				
				\addplot3[domain=0.5:0.75,samples=80,samples y=0,mark=none,black, opacity=0.5,thick, name path=M]({x},{x},{0});
				\addplot3[domain=0.5:0.75,samples=80,samples y=0,mark=none,black, opacity=0.5,thick, name path=N]({x},{0.5},{0});
				\addplot3[violet!15] fill between[of=M and N];
				
				\addplot3[domain=0.75:1,samples=80,samples y=0,mark=none,black, opacity=0.5,thick, name path=O]({x},{x},{0});
				\addplot3[domain=0.75:1,samples=80,samples y=0,mark=none,black, opacity=0.5,thick, name path=P]({x},{0.75},{0});
				\addplot3[violet!15] fill between[of=O and P];
								
			\end{axis}
		\end{tikzpicture}}
		\end{center}
		\caption{
			Sound \(b\)-bit blasting of \(xe^{\beta x}\). Prior densities for X and Y (shown in blue) when conditioned on {\(Y < X\)} (shown in violet) returns the posterior for X (shown in orange).
		}\label{fig:gammagrid}
	\end{subfigure}
	\caption{Key insight in bit blasting \(xe^{\beta x}\) probability density. \([.]_b\) refers to discretization of a continuous density into \(2^b\) intervals and \texttt{X}\(_b\) refers to the discretization of X. }
\end{figure}
What happens if we discretize the program in Figure~\ref{fig:gammaprogram} using \(b\) bits? We get the program in Figure~\ref{fig:gammaprogram2} where each continuous random variable has been replaced with its bit blasted counterpart (\texttt{X} replace by \(\texttt{X}_b\) and so on). We have already seen that for uniform and exponential distributions, \(b\)-equivalent Dice programs exist. But what about the other constructs? As Figure~\ref{fig:gammagrid} demonstrates, \texttt{observe (Y$_b$ < X$_b$)} incurs error over its continuous counterpart \texttt{observe (X < Y)}. The good news is that we can account for the error as shown by the following equations:
\begin{flalign*}
	 & \Pr({X}_b \mid Y \! < \! X) = \Pr({X}_b, {Y}_b \! < \! {X}_b \mid Y \! < \! X) + \Pr({X}_b, {Y}_b == {X}_b \mid Y \! < \! X)
	\\
	 & \qquad = \!\!\!\!\!\!\! \underbrace{\Pr({X}_b \mid {Y}_b < {X}_b)}_{\text{Output of the discrete program}} \!\!\!\!\!\!\! \cdot \Pr({Y}_b < {X}_b  \mid Y \! < \! X) + 
	\underbrace{\Pr({X}_b \mid {Y}_b == {X}_b, Y \! < \! X)}_{\text{Correction } \propto [\pdf{0}{\beta}]_b} \cdot \Pr({Y}_b == {X}_b \mid Y \! < \! X).
\end{flalign*}
The correction term in the above equations when computed algebraically turns out to be proportional to the exponential density (\(\pdf{0}{\beta}\)). So now, the resulting discrete probabilistic program after correction looks as shown in Figure~\ref{fig:gammaprogram3} where \(\theta = \mathit{Pr}(Y_b == X_b| Y < X)\). 

The rule~\ref{Trans-expo1} and Trans-expo1zero
(in the appendix) captures the above intuition. Here, \(\unifobs{\yvar{}}{b} = \prog{}\) is a helper judgment where \(\prog{}\) constructs a uniform distribution that it conditions through observe on being less than \(\yvar{}\).
\begin{align*}
	\inference[]{
		\texttt{fresh \(\yvar{1}\), \(\yvar{2}\), \(\yvar{3}\)}
		\;\;\;\;\;\;\;\;
		\beta \neq 0
		\\
		\pdf{0}{\beta} \comp_b \prog{1}
		\;\;\;\;\;\;\;\;
		\texttt{unifObs(\(\yvar{1}\), b)} = \prog{3}
		\;\;\;\;\;\;\;\;\; \theta = \frac{(e^{\beta \cdot 2^{-b}}(\beta \cdot 2^{-b} - 1) + 1)(1 - e^{\beta}) }{(1 - e^{\beta \cdot 2^{-b}})(e^{\beta}(\beta - 1) + 1)}
	}
	{\pdf{1}{\beta} \comp_b 
		\begin{array}{l}
			\texttt{let \(\yvar{1}\) = } \prog{1} \texttt{ in }
			\\ \texttt{let \_ = \(\prog{3}\) in}
			\\ \texttt{let \(\yvar{2}\) = } \prog{1} \texttt{ in}
			\\ \texttt{let \(\yvar{3}\) = flip\((\theta)\) in}
			\\ \texttt{if \(\yvar{3}\) then \(\yvar{2}\) else \(\yvar{1}\)}
		\end{array}
	}\label{Trans-expo1}~\tag{Expo1}
\end{align*}
We prove the \(b\)-equivalence and \(b\)-succinctnesss. The proof for \(b\)-equivalence is much more involved but it is easy to see that the Dice program in the above rule uses \(3b + 1\) coin flips: \(b\) coin flips in each occurrence of \(\prog{1}\), \(b\) coin flips in \(\prog{3}\) and \(1\) coin flip in the \(\Lite{}{}{}\) guard to create a mixture.

\begin{lemma}\label{lemma:alpha-1-density}
	\(\forall b \in \mathrm{Z}^+, \beta \neq 0 \in \mathrm{R}\), \texttt{p}, if \(\pdf{1}{\beta} \comp_b \texttt{p}\), then \(\pdf{1}{\beta}\) and \texttt{p} are \(b\)-equivalent.
\end{lemma}

\begin{lemma}\label{succ:alpha-1-density}
	\(\forall \beta \in \mathrm{R}, \comp_b\) is \(b\)-succinct for \(\pdf{1}{\beta}\)
\end{lemma}

\subsubsection{Generalized Gamma Distribution \(\pdf{\alpha}{\beta}\)} 
\begin{wrapfigure}{r}{0.27\textwidth}
	\begin{lstlisting}[mathescape=true]
Y ~ uniform(0, 1)
X ~ f(x)
observe(Y < X)
return X
	\end{lstlisting}
\end{wrapfigure}
The previous subsection shows how conditioning on the inequality (\texttt{Y} \(<\) \texttt{X}) introduces a linear factor to \(\pdf{0}{\beta}\) to obtain \(\pdf{1}{\beta}\). Note that conditioning on \texttt{(Y < X)} introduces a linear factor regardless of what initial probability density X had. That is, if X's initial probability density was $f(x)$, the resulting density of the following probabilistic program would be $xf(x)$. This implies that if we can bit blast $f(x)$, we can bit blast $xf(x)$. We still need to account for the error incurred by \texttt{observe (Y$_b$ < X$_b$)} i.e. $\Pr({X}_b | {Y}_b == {X}_b, Y < X)$. It turns out that the correction term is a mixture of gamma distributions which can be bit blasted soundly as well. We provide the judgement rule and proof of the following lemma in the appendix.

\begin{lemma}\label{lemma:alpha>1-density}
	\(\forall b, \alpha \in \mathrm{Z}^+, \beta \in \mathrm{R}\), \texttt{p}, if \(\pdf{\alpha}{\beta} \comp_b \texttt{p}\), then \(\pdf{\alpha}{\beta}\) and \texttt{p} are \(b\)-equivalent.
\end{lemma}

\begin{lemma}\label{succ:alpha>1-density}
	\(\forall \beta \in \mathrm{R}, \alpha \in \mathrm{Z}^+, \comp_b\) is \(b\)-succinct for \(\pdf{\alpha}{\beta}\)
\end{lemma}

\subsubsection{Mixture of Gamma Distributions \(\sum_{i} a_i \pdf{\alpha_i}{\beta_i}\)}

Since generalized gamma densities \(\pdf{\alpha}{\beta}\) can be bit blasted, mixed gamma densities can be bit blasted as well. One bit blasts each individual generalized gamma density and creates their mixture using \(\Lite{}{}{}\) constructs as follows. 
\begin{align*}
	\inference[]{
		\texttt{fresh \(\yvar{1}\), \(\yvar{2}\), \(\yvar{3}\)}		
		\\
		N > 1
		\;\;\;\;\;\;\;\;\;
		\pdf{\alpha_N}{\beta_N} \comp_b \prog{1}
		\;\;\;\;\;\;\;\;\;
		\sum_{i=1}^{N-1} \frac{a_i}{1 - a_N} \pdf{\alpha_i}{\beta_i} \comp_b \prog{2}
		\;\;\;\;\;\;\;\;\;
		\forall i, a_i \in [0, 1]
		\;\;\;\;\;\;\;\;\;
		\sum_{i = 1}^{N} a_i = 1
	}{
		\sum_{i=1}^{N} a_i \pdf{\alpha_i}{\beta_i} \comp_b
		\begin{array}{l}
			\texttt{let \(\yvar{1}\) = flip(\(a_N\)) in} \\
			\texttt{let \(\yvar{2}\) = \(\prog{1}\) in} \\
			\texttt{let \(\yvar{3}\) = \(\prog{2}\) in} \\
			\texttt{if \(\yvar{1}\) then \(\yvar{2}\) else \(\yvar{3}\)}
		\end{array}
	}\label{Trans-mix}~\tag{Trans-mix}
\end{align*}
We prove the following theorems with details in the appendix.

\begin{theorem}\label{theorem:semantics-preserving}
	\(\forall \Upsilon, b \in \mathrm{Z}^+, \texttt{p}\), if \(\Upsilon \comp_b \texttt{p}\) then \(\Upsilon\) and \texttt{p} are \(b\)-equivalent.
\end{theorem}

\begin{theorem}\label{succ:mixed-gamma}
	\(\comp_b\) is \(b\)-succinct for all mixed-gamma densities \(\Upsilon\)
\end{theorem}

Now that we have specified the rules for judgment \(\comp_b\), we specifically define a sound bit blasting function for all mixed-gamma densities \(\Upsilon\), that is \(\comp \circ \comp_b\).

\begin{theorem}\label{theorem:bitblasting}
	\(\comp \circ \comp_b\) is a sound \(b\)-bit blasting function
\end{theorem}

Earlier work~\cite{holtzen2020dice} defines the judgment \(\comp\) that takes as input a Dice program \(\prog{}\) and outputs a weighted Boolean formula \((\varphi, \gamma, w)\) that aligns with Definition~\ref{def:discrete-prob-closure} of a discrete probabilistic closure. And since \(\comp_b\) is \(b\)-succinct for all mixed-gamma densities by Theorem~\ref{succ:mixed-gamma}, \(\comp\) always outputs a \(w\) with \(\mathit{poly}(b)\) coin flips. Thus, \(\comp \circ \comp_b\) is a \(b\)-bit blasting function. Earlier work~\cite{holtzen2020dice} also proves the correctness of compilation to weighted Boolean formula with respect to the semantics of the Dice program. This fact combined with Theorem~\ref{theorem:semantics-preserving} concludes that \(\comp \circ \comp_b\) is a sound \(b\)-bit discretization function. Detailed proofs can be found in the appendix.

\subsubsection{Example: Laplace Distribution}

Previous sections described how mixed gamma densities can be bit blasted when they are confined to a unit interval. But how can distributions that are shifted or scaled to other finite intervals be bit blasted? We explain it through the example of a Laplace distribution. A Laplace distribution has two parameters: location (\(\mu\)) and scale (\(b\)) and has the probability density function as described below where \(x \in \mathrm{R}\).
\[\mathit{Laplace}(x | \mu, b) = \frac{1}{2b} e^{-\frac{|x - \mu|}{b}} = \begin{cases}
	\frac{1}{2b} e^{\frac{\mu}{b}} e^{-\frac{x}{b}} & x \geq \mu
\\  \frac{1}{2b} e^{-\frac{\mu}{b}} e^{\frac{x}{b}} & x < \mu
\end{cases}\]
Let us consider the Laplace distribution truncated at the interval \([\mu - r, \mu + r)\). We assume that \(r\) would be a suitable power of 2 allowing product with \(r\) (denoted by \(r \times\prog{}\)) to be just a decimal shift and \(\mu\) to be a \(b\)-bit representable number allowing precise shifting of \(\prog{}\). First we generate the exponentials scaled for an interval of width \(r\) instead of width 1:
\[\pdf{0}{-\frac{r}{b}} \comp_b \prog{1} \;\;\;\;\;\;\;\; \pdf{0}{\frac{r}{b}} \comp_b \prog{2}\]
And then we create a mixture of them: 
\[Laplace(x | \mu, b) \comp_b \Lite{\Lflip{0.5}}{\mu + r\times\prog{1}}{\mu - r + r\times\prog{2}}\]
Since \(\prog{1}\) and \(\prog{2}\) use \(b\) coin flips each, the program shown above uses \(2b\) coin flips and \(\comp \circ \comp_b\) is a sound bit blasting function for Laplace distributions as well. 

\subsection{How does bit blasting help in inference?}

We have demonstrated sound bit blasting functions for mixed gamma distributions. But how does that help in inference for probabilistic programs with these distributions? To answer this question, we focus on a particular inference strategy - that is knowledge compilation. We first describe the necessary preliminaries about knowledge compilation and then argue about how the programs obtained through \(\comp_b\) are efficient for knowledge compilation.

Knowledge compilation based approaches~\cite{holtzen2020dice, Fierens2015} for exact discrete probabilistic inference compile discrete probabilistic programs into weighted Boolean formulas that are represented using (reduced) ordered binary decision diagrams (OBDDs). These OBDDs are single rooted in case of a single Boolean random variable being returned and multi-rooted in case of a tuple of Boolean random variables being returned. By fixing a value for all the flips (biased coin flips with associated weights) in the program, and by traversing the OBDD following those values (solid line for true, dashed for false), we reach the terminal corresponding to the value of each bit. The operation of weighted model counting computes the probability of reaching the 1-terminal for each bit in the returned value. It is a dynamic programming algorithm that runs in time linear in the OBDD size. The size of an OBDD for a Boolean formula \(\phi\), denoted \(\mathit{OBDD}(\phi)\), is the number of nodes in the OBDD. Thus, if we can obtain a smaller OBDD representation for a distribution, we can efficiently compose it with other constructs in a discrete probabilistic program. 

We discuss and prove formally how every program obtained through the judgement \(\comp_b\) compiles into a weighted Boolean formula that compiles to a multi-rooted OBDD that grows linearly in the number of bits as opposed to the worst case exponentially. Recall that Dice programs compile to weighted Boolean formula \((\varphi, \gamma, w)\) where \(\varphi\) is the (tuple of) Boolean formulae corresponding to the return value of the program, \(\gamma\) is the accepting Boolean formula to encode observations and \(w\) is the weight function with flip probabilities.

\begin{theorem}\label{theorem:bddsize}
	\(\forall \Upsilon, \prog{}, \varphi, \gamma, w, \exists k, \forall b\), if \(\Upsilon \comp_b \prog{}\) and \(\prog{} \comp (\varphi, \gamma, w)\), then there exists a variable order \(\Pi\) of Boolean random variables in \(w\) such that OBDD(\(\varphi\)) + OBDD(\(\gamma\)) \(\leq kb\)  
\end{theorem}

We now provide intuition for the proof of the above theorem. Note that in the programs obtained through the judgement \(\comp_b\), there are only two constructs that depend on the number of bits: (1) construction of exponential distribution \(\pdf{0}{\beta}\), and (2) conditioning on inequality between an exponential and a uniform distribution through \(\unifobs{\yvar{}}{b}\). We provide intuition how the \(\mathit{OBDD}\) size for these constructs increase linearly in the number of bits, \(b\).

\subsubsection{Exponential Distribution}

In Figure~\ref{fig:expo_bdd}, we 3-bit blast an exponential distribution, i.e.\ we have a 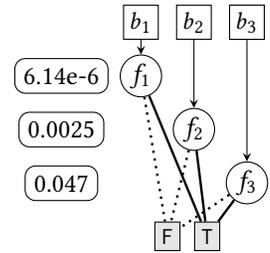
\begin{wrapfigure}{r}{0.25\textwidth}
	\begin{tikzpicture}
		\def\lvl{20pt}

		\node (f1) at (0, 0) [bddnode] {$f_1$};
		\node (f2) at ($(f1) + (20bp, -\lvl)$) [bddnode] {$f_2$};
		\node (f3) at ($(f2) + (20bp, -\lvl)$) [bddnode] {$f_3$};
		
		\node[draw, rounded corners] at ($(f1) + (-30bp, 0bp)$)  {6.14e-6};
		\node[draw, rounded corners] at ($(f1) + (-30bp, -20bp)$)  {0.0025};
		\node[draw, rounded corners] at ($(f1) + (-30bp, -40bp)$)  {0.047};

		\node (false) at ($(f3) + (-30bp, -1*\lvl)$) [bddterminal] {$\false$};
		
		\node (true) at ($(f3) + (-15bp, -1*\lvl)$) [bddterminal] {$\true$};

		\node (b2) at ($(f1) + (0bp, 20bp)$) [bddroot] {$b_1$};
		\node (b1) at ($(b2) + (20bp, 0bp)$) [bddroot] {$b_2$};
		\node (b0) at ($(b1) + (20bp, 0bp)$) [bddroot] {$b_3$};

		\begin{scope}[on background layer]
			\draw [highedge] (f1) -- (true);
			\draw [lowedge] (f1) -- (false);
			\draw [highedge] (f2) -- (true);
			\draw [lowedge] (f2) -- (false);
			\draw [highedge] (f3) -- (true);
			\draw [lowedge] (f3) -- (false);

			\draw[-stealth] (b2) -- (f1);
			\draw[-stealth] (b1) -- (f2);
			\draw[-stealth] (b0) -- (f3);

		\end{scope}
	\end{tikzpicture}
	\caption{
		Compiled BDD for exponential (-3) $e^{-3x}$.
	}\label{fig:expo_bdd}
\end{wrapfigure} discrete exponential distribution over 8 values (0, 0.125, 0.25, $\ldots$, 0.875). The figure shows a 3-rooted OBDD where each root labeled as $b_1, b_2$ and $b_3$ represents a bit in the returned value of 3 bits. Consider that we fix the value of the flip corresponding to the node $f_1$ to be true, then for $b_1$, we would reach the terminal 1 and its value would be assigned 1. The operation of weighted model counting (WMC) would calculate the probability of $b_1$ to be 1 as \(6.14e^{-6}\) as node $f_1$ has probability \(6.14e^{-6}\) to be true. Similarly WMC can be used for other roots of this OBDD.\@ Since each bit needs one OBDD node, the overall OBDD size grows linearly with the number of bits. Since WMC runs in time linear in the OBDD size, probabilistic inference for an exponential distribution would run linear in the number of bits $\mathcal{O}(b)$. Another example of OBDD for a uniform distribution can be found in the appendix.

For all programs \(\prog{}\) obtained through the judgement \(\comp_b\), if \(\prog{} \comp (\varphi, \gamma, w)\), then \(\varphi\) is the tuple of Boolean formulae representing the return value of the program. We argue that the return value of \(\prog{}\) is always a mixture of exponential distributions making its OBDD size linear in the number of bits.

\subsubsection{Conditioning on inequality between an exponential  and a uniform distribution} 

The rules for judgment \(\comp_b\) use the helper judgment \(\unifobs{\yvar{}}{b} = \prog{}\) to condition on an inequality between binary representations of a uniform distribution and an exponential distribution. 

\begin{definition}[inequality function]\label{def:inequality_wbf}
	A \(b\)-bit inequality function, \(\mathit{LT}_b:\{0,1\}^b \times \{0, 1\}^b \rightarrow \{0, 1\}\) takes as input two \(b\)-bit numbers \(x, y\) and outputs $1$ if $ x < y$ and $0$ otherwise. Thus,
	\[\mathit{LT}_b((x_1, \ldots x_b), (y_1, \ldots y_b)) = 
	\begin{cases}
		\neg x_1 y_1 & b = 1
		\\	\neg x_1 y_1 + (\neg x_1 \neg y_1 + x_1y_1)\mathit{LT}_{b-1}((x_2, \ldots, x_b), (y_2, \ldots, y_b)) & b > 1
	\end{cases}\]
\end{definition}

We prove the following lemma which states that the OBDD size for the inequality function grows linearly with the number of bits.

\begin{lemma}\label{lemma:less-than-bdd-size}
	\(\exists k, \forall b,\) for the variable order \(x_1, y_1, x_2, y_2, \ldots, x_b, y_b\), the size of the OBDD, that is  \(\mathit{OBDD}(\mathit{LT}((x_1, x_2, \ldots, x_b), (y_1, y_2, \ldots, y_b))) \leq kb\).
\end{lemma}

Since the only constructs that depend on the number of bits (exponential distributions and inequalities) grow linearly with the number of bits, Theorem~\ref{theorem:bddsize} holds intuitively. We provide a formal proof in the appendix.

\section{\hybit{}: A Probabilistic Programming System}\label{sec:hybit}

\begin{figure}[b]
	\begin{lstlisting}[mathescape=true]
$\tau$ ::= Bool | DistFix{$n, n$} 
v ::= T | F | DistFix{$n, n$}($r$)
e ::= $x$ | v | flip $\theta$ | general_gamma($n, n, n, r, r, r$) | bitblast($n, \mathit{pdf}, n, b, r, r$) 
		| if e then e else e | observe e | op$^n$(e$_1$, $\ldots$, e$_n$)
	\end{lstlisting}
	\caption{Syntax for the core \hybit{} expressions. \texttt{DistFix\{$n, n$\}} refers to the type of fixed-point numbers. The metavariable \(r\) ranges over real numbers, \(n\) over integers, \(b\) over Booleans, \(x\) over variable names, and \(\theta\) over real numbers in the range \([0, 1]\). The metavariable \texttt{pdf} ranges over continuous density functions \(\mathit{pdf}: \mathrm{R} \rightarrow \mathrm{R}\). \texttt{op}$^n$ ranges over arithemtic operations with \(n\)-arity.  We explain the API \texttt{DistFix}, \texttt{general\_gamma} and \texttt{bitblast} and their arguments in Figure~\ref{API}.}
	\label{fig:syntax}
\end{figure}

The previous section described how to bit blast mixed-gamma distributions. We further use it to build a probabilistic program system \hybit{} for hybrid probabilistic programs. This section describes its syntax and implementation and elaborates on two important aspects: piece-wise approximations of continuous distributions and advantages of a binary representation.

\subsection{\hybit{} --- Syntax and Implementation Details}

We build a probabilistic programming system \hybit{} around sound bit blasting of mixed-gamma densities and approximate bit blasting of other continuous distributions. \hybit{} has been implemented as a shallow embedded domain specific language in Julia~\cite{Julia-2017}.

\begin{figure}
		\raggedright{}
		\HRule[8pt]{8pt}
		\begin{verbatim}
			DistFix{W, F} 
		\end{verbatim}
		\textbf{Parameters:} %
		\texttt{W}: total number of bits being used; 
\hspace{0.3em} \texttt{F}: number of bits after the binary point
		\HRule[8pt]{8pt}
		\begin{verbatim}
			DistFix{W, F} general_gamma(int W, int F, int alpha, float beta, float ll, float ul)
		\end{verbatim}
		\textbf{Parameters:}
		\texttt{W, F}: number of bits for bit blasting; \hspace{0.3em}
		\texttt{alpha, beta}: parameters of the density \(\pdf{\alpha}{\beta}\)
		\\
		\hspace{5.4em} \texttt{ll, ul}: range of the continuous density
		\\
		\textbf{Returns}: Sound bit blasted distribution of type \texttt{DistFix\{W, F\}} 
		\HRule[8pt]{8pt}
		\begin{verbatim}
			DistFix{W, F} bitblast(int W, int F, function pdf, int pieces, bool dist, 
			float ll, float ul)
		\end{verbatim}
		
		\textbf{Parameters:} 
\texttt{W, F}: number of bits for bit blasting;\hspace{0.3em}
\texttt{pdf}: continuous density function
\\
\hspace{5.4em} \texttt{pieces}: number of pieces; \hspace{0.3em}\texttt{ll, ul}: range of the continuous density
\\
\hspace{5.4em} \texttt{dist}: Boolean indicating whether to use linear or exponential pieces
		
		\textbf{Returns:} Bit blasted distribution of type \texttt{DistFix\{W, F\}}
		\HRule[8pt]{8pt}
		\begin{verbatim}
			Dict{float, float} pr(DistFix{W, F} var)
			float expectation(DistFix{W, F} var)
			float variance(DistFix{W, F} var)
		\end{verbatim}
		\textbf{Parameters:} \texttt{var}: random variable
		
		\textbf{Returns:} Probability distribution / expectation / variance of \texttt{var}. 
		\HRule[8pt]{8pt}
		\caption{API for \hybit{}.}\label{API}
	\end{figure}

The core syntax of \hybit{} expressions is given in Figure~\ref{fig:syntax}. 
It provides support for distributions over Booleans (\texttt{flip \(\theta\)}) and fixed-point numbers (\texttt{general\_gamma} and \texttt{bitblast}). It supports Boolean operations (\(\neg, \wedge, \vee\)) and arithmetic operations (+, -, *, /, \%, <, ==) over these distributions as well as hard observations for probabilistic conditioning (\texttt{observe}). For all the constructs in Figure~\ref{fig:syntax}, \hybit{} performs a non-standard execution and compiles them to OBDDs to perform probabilistic inference. Since \hybit{} has been implemented as a library in Julia, HyBit programmers can also make use of Julia constructs such as (bounded) loops, tuples and functions. As an example, a \texttt{for} loop from Julia can be used with \hybit{} constructs in the loop body to build a probabilistic model. \hybit{} is available as an open source repository with a comprehensive set of examples.
\footnote{\href{https://github.com/Tractables/Dice.jl/tree/hybit}{https://github.com/Tractables/Dice.jl/tree/hybit}}.

Figure~\ref{API} contains more details on the API of \hybit{}. \texttt{DistFix\{W, F\}} is the type of fixed point numbers of bitwidth \texttt{W} with \texttt{F} bits after the binary point. The function \texttt{general-gamma} performs sound bit blasting of a specified generalized gamma density to the given fixed-point \texttt{W} and \texttt{F}. Sound bit blasting of mixed-gamma densities is achieved by using the \texttt{if-then-else} construct over generalized gamma densities. The function \texttt{bitblast} is used for bit blasting any arbitrary continuous distribution using piece-wise approximation, employing the bits and pieces specified using the parameters \texttt{W}, \texttt{F}, and \texttt{pieces}. The API also allows the user to choose the type of discrete distribution $-$ linear or exponential $-$ for the piece-wise approximation. The parameters of linear (slope) and exponential (\(\beta\)) pieces are automatically chosen such that the ratio of probabilities of the first and last interval is the same as that for the na\"ive discretization. Finally, the API also provides functions for querying the probability distribution, expectation and the variance of a random variable. The next subsections describe piece-wise approximations and the computation of expectation and variance in more detail.

\subsection{Piece-wise Approximations}

Even though mixed-gamma distributions capture many natural distributions, there are other commonly occurring ones, such as the Gaussian. It is an open problem as to whether Gaussians admit a sound bit blasting function, let alone one that  compiles to a compact OBDD. For such distributions, one can instead use a piece-wise approximation.

\begin{wrapfigure}{r}{0.35\textwidth}
	\centering
	\includegraphics[width = 0.35\textwidth]{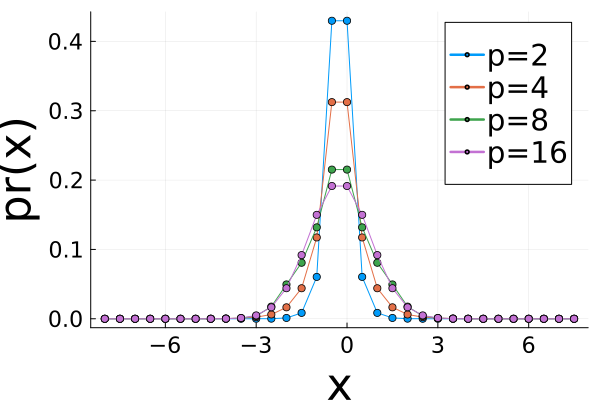}
	\caption{Bit blasted Gaussian distribution using 5 bits in the range \([-8, 8)\) using exponential pieces \(p\) --- for 2, 4, 8 and 16 pieces.}
	\label{fig:gaussian}
\end{wrapfigure}

Let \(C\) be an arbitrary continuous probability distribution over the interval \([l, u)\). To bit blast \(C\) using a piece-wise distribution with \(t\) pieces, C is approximated using a mixture of \(t\) discrete probability distributions over disjoint intervals. For every piece, one creates a shifted and scaled instance of a bit blasted mixed-gamma density and then creates a mixture of them. Note that since each piece uses \(\mathcal{O}(b)\) coin flips, a piece-wise distribution with \(t\) pieces uses \(\mathcal{O}(tb)\) coin flips. Section~\ref{sec:experiments} shows empirical advantages of this approach. This piece-wise approximation using linear or exponential pieces can be easily built using the \texttt{bitblast} API available in \hybit{} (Figure~\ref{API}). Figure~\ref{fig:gaussian} shows bit blasting of a Gaussian distribution using 2, 4, 8 and 16 pieces, where each piece is a bit blasted exponential. This provides the user with the conventional trade off between accuracy and performance. We elaborate more on this in Section~\ref{eval-lp}.\begin{wrapfigure}[23]{r}{0.35\textwidth}
	\begin{subfigure}{0.35\textwidth}
		\includegraphics[width=\textwidth]{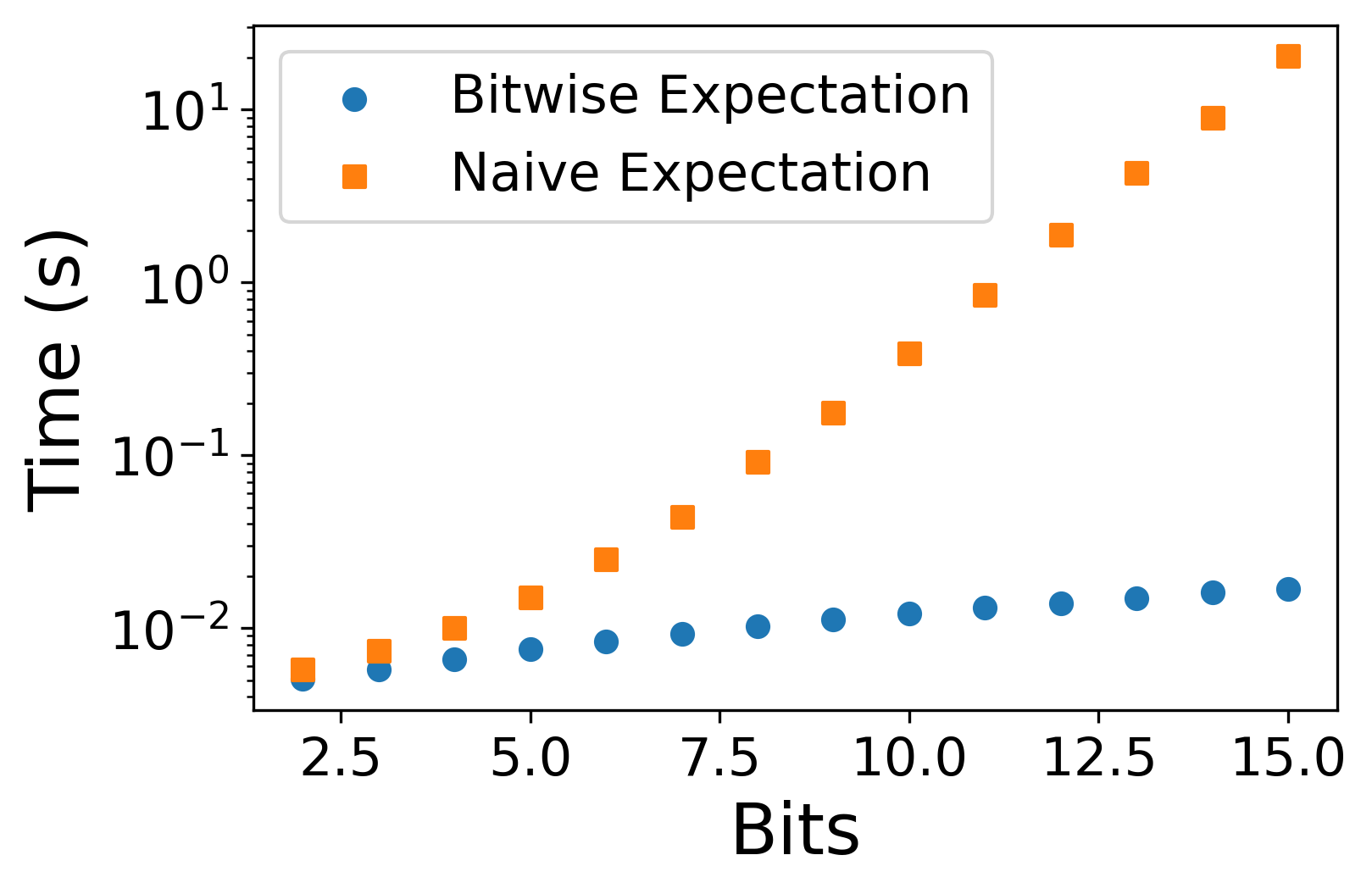}
		\caption{Expectation}\label{figure9a}
	\end{subfigure}
	\hfill
	\begin{subfigure}{0.35\textwidth}
		\includegraphics[width=\textwidth]{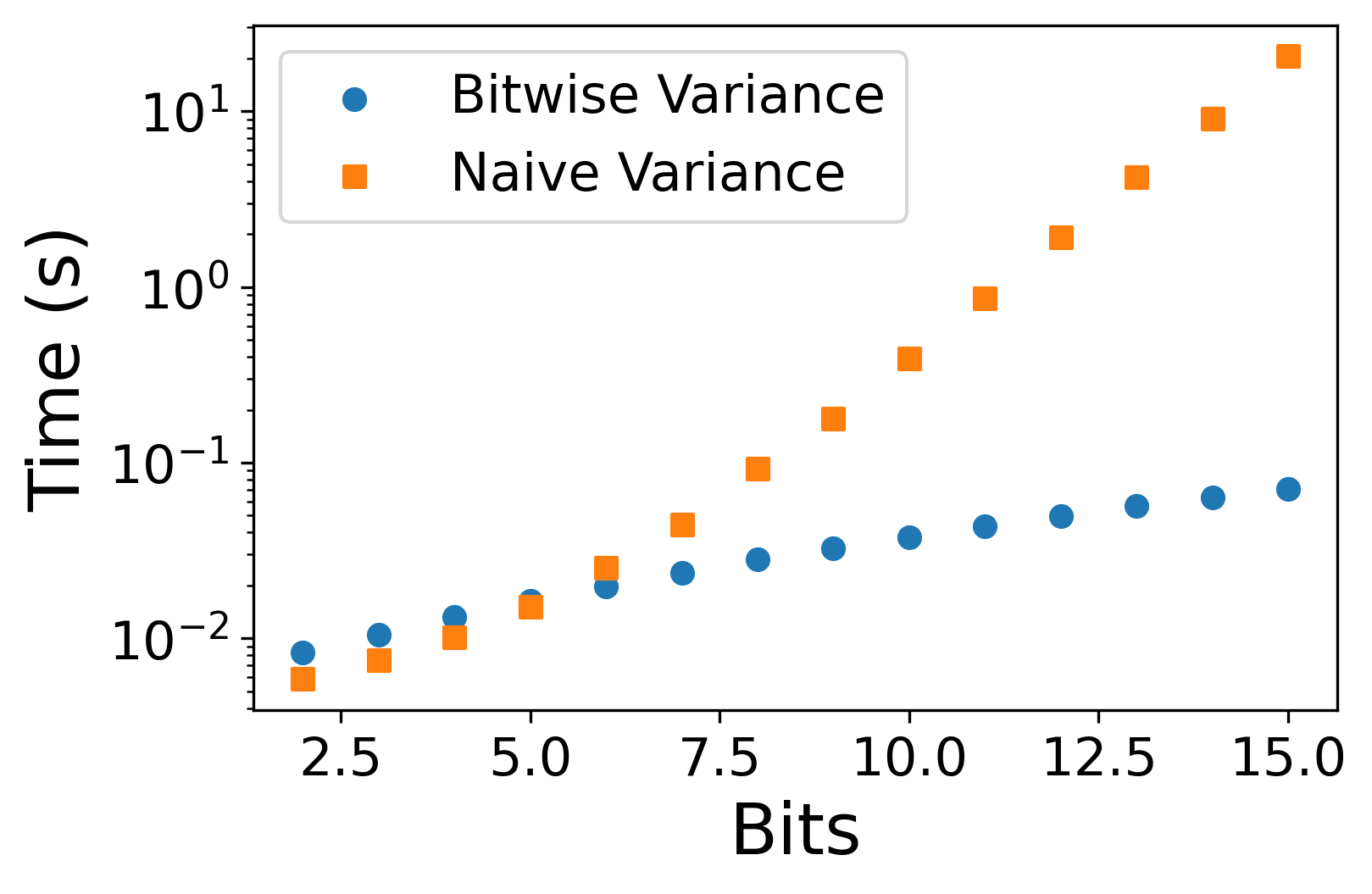}
		\caption{Variance}\label{figure9b}
	\end{subfigure}
\caption{Speedup in computing expectation and variance}\label{figure9}
\end{wrapfigure}
	
\subsection{Advantages of the Binary Representation}

The binary representation has important advantages for probabilistic reasoning beyond the succinctness that bit blasting provides.
First, many hybrid probabilistic models involve arithmetic operation on continuous random variables. Since we use a binary representation of fixed point numbers, arithmetic operations such as +, *, /, < are compiled as Boolean formulas over binary numbers (similar to ALU circuits in architecture). This representation allows probabilistic inference (specifically the knowledge compilation approach that we employ) to identify and exploit the structure that exists in arithmetic, such as conditional independences among the resulting bits in a computation. Recent work~\cite{cao2023scaling} described this compilation and showed its advantage empirically for integers; \hybit{} leverages these advantages for computations over fixed-point numbers.

The binary representation also enables an optimized computation of expectation and variance. Na\"ive computation of expectation and variance for a distribution over \(2^b\) values requires one to compute probability of \(2^b\) values. Bitwise representation allows one to achieve this computation by only computing probability of \(b\) bits which gives an exponential improvement. Note that in the worst case for an arbitrary hybrid probabilistic program, getting the corresponding OBDD for the binary representation can itself be exponential in the number of bits. But for the class of mixed-gamma distributions, this conversion is linear in the number of bits (Theorem~\ref{theorem:bddsize}). We formalize the computation of expectation and variance in the following two theorems and provide proofs in the appendix.
	
\begin{theorem}\label{expectation}
	Let D be a discrete probability distribution over the interval $[0, 2^n)$ represented as a distribution over $n$ bits as $(b_n, b_{n-1}, \ldots, b_1)$, then the expectation of D can be computed using linearity of expectation as follows:
	\[\mathrm{E}[D] = \sum_{i = 0}^{2^n-1} i \cdot \mathit{pr}(i) = \mathrm{E}(\sum_{j = 1}^{n} 2^{j-1}b_j) = \sum_{j=1}^{n}2^{j-1} \cdot \mathit{pr}(b_j) \]
\end{theorem}
	
\begin{theorem}\label{variance}
	Let D be a discrete probability distribution over the interval $[0, 2^n)$ represented as a distribution over $n$ bits as $(b_n, b_{n-1}, \ldots, b_1)$, then the variance of D can be computed as follows:
	\[\mathit{Var}[D] = \sum_{i=0}^{2^n - 1} i^2 \mathit{pr}(i) - (\mathrm{E}(D))^2 = \mathit{Var}(\sum_{j = 1}^{n} 2^{j-1}b_j) = \sum_{k = 1}^{n} \sum_{l = 1}^{n} 2^{l+k-2} [\mathit{pr}(b_l \wedge b_k) - \mathit{pr}(b_l)\mathit{pr}(b_k)] \]
\end{theorem}
	
\textit{Example.} Consider  a discrete uniform distribution \(U_4\) over the integers $\{0, 1, 2, 3\}$ represented using two bits, $(X_2, X_1)$. The direct way of calculating the expectation and variance of this uniform distribution requires inferring the probability of all the integers in the domain.
But Theorems~\ref{expectation} and~\ref{variance} allow us to compute these quantities by using only the probabilities of the individual bits.
\[\mathrm{E}[U_4] = \sum_{j = 1}^{2}2^{j-1}\cdot\mathit{pr}(b_j) = 1.5 \;\;\;\;\;
\mathit{Var}[U_4] = \sum_{k = 1}^{2} \sum_{l = 1}^{2} 2^{l+k-2} [\mathit{pr}(b_l \wedge b_k) - \mathit{pr}(b_l)\mathit{pr}(b_k)] = 1.25\]
Figure~\ref{figure9} empirically shows the performance benefits in computing expectation and variance of a distribution as we increase the number of bits in bit blasting a standard normal distribution.

\section{Empirical Evaluation}\label{sec:experiments}
	
We evaluate the practicality of bit blasting on real-life probabilistic programs. We have carried out relevant experiments to investigate the following questions:
	\begin{description}[labelindent=0.5cm]
		\item Q1: How does \hybit{} perform in comparison to existing inference algorithms? Section~\ref{baselines1}
		
		\item Q2: How effective is the piece-wise approximation? Section~\ref{eval-lp}
	\end{description}
	
	\begin{table}[b]%
		\centering
		\caption{\textit{Comparison of HyBit against other approximate inference algorithms.} Each row consists of one entry in bold indicating the lowest absolute error achieved among all inference algorithms. A '\xmark{}' denotes that the baseline does not support inference for the benchmark. A '\(\phi\)' denotes timeout. A '\(\infty\)' denotes infinite bounds.}\label{table1}
		\scriptsize
		\begin{tabular}{lrrrrrrrrrrr}\toprule
			Benchmarks &\multicolumn{3}{c}{HyBit} &AQUA &\multicolumn{3}{c}{WebPPL} &Stan &GuBPI \\\cmidrule{1-10}
			& &Bit &Pieces & &rejection &MCMC &SMC &  \\\midrule
			Pi~\cite{pi}                    & 1.05E-04             & 14                       & $-$                          & \xmark{}                         & 8.30E-05                      & 9.66E-05                    & 1.38E-03               & \textbf{4.84E-05}     & \xmark{}                                  \\
			weekend~\cite{gehr2016psi}                & \textbf{2.08E-08}    & 24                       & 4096                       & \xmark{}                         & 1.57E-02                      & 1.57E-02                    & 1.66E-02                        & \multicolumn{1}{r}{\xmark{}} & 2.50E-05                                   \\
			spacex~\cite{spacex}                 & 6.94E-04             & 19                       & 32                         & \xmark{}                         & 9.06E-04                      & 3.24E-03                    & 1.88E-02               & \textbf{1.15E-04} & \(\phi\)                                       \\
			GPA~\cite{https://doi.org/10.48550/arxiv.1806.02027}                    & \textbf{2.22E-16}    & 25                       & 4096                       & \multicolumn{1}{r}{3.62E-01}          & 1.70E-02                      & 9.39E-03                    & 1.51E-02                        & \multicolumn{1}{r}{\xmark{}} & 3.88E-01                       \\
			Tug of war~\cite{huang2021aqua}             & \textbf{4.50E-07}    & 22                       & 16                         & \xmark{}                         & 6.93E-04                      & 6.94E-04                    & 2.35E-03                        & 4.51E-05                           & \(\phi\)                     \\
			altermu2~\cite{https://doi.org/10.48550/arxiv.1312.5386}                & 3.48E-06             & 17                       & 256                        & \multicolumn{1}{r}{\textbf{3.41E-07}} & \multicolumn{1}{r}{\(\phi\)}   & 4.61E-01                    & 4.38E-01                        & 1.68E-03       & 1.57E-02                                         \\
			conjugate gaussians
			\\ \hspace{2pt}\cite{conjugategaussians}     & \textbf{4.92E-06}    & 23                       & 16                         & \multicolumn{1}{r}{0.99}              & 2.19E-04                      & 3.53E-04                    & 3.18E-03                        & 1.06E-04                 & 1.09E-03                               \\
			normal\_mix ($\theta$)
			\\ \hspace{2pt}\cite{huang2021aqua} & 5.49E-05             & 9                        & 64                         & \multicolumn{1}{r}{\textbf{4.13E-07}} & \multicolumn{1}{r}{\(\phi\)}   & 3.90E-04                    & 5.30E-03                        & 4.29E-01 & \(\infty\)                                                \\
			normal\_mix ($\mu_1$)
			\\ \hspace{2pt}\cite{huang2021aqua}   & 5.20E-03             & 9                        & 16                         & \multicolumn{1}{r}{\textbf{7.55E-06}} & \multicolumn{1}{r}{\(\phi\)}   & 1.36E-03                    & 2.00E-02                        & 1.87E+01 & 9.21E+00                                               \\
			normal\_mix ($\mu_2$)
			\\ \hspace{2pt}\cite{huang2021aqua}   & 3.92E-03             & 9                        & 32                         & \multicolumn{1}{r}{\textbf{8.65E-06}} & \multicolumn{1}{r}{\(\phi\)}   & 7.11E-04                    & 1.15E-02                        & 1.77E+01 & 9.44E+00                                                \\
			zeroone (w1)~\cite{Bissiri_2016}          & \textbf{9.40E-05}    & 16                       & $-$                          & \multicolumn{1}{r}{5.66E-02}          & \multicolumn{1}{r}{\(\phi\)}   & \multicolumn{1}{r}{\(\phi\)} & \multicolumn{1}{r}{\(\phi\)} & 1.73E-01 & \(\infty\)                                                \\
			zeroone (w2)~\cite{Bissiri_2016}           & \textbf{4.51E-04}    & 19                       & $-$                          & \multicolumn{1}{r}{3.69E+00}          & 1.64E+00                      & 1.64E+00                    & 1.66E+00                        & 2.38E-01 &\(\infty\)                                               \\
			coinBias~\cite{gehr2016psi}               & \textbf{2.02E-07}    & 22                       & 4096                       & \multicolumn{1}{r}{6.25E-02}          & 1.69E-05                      & 7.73E-05                    & 1.22E-03                        & 1.18E-05  & 4.01E-03                                              \\
			Addfun/sum~\cite{gehr2016psi}              & \textbf{3.81E-06}    & 23                       & 16                         & \xmark{}                         & 4.41E-04                      & 1.69E-03                    & 5.41E-03                        & 8.45E-05 & 3.12E-02                                                \\
			ClickGraph~\cite{gehr2016psi}             & 1.75E-03             & 10                       & $-$                          & \xmark{}                         & 7.29E-04                      & 1.22E-03                    & 3.41E-03               & \textbf{2.80E-05} & \(\phi\)                                       \\
			trueskill~\cite{gehr2016psi}              & 3.05E-03             & 10                       & 16                         & \xmark{}                         & 1.81E-04             & 4.22E-04                    & 1.68E-03                        & \multicolumn{1}{l}{\textbf{6.88E-05}} & \(\phi\) \\
			clinicaltrial1~\cite{gehr2016psi}         & \textbf{5.27E-16}    & 8                        & $-$                          & \xmark{}                         & 1.51E-01                      & 1.53E-01                    & 1.49E-01                        & 9.27E-04 	& \(\phi\)                                               \\
			clinicaltrial2~\cite{gehr2016psi}         & \textbf{6.81E-07}    & 12                       & $-$                          & \xmark{}                         & 1.42E-01                      & 1.43E-01                    & 1.42E-01                        & 4.54E-05 & 2.86E-01                                               \\
			addfun/max~\cite{gehr2016psi}             & \textbf{2.93E-07}    & 23                       & 128                        & \xmark{}                         & 4.38E-04                      & 6.11E-04                    & 3.26E-03                        & 1.19E-04   & 8.56E-01    \\
			\bottomrule
		\end{tabular}
	\end{table}

		\subsection{Comparison with existing inference algorithms}\label{baselines1} 
		
		\subsubsection{Approximate Inference Algorithms}
		We evaluate \hybit{} against two classes of approximate inference algorithms.
		
\textbf{Sampling Methods} We compare against WebPPL rejection sampling, MCMC sampling (with a Metropolis Hastings kernel), SMC sampling and Stan HMC as representatives of this class.

\textbf{Discretization Methods}  We compare against AQUA and GuBPI in this class of inference algorithms~\cite{huang2021aqua, https://doi.org/10.48550/arxiv.2204.02948}.

		Comparing performance of different probabilistic programming systems is a challenging task since performance is directly affected by the structure of the program. We write equivalent programs for these benchmarks in each system and put in our best effort to optimize them. The tables in this section and subsequent sections report the mean value of absolute error over 10 runs for stochastic algorithms. For other inference algorithms, we report output of a single run. All experiments were single-threaded and were carried out on a server with \(2.4\) GHz CPU and \(512\) GB RAM.\@
		
		Table~\ref{table1} reports the results of performance evaluation of \hybit{} against other approximate inference algorithms. We take all the hybrid and continuous benchmarks on which \textit{Psi}~\cite{gehr2016psi} was evaluated and a few more relevant benchmarks from existing work~\cite{huang2021aqua}. We put in our best effort to compute ground truth for these benchmarks either analytically or using computer algebra systems. We include only those benchmarks in our evaluation for which we were able to compute the ground truth reliably. We report the absolute error with respect to the ground \begin{wraptable}[23]{r}{0.34\textwidth}%
			\caption{Comparison of HyBit with Psi, an exact inference PPL. '\(\phi\)' denotes a timeout, '\xmark{}' denotes result unsolved by Mathematica}\label{table2}
			\footnotesize
			\begin{tabular}{lrrr}\toprule
				Benchmarks &HyBit &Psi \\\midrule
				Pi &\cmark{} &\xmark{} \\
				weekend&\cmark{} &\cmark{} \\
				spacex&\cmark{} &\xmark{} \\
				GPA&\cmark{} &\cmark{} \\
				Tug of war &\cmark{} &\xmark{} \\
				altermu2 &\cmark{} &\cmark{} \\
				conjugate gaussians &\cmark{} &\cmark{} \\
				normal\_mixture (\(\theta\)) &\cmark{} & \(\phi\) \\
				normal\_mixture (\(\mu_1\)) &\cmark{} &\(\phi\)  \\
				normal\_mixture (\(\mu_2\)) &\cmark{} & \(\phi\)  \\
				zeroone (w1) &\cmark{} & \(\phi\)  \\
				zeroone (w2) &\cmark{} & \(\phi\)  \\
				coinBias &\cmark{} &\cmark{} \\
				Addfun/sum &\cmark{} &\cmark{} \\
				ClickGraph &\cmark{} &\cmark{} \\
				trueskill  &\cmark{} &\xmark{} \\
				clinicaltrial1  &\cmark{} & \(\phi\)  \\
				clinicaltrial2  &\cmark{} &\cmark{} \\
				addfun/max &\cmark{} &\cmark{} \\
				\bottomrule
			\end{tabular}
		\end{wraptable}truth for all the benchmarks. For benchmarks that returned a non-boolean value, we compute the absolute error of expectation for each of the approaches. We report the minimum error achieved by an inference algorithm within a timeout of 20 minutes.

		For all the benchmarks, \hybit{} replaced mixed-gamma distributions with their sound bit blasted distribution and all other distributions with their linear piece-wise approximation \(\pdf{1}{0}\). The employed bits and pieces for each benchmark are reported in Table~\ref{table1}. To run Stan on these benchmarks, we make use of SlicStan~\cite{gorinova2019automatic} to get the Stan program with marginalized discrete random variables. For all WebPPL baselines, default settings were used for all the sampling algorithms with maximum number of samples within 20~minutes.
		
		As Table~\ref{table1} shows, \hybit{} with bit blasting is comparable with the existing approaches on all the benchmarks, even better on 11/19 of them. For the other 8 benchmarks, \hybit{} achieves a very close accuracy. AQUA performs better on only four of the benchmarks and GuBPI fails to obtain good accuracy. This is primarily because their enumerative discretization does not scale well for higher precision. WebPPL and Stan (equipped with automated marginalization through SlicStan) support most of the benchmarks but do not achieve good accuracy within the threshold time. This is because sampling based algorithms are stochastic and cannot obtain sufficiently many samples from the true posterior in limited time.

		\subsubsection{Exact Inference Algorithms}\label{baselines2}
		Table~\ref{table2} compares \hybit{} against a probabilistic programming system that performs exact inference using algebraic methods i.e.\ Psi~\cite{gehr2016psi}. We put in our best effort to translate the benchmarks for optimal performance in Psi. It would often output a symbolic expression which we would feed to Mathematica for further simplification. Computing and simplifying these algebraic expressions is not a trivial task and hence, Psi timed out on 6 of these benchmarks and Mathematica failed to simplify 4 of these benchmarks. \hybit{} works for all 19 benchmarks as it reduces the computation to discrete inference on Boolean random variables and approximates the inference query.
		
		\begin{figure}[t]
			\centering
			\begin{tabular}[c]{cccc}
				\begin{subfigure}[c]{0.22\textwidth}
					\includegraphics[width=\textwidth]{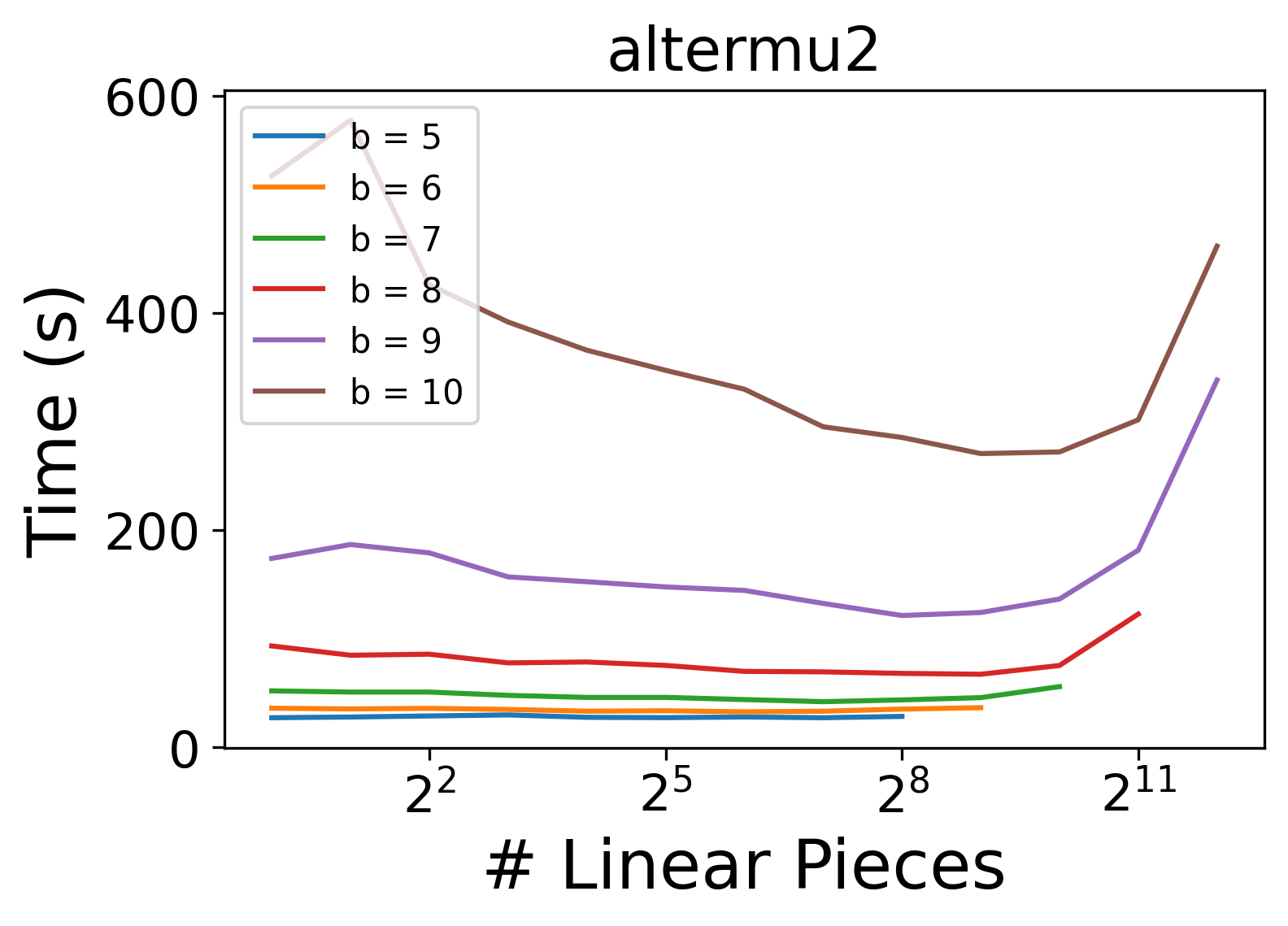}\label{fig:ceob}
				\end{subfigure}&
				\begin{subfigure}[c]{0.22\textwidth}
					\includegraphics[width=\textwidth]{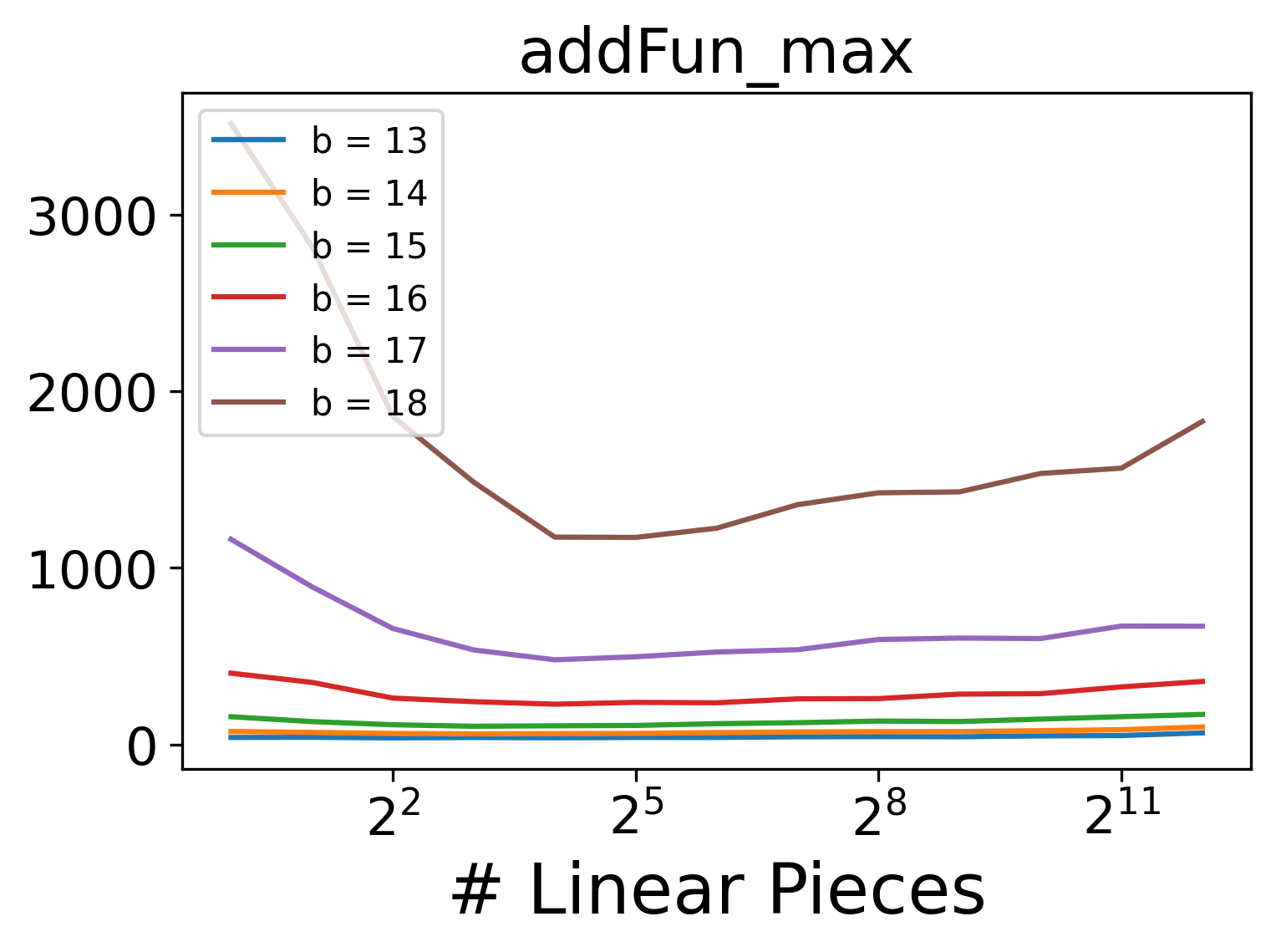}\label{fig:ceoc}
				\end{subfigure}&
				\begin{subfigure}[c]{0.22\textwidth}
					\includegraphics[width=\textwidth]{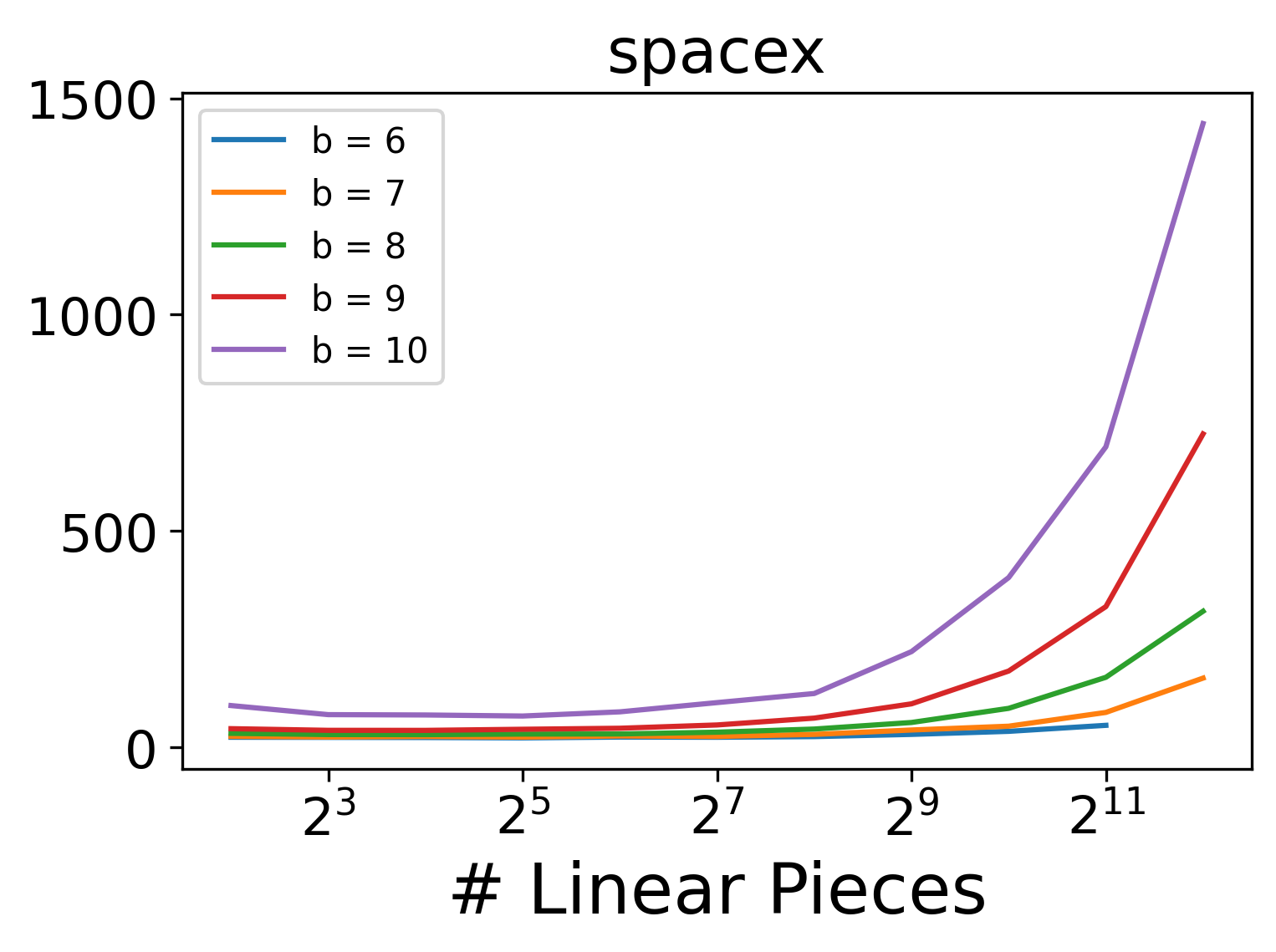}\label{fig:ceoe}
				\end{subfigure}&
				\begin{subfigure}[c]{0.22\textwidth}
					\includegraphics[width=\textwidth]{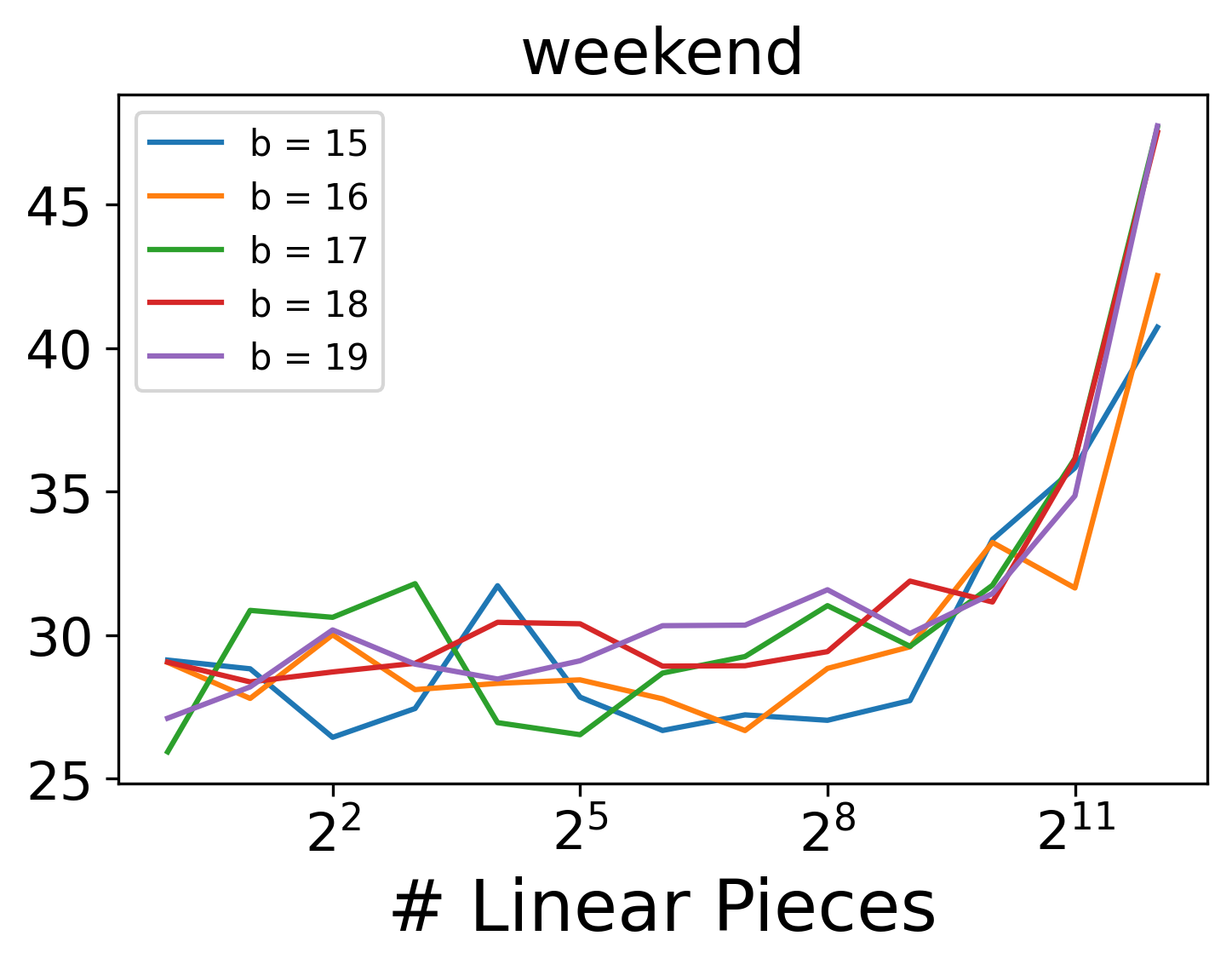}\label{fig:cdfeod}
				\end{subfigure}\\
				
				\begin{subfigure}[c]{0.22\textwidth}
					\includegraphics[width=\textwidth]{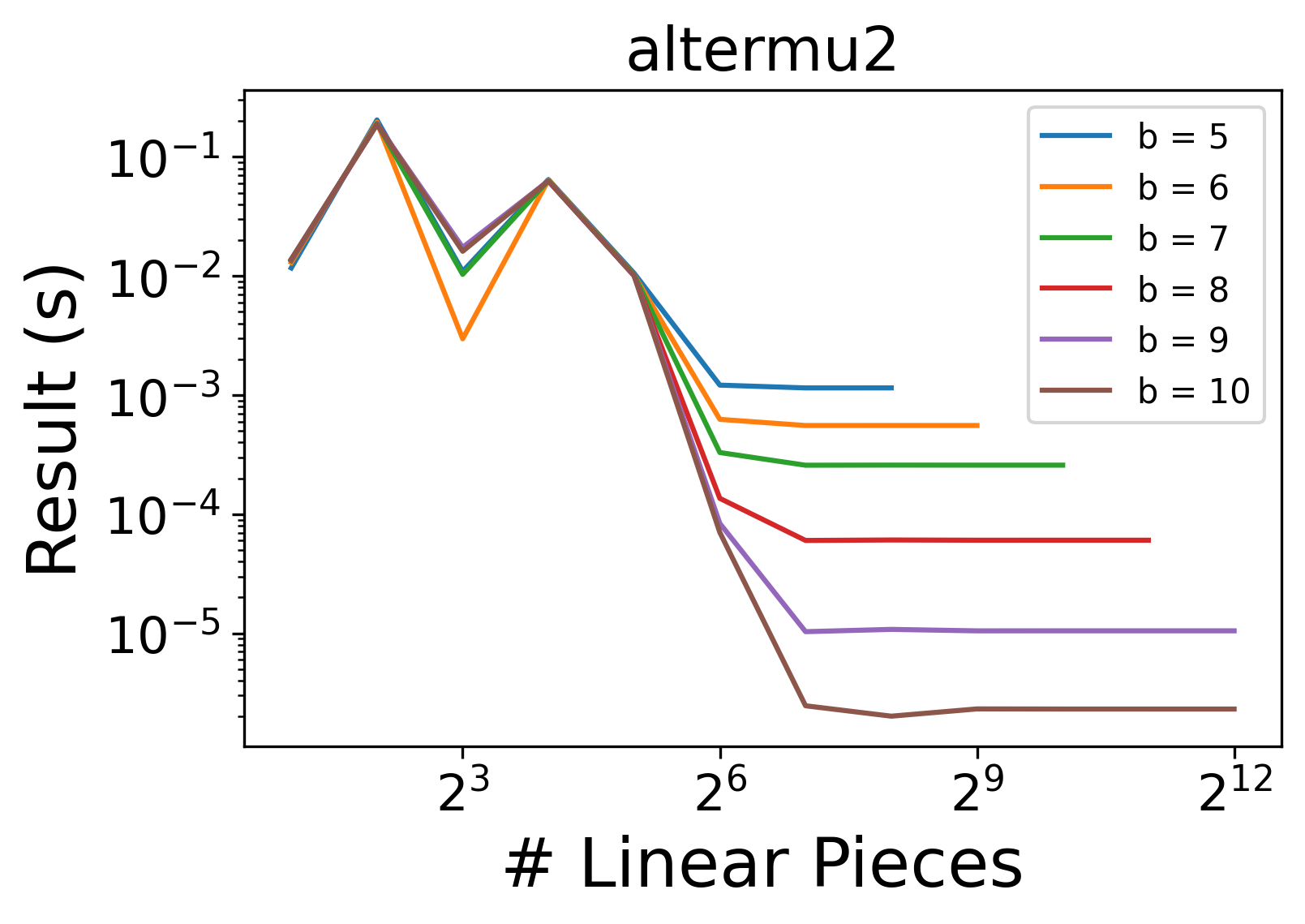}\label{fig:caeob}
				\end{subfigure}&
				\begin{subfigure}[c]{0.22\textwidth}
					\includegraphics[width=\textwidth]{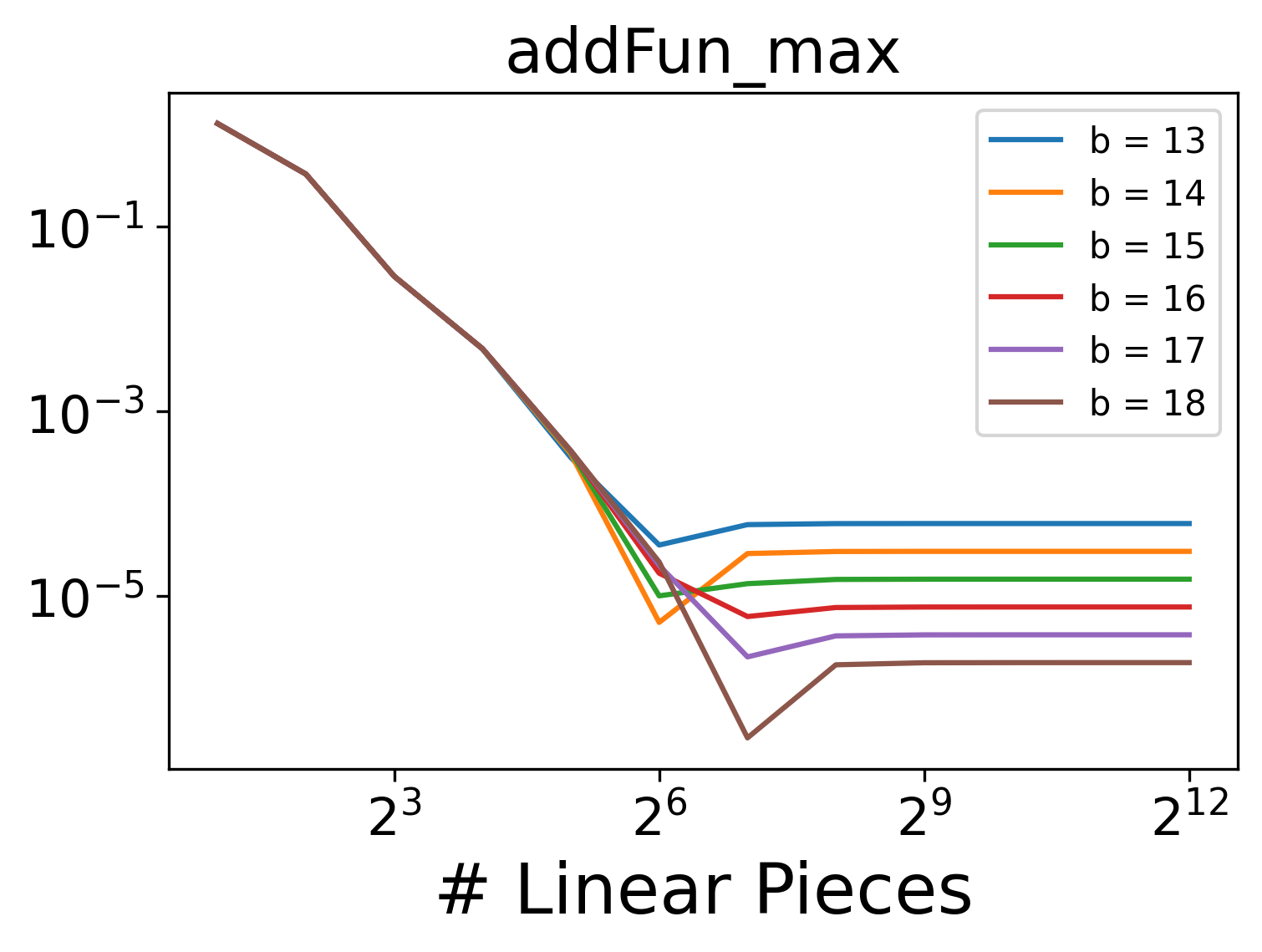}\label{fig:ceoac}
				\end{subfigure}&
				\begin{subfigure}[c]{0.22\textwidth}
					\includegraphics[width=\textwidth]{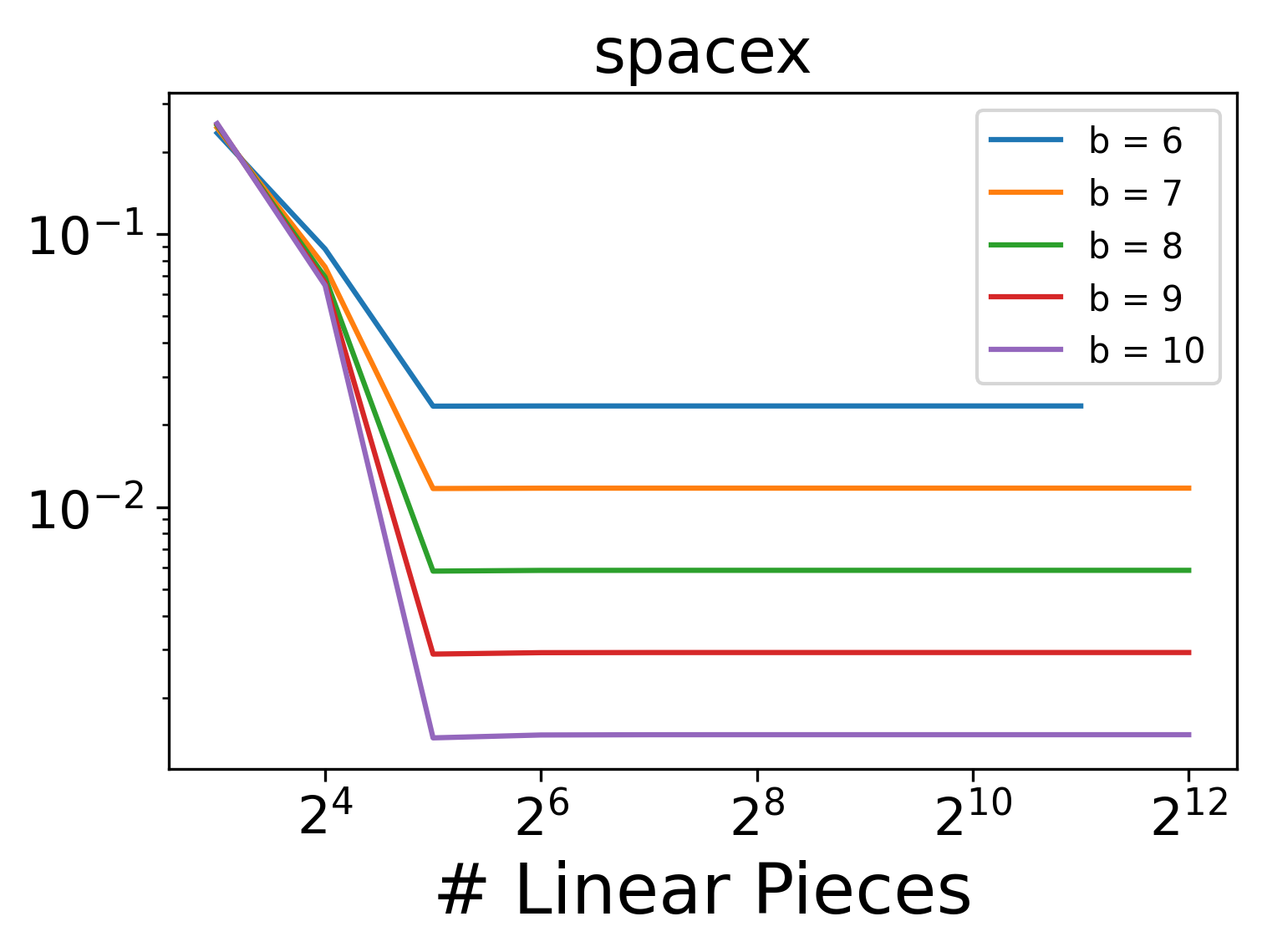}\label{fig:ceose}
				\end{subfigure}&
				\begin{subfigure}[c]{0.22\textwidth}
					\includegraphics[width=\textwidth]{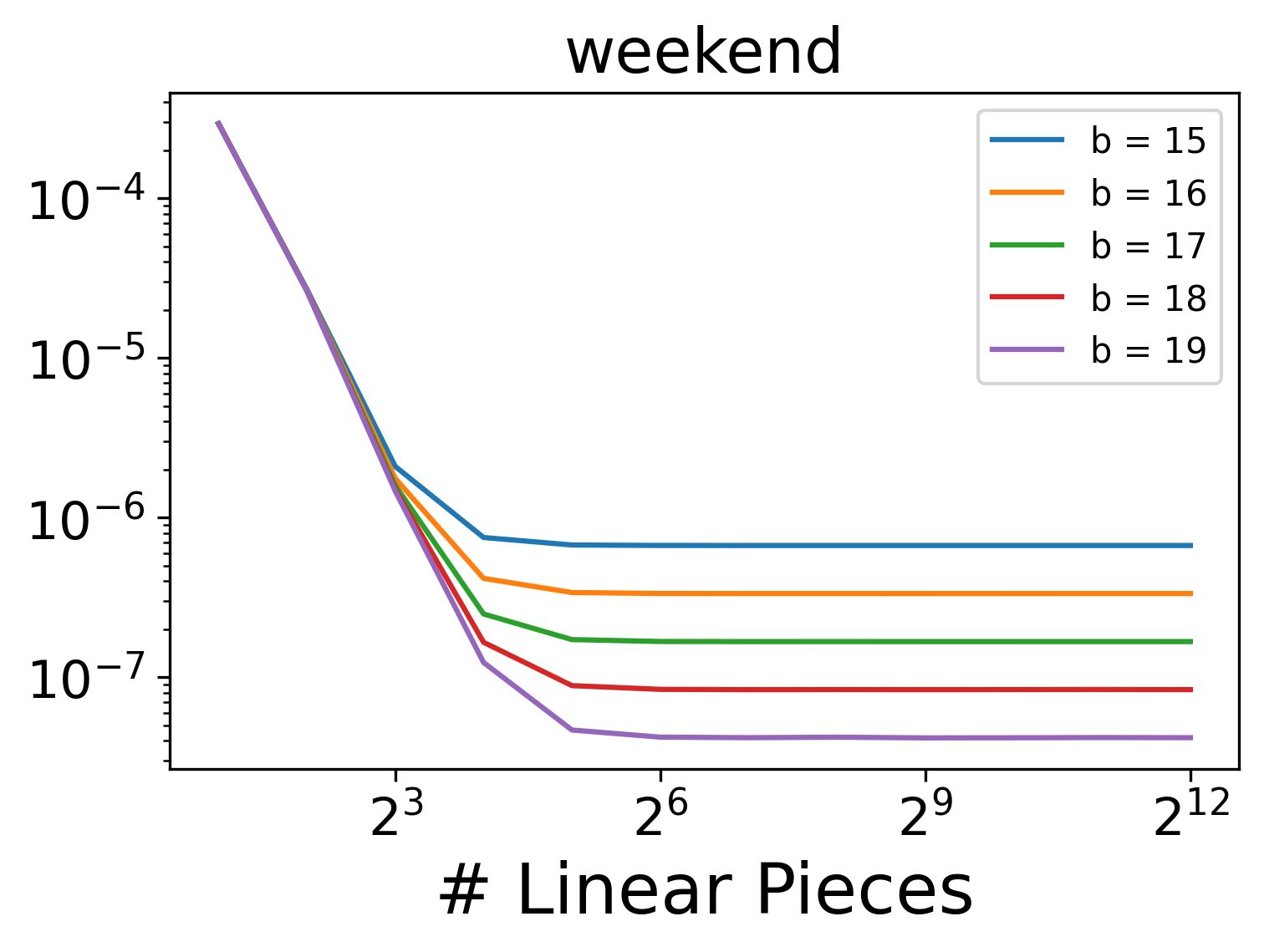}\label{fig:ceod}
				\end{subfigure}\\
			\end{tabular}    
			\caption{Runtime and Accuracy trends with respect to linear pieces for different bitwidths 
			}\label{figure8}
		\end{figure}
		
		\subsection{How effective is piece-wise approximation?}\label{eval-lp}

We analyze the tradeoff between performance and accuracy when using different numbers of pieces to approximate the continuous distribution. Figure~\ref{figure8} demonstrates the trends of runtime and accuracy with the increase in pieces for different bitwidths for four benchmarks. As the number of linear pieces increases, runtime tends to first decrease and then increase. As the number of pieces increases, the accuracy tends to improve as shown by the lower four plots. This is because as we increase the number of pieces, continuous distributions are replaced with more accurate bit blasted distributions. That accuracy improvement comes at the cost of increased runtime after a certain sweet spot. The appendix provides additional experiments that also justify the usage of piece-wise approximations over an approach based on the central limit theorem.

		\section{Related Work}
		Probabilistic programming has been an an active area of research both from the perspective of semantics and inference~\cite{10.5555/1642293.1642508, dahlqvist2023deterministic}. This section positions \hybit{} with respect related work. At a high level, the key distinction in \hybit{} is the development of bit blasting for succinct discretization of hybrid probabilistic programs.
		
		\emph{Discretization approaches.} Prior approaches that discretize continuous or hybrid probabilistic programs estimate the posterior by exhaustively enumerating all of the discretized values~\cite{huang2021aqua}, which does not scale to provide sufficient accuracy in many cases. One prior discretization technique also employs a bit representation~\cite{10.1145/2491411.2491423}. However, their approach is not a form of bit blasting, since it is not succinct and still in general produces a representation that is proportional to the number of discretized points.
		Finally, a recent approach uses discretization to produce   
		 upper and lower bounds on the posterior of a probabilistic program~\cite{https://doi.org/10.48550/arxiv.2204.02948}.
		
		\emph{Inference algorithms for hybrid probabilistic programs.} Other research specifically targets hybrid probabilistic programs. Leios~\cite{10.1007/978-3-030-44914-814} \emph{continualizes} the hybrid probabilistic program in order to harness the power of existing continuous inference algorithms. \hybit{}, on the other hand discretizes the hybrid programs which helps in scaling inference for hybrid programs specifically with respect to the discrete structure. SPPL supports hybrid programs by translating them to specific representations for inference~\cite{saad2021sppl}. However, these representations constrain the hybrid programs that can be supported. For instance, SPPL does not support arithmetic on continuous random variables while \hybit{} can. Finally, probabilistic logic programming languages have been extended to support hybrid models using interval traces~\cite{10.1007/978-3-642-21295-6_12}.
		
		\emph{Algebraic approaches.} Some PPL inference algorithms produce closed form algebraic expressions to encode probability distributions and then use symbolic techniques to perform exact inference \citep{gehr2016psi, narayanan2016probabilistic, Hur2014}. However, these systems are necessarily limited in their expressivity and the programs that they can handle, as shown in Table~\ref{table2}.
		
		\emph{Path based inference algorithms.} A common class of inference algorithms for PPLs are \emph{operational}: they record traces on the program by using concrete values of the random variables. This includes sampling algorithms and variational approximations~\cite{JSSv076i01, Hur2015, pfeffer2007general, chaganty2013efficiently, wood2014new, van2015particle, mansinghka2013approximate, goodman:uai08, saad2016, Mansinghka2018, bingham2019pyro, dillon2017tensorflow,
			wingate2013automated, kucukelbir2015automatic, InferNET14, 10.5555/2969033.2969117}. 
		Sampling algorithms like rejection sampling and MCMC methods are universal but have known limitations such as difficulty in handling multi-modality and low-probability evidence, as described in Section~\ref{sec:motivation}. More sophisticated techniques like Hamiltonian Monte Carlo and variational approximation address these limitations but impose constraints of continuity and almost-everywhere differentiability, so they must resort to marginalizing out all discrete structure.
		
		\emph{Use of a binary representation.} Bit blasting has been a widespread technique in software verification, used in constraint solvers to reason about arithmetic using a bit representation~\cite{5361322}. Recent work in scaling inference for probabilistic programs over integers also employs a binary representation for numbers~\cite{cao2023scaling}, in order to exploit conditional independences in that representation. The bit blasting in \hybit{} is inspired by these techniques but has a different purpose and hence a very different technical approach: to develop succinct, and in many cases provably sound, approximations of continuous probability distributions.
		
		\section{Conclusion and Future Work}
		In this work, we motivated the need for new inference methods for hybrid probabilistic programs. We described \emph{bit blasting}, whereby hybrid probabilistic programs are succinctly discretized and then analyzed using algorithms for discrete inference.  We characterized a class of continuous densities --- mixed-gamma densities --- for which bit blasting is not only succinct but also sound relative to an explicit discretization approach as well as provably efficient to analyze.  We then presented a new PPL \hybit{} that employs a novel inference algorithm for hybrid programs based on bit blasting. We demonstrated the performance benefits of \hybit{} over existing approximate inference algorithms.
		
		In future work, we hope to expand the class of distributions that can be bit blasted soundly. We plan to investigate how \hybit{} can be extended to support hierarchical Bayesian models. We plan to enhance its usability by not requiring user to specify the hyperparameters for every probabilistic program. We are also interested to explore the integration of \hybit{} with other inference approaches, to leverage their relative strengths for support of a wider range of hybrid probabilistic programs.

		\begin{acks}                            %
		The authors would like to thank previous and current members of Star AI lab for helpful discussions and emotional support. This work was funded in part by the DARPA PTG Program under award HR00112220005, the DARPA ANSR program under award FA8750-23-2-0004, 
		NSF grants \#IIS-1943641, \#IIS-1956441, \#CCF-1837129, and a gift from RelationalAI. GVdB discloses a financial interest in RelationalAI.
		\end{acks}
		
		\section*{Artifact}
		
		\hybit{} probabilistic programming system is available as an open source repository on GitHub at \href{https://github.com/Tractables/Dice.jl/tree/hybit}{https://github.com/Tractables/Dice.jl/tree/hybit} with thorough documentation for reusability. It is also available as an archived version on Zenodo~\cite{garg_2024_10901544} with comprehensive instructions and scripts for reproducibility of experiments reported in the paper.

		\bibliographystyle{ACM-Reference-Format}
		\bibliography{references}
		
		\newpage
				\appendix
		
		\section{Syntax of Dice}\label{section:Dicesyntax}
		Subset of syntax of \dice{} language.  The metavariable $y$ ranges over variable names, and $\theta$ over real numbers in the range $[0,1]$.
		\begin{lstlisting}[mathescape=true]
			$\tau$ ::= $\mathbf{Bool}$ | $\tau_1 \times \tau_2$
			$v$ ::= T | F | ($v$, $v$)
			aexp ::= $y$ | $v$
			exp ::=  aexp | fst aexp | snd aexp | (aexp, aexp) | let $y$ = exp in exp | flip $\theta$ 
			| if aexp then exp else exp | observe aexp
			p ::= exp
		\end{lstlisting}
		
		\section{Semantics of Dice}

		V is the set of all Dice values. A (normalized) discrete probability distribution on V is a function \(\text{pr}: V \rightarrow [0, 1]\) such that \(\sum_{v \in V} \text{pr}(v) = 1\). An unnormalized probability distribution on V is a function \(\text{upr}: V \rightarrow [0, \infty)\).
		
		The semantic function \(\dbracket{.}\) maps Dice programs to unnormalized probability distribution.
		
		\[\dbracket{.} : \texttt{p} \rightarrow (V \rightarrow [0, \infty))\]
		
		The function $\delta(v)$ is a
		probability distribution that assigns a probability of 1 to the value $v$ and 0
		to all other values.
		\begin{align*}
			\dbracket{v_1}(v) \defeq \big(\delta(v_1)\big)(v)
			\label{Dice-value}~\tag{Dice-value}
		\end{align*}
		\begin{align*}
			\dbracket{\Lfst{(v_1, v_2)}}(v) \defeq \big(\delta(v_1)\big)(v)
			\label{Dice-tuple-fst}~\tag{Dice-tuple-fst}
		\end{align*}
		\begin{align*}
			\dbracket{\Lsnd{(v_1, v_2)}}(v) \defeq \big(\delta(v_2)\big)(v)
			\label{Dice-tuple-snd}~\tag{Dice-tuple-snd}
		\end{align*}
		\begin{align*}
			\dbracket{\Lite{v_g}{\te_1}{\te_2}}(v) \defeq
			\begin{cases}
				\dbracket{\te_1}(v) ~& \text{if } v_g = \true\\  
				\dbracket{\te_2}(v) ~& \text{if } v_g = \false\\
				0 \quad& \text{otherwise}
			\end{cases}
			\label{Dice-Ite}~\tag{Dice-Ite}
		\end{align*}
		\begin{align*}
			\dbracket{\Lflip{\theta}}(v) \defeq& \begin{cases}
				\theta ~& \text{if }v = \true\\
				1-\theta ~& \text{if }v=\false\\
				0 ~& \text{otherwise}
			\end{cases}
			\label{Dice-flip}~\tag{Dice-flip}
		\end{align*}
		\begin{align*}
			\dbracket{\Lobs{v_1}}(v) \defeq
			\begin{cases}
				1 ~& \text{if } v_1 = \true \text{ and } v = \true,\\
				0 ~& \text{otherwise}\\
			\end{cases}
			\label{Dice-observe}~\tag{Dice-observe}
		\end{align*}
		\begin{align*}
			\dbracket{\Llet{\yvar{} = \te_1}{\te_2}}(v)\defeq
			\sum_{v'}\dbracket{\te_1}(v') \times \dbracket{\te_2[\yvar{} \mapsto v']}(v)
			\label{Dice-let}~\tag{Dice-let} 
		\end{align*}
		
		The distributional semantics function \(\dbracket{.}_D : \texttt{p} \rightarrow V \rightarrow [0, 1]\) takes as input a Dice program \texttt{p} and outputs a normalized probability distribution.
		\begin{flalign*}
			\dbracket{\texttt{p}}_D(v') = \frac{\dbracket{\texttt{p}}(v')}{\sum_{v \in V} \dbracket{\texttt{p}}(v)}
			~\tag{normalized-pr}\label{normalized-pr}
		\end{flalign*}
		
		\section{Proofs of \(b\)-equivalence}
		
		In the following proofs, we treat \texttt{T} as 1 and \texttt{F} as 0.
		
		The compilation judgement for mixed-gamma densities are of the form \(\Upsilon \comp_b \texttt{p}\) where \texttt{p} are Dice programs.

		Further, we give the rules of the compilation judgement inductively and alongside prove theorem~\ref{theorem:semantics-preserving} inductively. We first prove theorem~\ref{theorem:semantics-preserving} for \(\pdf{0}{\beta}\), then for \(\pdf{1}{\beta}\) and finally more generally for \(\pdf{\alpha}{\beta}\). We further extend it to any mixed-gamma density \(\Upsilon\).

		\subsection{Proof of Lemma~\ref{lemma:alpha-0-density}}
		
		\begin{proof}
			By \ref{Trans-expo0}, if \(\pdf{0}{\beta} \comp_b \texttt{p}\), then 
			
			$\texttt{p} = \begin{array}{l}
				\texttt{let (\(\yvar{1}\) = flip($\theta_1$) in}
				\\ \texttt{let (\(\yvar{2}\) = flip($\theta_2$) in}
				\\ \ldots
				\\ \texttt{let (\(\yvar{b}\) = flip($\theta_b$) in}
				\\ \texttt{((\(\yvar{1}\), ((\(\yvar{2}\), (\ldots, (\(\yvar{b}\)))\ldots))}
			\end{array}$ where $\theta_i = $ \texttt{flip\_param}(\(\beta, i\))
			
			Now, to prove that \(\pdf{0}{\beta}\) and \texttt{p} are \(b\)-equivalent, we need to prove the following by Definition~\ref{def:b-equivalence}.
			
			\[\forall r \in [0, 1]_b, \;\;\;\;\; \int_{r}^{r + \frac{1}{2^b}} \pdf{0}{\beta}(y) \; dy = \dbracket{\texttt{p}}_D(\bin{r}{b})\]
			
			Evaluating the left hand side,
			
			For all \(r \in [0, 1]_b\)
			
			\begin{flalign*}
				\int_{r}^{r + \frac{1}{2^b}} \pdf{0}{\beta}(y) \; dy =& \int_{r}^{r + \frac{1}{2^b}} \frac{e^{\beta y}}{\int_{0}^{1} e^{\beta z} \; dz} \; dy
				~\tag{Definition~\ref{def:general-gamma}}
				\\
				=& \begin{cases}
					\frac{e^{\beta \cdot 2^{-b}} -1}{e^{\beta} - 1}e^{\beta r} ~& \text{ if }\beta \neq 0
					\\  \frac{1}{2^b} ~& \text{ if }\beta = 0
				\end{cases}
			\end{flalign*}

			To evaluate the right hand side, we first evaluate \(\dbracket{\texttt{p}}\)
			
			For any \(r \in [0, 1]_b\), if \(\bin{r}{b} = (v_1, (v_2 (\ldots, v_b))\ldots)\) where \(v_i \in \{0, 1\}\)
			
			\begin{flalign*}
				\dbracket{\texttt{p}}(\bin{r}{b}) &= \dbracket{\texttt{p}}(v_1, (v_2 (\ldots, v_b))\ldots)
				\\
				=& \sum_{v'_1} \dbracket{\Lflip{\theta_1}}(v'_1) \times \dbracket{\begin{array}{l}
						\texttt{let (\(\yvar{2}\) = flip($\theta_2$) in}
						\\ \ldots
						\\ \texttt{let (\(\yvar{b}\) = flip($\theta_b$) in}
						\\ \texttt{(\(v'_1\), ((\(\yvar{2}\), (\ldots, (\(\yvar{b}\)))\ldots))}
				\end{array}}(v_1, (v_2 (\ldots, v_b))\ldots)
				~\tag{\ref{Dice-let}}
				\\
				=& \sum_{v'_1} \dbracket{\Lflip{\theta_1}}(v'_1) \times \sum_{v'_2} \dbracket{\Lflip{\theta_2}}(v'_2)
				\\
				&\times \dbracket{\begin{array}{l}
						\texttt{let (\(\yvar{3}\) = flip($\theta_2$) in}
						\\ \ldots
						\\ \texttt{let (\(\yvar{b}\) = flip($\theta_b$) in}
						\\ \texttt{(\(v'_1\), (\(v'_2\), (\ldots, (\(\yvar{b}\)))\ldots))}
				\end{array}}(v_1, (v_2 (\ldots, v_b))\ldots)
				~\tag{\ref{Dice-let}}
				\\
				=&  \sum_{v'_1} \dbracket{\Lflip{\theta_1}}(v'_1) \times \sum_{v'_2} \dbracket{\Lflip{\theta_2}}(v'_2) \ldots \times \sum_{v'_b} \dbracket{\Lflip{\theta_b}}(v'_b) 
				\\
				&\times \delta((v'_1 (v'_2 \ldots v'_b)) (v_1, (v_2 (\ldots, v_b))\ldots)
				~\tag{\ref{Dice-let}}
				\\
				=&  \theta_1 \times \sum_{v'_2} \dbracket{\Lflip{\theta_2}}(v'_2) \ldots \times \sum_{v'_b} \dbracket{\Lflip{\theta_b}}(v'_b) 
				\times \delta((\true (v'_2 \ldots v'_b)) (v_1, (v_2 (\ldots, v_b))\ldots) 
				\\
				&+ (1 - \theta_1) \times \sum_{v'_2} \dbracket{\Lflip{\theta_2}}(v'_2) \ldots \times \sum_{v'_b} \dbracket{\Lflip{\theta_b}}(v'_b) 
				\times \delta((\false (v'_2 \ldots v'_b)) (v_1, (v_2 (\ldots, v_b))\ldots)
				~\tag{~\ref{Dice-flip}}
				\\
				=& \frac{e^{\frac{\beta}{2}}}{1 + e^{\frac{\beta}{2}}} \times \sum_{v'_2} \dbracket{\Lflip{\theta_2}}(v'_2) \ldots \times \sum_{v'_b} \dbracket{\Lflip{\theta_b}}(v'_b) 
				\times \delta((\true (v'_2 \ldots v'_b)) (v_1, (v_2 (\ldots, v_b))\ldots) 
				\\
				&+ \frac{1}{1 + e^{\frac{\beta}{2}}} \times \sum_{v'_2} \dbracket{\Lflip{\theta_2}}(v'_2) \ldots \times \sum_{v'_b} \dbracket{\Lflip{\theta_b}}(v'_b) 
				\times \delta((\false (v'_2 \ldots v'_b)) (v_1, (v_2 (\ldots, v_b))\ldots)
				~\tag{Definition~\ref{def:flip-parameter}}
				\\
				=& \frac{e^{\frac{\beta \cdot 1}{2}}}{1 + e^{\frac{\beta}{2}}} \times \sum_{v'_2} \dbracket{\Lflip{\theta_2}}(v'_2) \ldots \times \sum_{v'_b} \dbracket{\Lflip{\theta_b}}(v'_b) 
				\times \delta((\true (v'_2 \ldots v'_b)) (v_1, (v_2 (\ldots, v_b))\ldots) 
				\\
				&+ \frac{e^{\frac{\beta \cdot 0}{2}}}{1 + e^{\frac{\beta}{2}}} \times \sum_{v'_2} \dbracket{\Lflip{\theta_2}}(v'_2) \ldots \times \sum_{v'_b} \dbracket{\Lflip{\theta_b}}(v'_b) 
				\times \delta((\false (v'_2 \ldots v'_b)) (v_1, (v_2 (\ldots, v_b))\ldots)
				~\tag{\(\frac{1}{1 + e^{\frac{\beta}{2^b}}} = \frac{e^{\frac{\beta \cdot 0}{2^b}}}{1 + e^{\frac{\beta}{2^b}}}\)}
				\\
				=& \sum_{v'_1}\frac{e^{\frac{\beta \cdot v'_1}{2}}}{1 + e^{\frac{\beta}{2}}} \sum_{v'_2} \dbracket{\Lflip{\theta_2}}(v'_2) \ldots \sum_{v'_b} \dbracket{\Lflip{\theta_b}}(v'_b) 
				\times \delta((v'_1 (v'_2 \ldots v'_b)) (v_1, (v_2 (\ldots, v_b))\ldots) 
				~\tag{considering \true as 1, \false as 0}
				\\
				=& \sum_{v'_1}\frac{e^{\frac{\beta \cdot v'_1}{2}}}{1 + e^{\frac{\beta}{2}}} \sum_{v'_2} \frac{e^{\frac{\beta \cdot v'_2}{2^2}}}{1 + e^{\frac{\beta}{2^2}}} \ldots \sum_{v'_b} \dbracket{\Lflip{\theta_b}}(v'_b) 
				\times \delta((v'_1 (v'_2 \ldots v'_b)) (v_1, (v_2 (\ldots, v_b))\ldots) 
				~\tag{repeating the last 4 steps for \(v'_2\)}		
			\end{flalign*}
			\begin{flalign*}
				=& \sum_{v'_1}\frac{e^{\frac{\beta \cdot v'_1}{2}}}{1 + e^{\frac{\beta}{2}}} \sum_{v'_2} \frac{e^{\frac{\beta \cdot v'_2}{2^2}}}{1 + e^{\frac{\beta}{2^2}}} \ldots \sum_{v'_b} \frac{e^{\frac{\beta \cdot v'_b}{2^b}}}{1 + e^{\frac{\beta}{2^b}}} 
				\times \delta((v'_1 (v'_2 \ldots v'_b)) (v_1, (v_2 (\ldots, v_b))\ldots) 
				~\tag{repeating the last 4 steps for all \(v'_i\)}	
			\end{flalign*}
			In the above sum, the only non zero term would be one where \(v'_1 = v_1, \ldots v'_b = v_b\), thus
			\begin{flalign*}
				\dbracket{\texttt{p}}(\bin{r}{b}) =& \frac{e^{\frac{\beta v_1}{2}}}{1 + e^{\frac{\beta}{2}}} \cdot \frac{e^{\frac{\beta v_2}{4}}}{1 + e^{\frac{\beta}{4}}} \ldots \frac{e^{\frac{\beta v_b}{2^b}}}{1 + e^{\frac{\beta}{2^b}}}
				\\
				=& \begin{cases}
					\frac{e^{\beta \cdot 2^{-b}} -1}{e^{\beta} - 1}e^{\beta r} ~& \text{ if }\beta \neq 0
					\\  \frac{1}{2^b} ~& \text{ if }\beta = 0
				\end{cases}
				~\tag{Definition~\ref{def:binarize}}
			\end{flalign*}
			
			Evaluating the right hand side now,
			
			\begin{flalign*}
				\dbracket{\texttt{p}}_D(\bin{r}{b}) =& \frac{\dbracket{\texttt{p}}(\bin{r}{b})}{\sum_{r' \in [0, 1]_b} \dbracket{\texttt{p}}(\bin{r'}{b})} = \begin{cases}
					\frac{e^{\beta \cdot 2^{-b}} -1}{e^{\beta} - 1}e^{\beta r} ~& \text{ if }\beta \neq 0
					\\  \frac{1}{2^b} ~& \text{ if }\beta = 0
				\end{cases}
			\end{flalign*}
		\end{proof}
		
		\subsection{Proof of Lemma~\ref{lemma:alpha-1-density}}
		
		Now we describe the compilation judgement for \(\pdf{1}{\beta}\). To do so, we first describe the two helper judgements: \lessthan{.}{.}{.} and \unifobs{.}{.} and prove relevant lemmas about them.
		
		\subsubsection{Less Than} helper judgement is of the following form where \(\yvar{1}\) and \(\yvar{2}\) are Dice variables and \(b > 0 \in \mathbb{Z}^+\):
		
		\[\lessthan{\(\yvar{1}\)}{\(\yvar{2}\)}{b} = \texttt{p}\]

		In the rules for the helper judgement \lessthan{.}{.}{.}, Boolean operators \(\wedge, \vee, \neg\) are being used as syntactic sugars as follows
		
		\begin{flalign*}
			\yvar{1} \wedge \yvar{2} \rightarrow \Lite{\yvar{1}}{\yvar{2}}{\false}	
			~\tag{Dice-and}\label{Dice-and}
			\\
			\yvar{1} \vee \yvar{2} \rightarrow \Lite{\yvar{1}}{\true}{\yvar{2}}	
			~\tag{Dice-or}\label{Dice-or}
			\\
			\neg \yvar{1} \rightarrow \Lite{\yvar{1}}{\false}{\true}	
			~\tag{Dice-not}\label{Dice-not}
		\end{flalign*}
		
		Now we describe the rules for the \lessthan{.}{.}{.} judgement.
		
		\begin{align*}
			\inference[]{
			}{
				\texttt{less\_than(\(\yvar{1}\), \(\yvar{2}\), 1) = ((\(\yvar{2}\) $\wedge$ $\neg$(\(\yvar{1}\))}
			}~\label{Less-Than-1}~\tag{Less-Than-1}
		\end{align*}
		
		\begin{align*}
			\inference[]{
				\texttt{fresh \(\yvar{1f}\), \(\yvar{2f}\), \(\yvar{1wins}\), \(\yvar{2wins}\), \(\yvar{1s}\), \(\yvar{2s}\)}
				\\
				\texttt{b} > 1 \;\;\;\;\;\;\;\;\;\;\;\;
				\texttt{less\_than(\(\yvar{1s}\), \(\yvar{2s}\), b-1)} = \prog{+}
			}{		
				\texttt{less\_than(\(\yvar{1}\), \(\yvar{2}\), b)} = 
				\begin{array}{l}
					\texttt{let \(\yvar{1f}\) = fst(\(\yvar{1}\)) in} \\
					\texttt{let \(\yvar{2f}\) = fst(\(\yvar{2}\)) in} \\
					\texttt{let \(\yvar{1wins}\) = \(\yvar{1f}\) \(\wedge \neg\) \(\yvar{2f}\) in} \\
					\texttt{let \(\yvar{2wins}\) = \(\yvar{2f}\) \(\wedge \neg\) \(\yvar{1f}\) in} \\
					\texttt{if \(\yvar{1wins}\) then \false else} \\
					\texttt{if \(\yvar{2wins}\) then \true else} \\
					\texttt{let \(\yvar{1s}\) = snd(\(\yvar{1}\)) in} \\
					\texttt{let \(\yvar{2s}\) = snd(\(\yvar{2}\)) in} \\
					\prog{+}
				\end{array} 
			}~\label{Less-Than>1}~\tag{Less-Than>1}
		\end{align*}
		
		\begin{lemma}\label{lemma:lessthan}
			For any positive integer \(b\) such that \(\lessthan{\(\yvar{1}\)}{\(\yvar{2}\)}{b} = \texttt{p}\) and values \(r, r' \in [0, 1]_b\), the following holds
			\[\forall v \;\;\; \dbracket{\texttt{p}[\yvar{1} \mapsto \bin{r}{b}, \yvar{2} \mapsto \bin{r'}{b}]}(v) = \begin{cases}
				(\delta(\true))(v) ~& \text{ if } r < r'
				\\	(\delta(\false))(v) ~& \text{ otherwise}
			\end{cases}\]
		\end{lemma}
		
		\begin{proof}
			The proof is by induction on the number of bits, \(b\).\\
			\underline{Base Case:} \(b = 1\) \\
			Then \texttt{p} = (\(\yvar{2} \wedge \neg \yvar{1}\)) by \ref{Less-Than-1} 
			\\
			Also \(r, r' \in [0, 1]_1 \implies r, r' \in \{0, 0.5\}\) by Definition~\ref{def:discretize-unit-interval}. 
			\\
			We also have by Definition~\ref{def:binarize}, \(\bin{0}{1} = (\false)\) and \(\bin{0.5}{1} = (\true)\).
			\\
			Then, by case analysis
			\begin{itemize}
				\item \(r = 0, r' = 0\) and \(r < r'\) does not hold. 
				\begin{flalign*}
					\dbracket{\yvar{2} \wedge \neg \yvar{1}[\yvar{1} \mapsto \bin{0}{1}, \yvar{2} \mapsto \bin{0}{1}]}(v) =& \dbracket{\yvar{2} \wedge \neg \yvar{1}[\yvar{1} \mapsto \false, \yvar{2} \mapsto \false]}(v)
					~\tag{Definition~\ref{def:binarize}}
					\\
					=& \dbracket{\false \wedge \neg \false}(v) 
					~\tag{substitution}
					\\
					=& \dbracket{\Lite{\false}{\neg \false}{\false}}(v)
					~\tag{\ref{Dice-and}}
					\\
					=& \dbracket{\false}(v) = \delta(\false)(v)
					~\tag{\ref{Dice-Ite} and \ref{Dice-value}}
				\end{flalign*}
				
				\item \(r = 0, r' = 0.5\) and \(r < r'\) holds. 
				\begin{flalign*}
					\dbracket{\yvar{2} \wedge \neg \yvar{1}[\yvar{1} \mapsto \bin{0}{1}, \yvar{2} \mapsto \bin{0.5}{1}]}(v) =& \dbracket{\yvar{2} \wedge \neg \yvar{1}[\yvar{1} \mapsto \false, \yvar{2} \mapsto \true]}(v)
					~\tag{Definition~\ref{def:binarize}}
					\\
					=& \dbracket{\true \wedge \neg \false}(v) 
					~\tag{substitution}
					\\
					=& \dbracket{\Lite{\true}{\neg \false}{\false}}(v)
					~\tag{\ref{Dice-and}}
					\\
					=& \dbracket{\neg \false}(v) = \dbracket{\Lite{\false}{\false}{\true}}(v)
					~\tag{\ref{Dice-Ite} and \ref{Dice-not}}
					\\
					=& \dbracket{\true}(v) = \delta(\true)(v)
					~\tag{\ref{Dice-Ite} and \ref{Dice-value}}
				\end{flalign*}
				
				\item \(r = 0.5, r' = 0\) and \(r < r'\) does not hold. 
				\begin{flalign*}
					\dbracket{\yvar{2} \wedge \neg \yvar{1}[\yvar{1} \mapsto \bin{0.5}{1}, \yvar{2} \mapsto \bin{0}{1}]}(v) =& \dbracket{\yvar{2} \wedge \neg\yvar{1}[\yvar{1} \mapsto \true, \yvar{2} \mapsto \false]}(v)
					~\tag{Definition~\ref{def:binarize}}
					\\
					=& \dbracket{\false \wedge \neg \true}(v) 
					~\tag{substitution}
					\\
					=& \dbracket{\Lite{\false}{\neg \true}{\false}}(v)
					~\tag{\ref{Dice-and}}
					\\
					=& \dbracket{\false}(v) = \delta(\false)(v)
					~\tag{\ref{Dice-Ite} and \ref{Dice-value}}
				\end{flalign*}	
				
				\item \(r = 0.5, r' = 0.5\) and \(r < r'\) does not hold. 
				\begin{flalign*}
					\dbracket{\yvar{2} \wedge \neg\yvar{1}[\yvar{1} \mapsto \bin{0.5}{1}, \yvar{2} \mapsto \bin{0.5}{1}]}(v) =& \dbracket{\yvar{2} \wedge \neg\yvar{1}[\yvar{1} \mapsto \true, \yvar{2} \mapsto \true]}(v)
					~\tag{Definition~\ref{def:binarize}}
					\\
					=& \dbracket{\true \wedge \neg \true}(v) 
					~\tag{substitution}
					\\
					=& \dbracket{\Lite{\true}{\neg \true}{\false}}(v)
					~\tag{\ref{Dice-and}}
					\\
					=& \dbracket{\neg \true}(v) = \dbracket{\Lite{\true}{\false}{\true}}(v)
					~\tag{\ref{Dice-Ite} and \ref{Dice-not}}
					\\
					=& \dbracket{\false}(v) = \delta(\false)(v)
					~\tag{\ref{Dice-Ite} and ~\ref{Dice-value}}
				\end{flalign*}		
			\end{itemize}
			\underline{Induction Hypothesis:} \(\forall b' < b\), \(\forall r'', r''' \in [0, 1]_{b'}\) if \(\lessthan{\(\yvar{1s}\)}{\(\yvar{2s}\)}{b'} = \texttt{p'}\) then the following holds:
			\[\dbracket{\texttt{p'}[\yvar{1s} \mapsto \bin{r''}{b'}, \yvar{2s} \mapsto \bin{r'''}{b'}]}(v) = \begin{cases}
				\delta(\true)(v) ~& \text{ if } r'' < r'''
				\\	\delta(\false)(v) ~& \text{ otherwise}
			\end{cases}\]
			\\
			\underline{Inductive Step:}\\
			
			Let \(\lessthan{\(\yvar{1}\)}{\(\yvar{2}\)}{b} = \texttt{p}\) and \(r, r' \in [0, 1]_b\).
			
			Also, let \lessthan{\(\yvar{1'}\)}{\(\yvar{2'}\)}{b-1} \(=\) \texttt{p'}.
			Let \(\bin{r}{b} = (v_1, (v_2, (\ldots v_b)\ldots))\) and \(\bin{r'}{b} = (v'_1, (v'_2, (\ldots v'_b)\ldots))\) in accordance to Definition~\ref{def:binarize}
			
			\begin{flalign*}
				&\dbracket{\texttt{p}[\yvar{1} \mapsto \bin{r}{b}, \yvar{2} \mapsto \bin{r'}{b}]}(v) 
				\\
				=& \dbracket{\begin{array}{l}
						\texttt{let \(\yvar{1f}\) = fst(\(\yvar{1}\)) in} \\
						\texttt{let \(\yvar{2f}\) = fst(\(\yvar{2}\)) in} \\
						\texttt{let \(\yvar{1wins}\) = \(\yvar{1f}\) \(\wedge \neg\) \(\yvar{2f}\) in} \\
						\texttt{let \(\yvar{2wins}\) = \(\yvar{2f}\) \(\wedge \neg\) \(\yvar{1f}\) in} \\
						\texttt{if \(\yvar{1wins}\) then \false else} \\
						\texttt{if \(\yvar{2wins}\) then \true else} \\
						\texttt{let \(\yvar{1s}\) = snd(\(\yvar{1}\)) in} \\
						\texttt{let \(\yvar{2s}\) = snd(\(\yvar{2}\)) in} \\
						\prog{+}
					\end{array}  
					[\yvar{1} \mapsto \bin{r}{b}, \yvar{2} \mapsto \bin{r'}{b}]}(v)
				~\tag{\ref{Less-Than>1}}
			\end{flalign*}
			
			\begin{flalign*}
				=& \dbracket{\begin{array}{l}
						\texttt{let \(\yvar{1f}\) = fst(\((v_1, (v_2, (\ldots v_b)\ldots))\)) in} \\
						\texttt{let \(\yvar{2f}\) = fst(\((v'_1, (v'_2, (\ldots v'_b)\ldots))\)) in} \\
						\texttt{let \(\yvar{1wins}\) = \(\yvar{1f}\) \(\wedge \neg\) \(\yvar{2f}\) in} \\
						\texttt{let \(\yvar{2wins}\) = \(\yvar{2f}\) \(\wedge \neg\) \(\yvar{1f}\) in} \\
						\texttt{if \(\yvar{1wins}\) then \false else} \\
						\texttt{if \(\yvar{2wins}\) then \true else} \\
						\texttt{let \(\yvar{1s}\) = snd(\((v_1, (v_2, (\ldots v_b)\ldots))\)) in} \\
						\texttt{let \(\yvar{2s}\) = snd(\((v'_1, (v'_2, (\ldots v'_b)\ldots))\)) in} \\
						\prog{+}
				\end{array}}(v)
				~\tag{substitution}
				\\
				=& \sum_{v'} \dbracket{\Lfst{(v_1, (v_2, (\ldots v_b)\ldots))}}(v') \times \dbracket{\begin{array}{l}
						\texttt{let \(\yvar{2f}\) = fst(\((v'_1, (v'_2, (\ldots v'_b)\ldots))\)) in} \\
						\texttt{let \(\yvar{1wins}\) = \(v' \wedge \neg\) \(\yvar{2f}\) in} \\
						\texttt{let \(\yvar{2wins}\) = \(\yvar{2f}\) \(\wedge \neg v'\) in} \\
						\texttt{if \(\yvar{1wins}\) then \false else} \\
						\texttt{if \(\yvar{2wins}\) then \true else} \\
						\texttt{let \(\yvar{1s}\) = snd(\((v_1, (v_2, (\ldots v_b)\ldots))\)) in} \\
						\texttt{let \(\yvar{2s}\) = snd(\((v'_1, (v'_2, (\ldots v'_b)\ldots))\)) in} \\
						\prog{+}
				\end{array}}(v)
				~\tag{\ref{Dice-let}}
				\\
				=& \sum_{v'} \delta(v_1)(v') \times \dbracket{\begin{array}{l}
						\texttt{let \(\yvar{2f}\) = fst(\((v'_1, (v'_2, (\ldots v'_b)\ldots))\)) in} \\
						\texttt{let \(\yvar{1wins}\) = \(v' \wedge \neg\) \(\yvar{2f}\) in} \\
						\texttt{let \(\yvar{2wins}\) = \(\yvar{2f}\) \(\wedge \neg v'\) in} \\
						\texttt{if \(\yvar{1wins}\) then \false else} \\
						\texttt{if \(\yvar{2wins}\) then \true else} \\
						\texttt{let \(\yvar{1s}\) = snd(\((v_1, (v_2, (\ldots v_b)\ldots))\)) in} \\
						\texttt{let \(\yvar{2s}\) = snd(\((v'_1, (v'_2, (\ldots v'_b)\ldots))\)) in} \\
						\prog{+}
				\end{array}}(v)
				~\tag{\ref{Dice-tuple-fst}}
				\\
				=& \dbracket{\begin{array}{l}
						\texttt{let \(\yvar{2f}\) = fst(\((v'_1, (v'_2, (\ldots v'_b)\ldots))\)) in} \\
						\texttt{let \(\yvar{1wins}\) = \(v_1 \wedge \neg\) \(\yvar{2f}\) in} \\
						\texttt{let \(\yvar{2wins}\) = \(\yvar{2f}\) \(\wedge \neg v_1\) in} \\
						\texttt{if \(\yvar{1wins}\) then \false else} \\
						\texttt{if \(\yvar{2wins}\) then \true else} \\
						\texttt{let \(\yvar{1s}\) = snd(\((v_1, (v_2, (\ldots v_b)\ldots))\)) in} \\
						\texttt{let \(\yvar{2s}\) = snd(\((v'_1, (v'_2, (\ldots v'_b)\ldots))\)) in} \\
						\prog{+}
				\end{array}}(v)
				~\tag{definition of Delta distribution (\(v' = v_1\))}
				\\
				=& \sum_{v'} \dbracket{\Lfst{(v'_1, (v'_2, (\ldots v'_b)\ldots))}}(v') \times \dbracket{\begin{array}{l}
						\texttt{let \(\yvar{1wins}\) = \(v_1 \wedge \neg v'\) in} \\
						\texttt{let \(\yvar{2wins}\) = \(v' \wedge \neg v_1\) in} \\
						\texttt{if \(\yvar{1wins}\) then \false else} \\
						\texttt{if \(\yvar{2wins}\) then \true else} \\
						\texttt{let \(\yvar{1s}\) = snd(\((v_1, (v_2, (\ldots v_b)\ldots))\)) in} \\
						\texttt{let \(\yvar{2s}\) = snd(\((v'_1, (v'_2, (\ldots v'_b)\ldots))\)) in} \\
						\prog{+}
				\end{array}}(v)
				~\tag{\ref{Dice-let}}
			\end{flalign*}
			\begin{flalign*}
				=& \sum_{v'} \delta(v'_1)(v') \times \dbracket{\begin{array}{l}
						\texttt{let \(\yvar{1wins}\) = \(v_1 \wedge \neg v'\) in} \\
						\texttt{let \(\yvar{2wins}\) = \(v' \wedge \neg v_1\) in} \\
						\texttt{if \(\yvar{1wins}\) then \false else} \\
						\texttt{if \(\yvar{2wins}\) then \true else} \\
						\texttt{let \(\yvar{1s}\) = snd(\((v_1, (v_2, (\ldots v_b)\ldots))\)) in} \\
						\texttt{let \(\yvar{2s}\) = snd(\((v'_1, (v'_2, (\ldots v'_b)\ldots))\)) in} \\
						\prog{+} 
				\end{array}}(v)
				~\tag{\ref{Dice-tuple-fst}}
				\\
				=&  \dbracket{\begin{array}{l}
						\texttt{let \(\yvar{1wins}\) = \(v_1 \wedge \neg v'_1\) in} \\
						\texttt{let \(\yvar{2wins}\) = \(v'_1 \wedge \neg v_1\) in} \\
						\texttt{if \(\yvar{1wins}\) then \false else} \\
						\texttt{if \(\yvar{2wins}\) then \true else} \\
						\texttt{let \(\yvar{1s}\) = snd(\((v_1, (v_2, (\ldots v_b)\ldots))\)) in} \\
						\texttt{let \(\yvar{2s}\) = snd(\((v'_1, (v'_2, (\ldots v'_b)\ldots))\)) in} \\
						\prog{+} 
				\end{array}}(v)
				~\tag{\ref{Dice-value}}
				\\
				=& \sum_{v'} \dbracket{v_1 \wedge \neg v'_1}(v') \times
				\sum_{v''} \dbracket{v'_1 \wedge \neg v_1}(v'') \times \dbracket{\begin{array}{l}
						\texttt{if \(v'\) then \false else} \\
						\texttt{if \(v''\) then \true else} \\
						\texttt{let \(\yvar{1s}\) = snd(\((v_1, (v_2, (\ldots v_b)\ldots))\)) in} \\
						\texttt{let \(\yvar{2s}\) = snd(\((v'_1, (v'_2, (\ldots v'_b)\ldots))\)) in} \\
						\prog{+} 
				\end{array}}(v)
				~\tag{\ref{Dice-let}}
			\end{flalign*}
			
			We evaluate further by case analysis on \(r\) and \(r'\)
			
			\underline{Case 1:} \(r < 0.5, r' \geq 0.5\).
			
			This implies that \(v_1 = \false, v'_1 = \true\). Also, \(r < r'\)
			
			\begin{flalign*}
				&\dbracket{\texttt{p}[\yvar{1} \mapsto \bin{r}{b}, \yvar{2} \mapsto \bin{r'}{b}]}(v) 
				\\	
				=& \sum_{v'} \dbracket{\false \wedge \neg \true}(v') \times
				\sum_{v''} \dbracket{\true \wedge \neg \false}(v'') \times \dbracket{\begin{array}{l}
						\texttt{if \(v'\) then \false else} \\
						\texttt{if \(v''\) then \true else} \\
						\texttt{let \(\yvar{1s}\) = snd(\((v_1, (v_2, (\ldots v_b)\ldots))\)) in} \\
						\texttt{let \(\yvar{2s}\) = snd(\((v'_1, (v'_2, (\ldots v'_b)\ldots))\)) in} \\
						\prog{+} 
				\end{array}}(v)
				\\
				=& \sum_{v'} \delta(\false)(v') \times
				\sum_{v''} \delta(\true)(v'') \times \dbracket{\begin{array}{l}
						\texttt{if \(v'\) then \false else} \\
						\texttt{if \(v''\) then \true else} \\
						\texttt{let \(\yvar{1s}\) = snd(\((v_1, (v_2, (\ldots v_b)\ldots))\)) in} \\
						\texttt{let \(\yvar{2s}\) = snd(\((v'_1, (v'_2, (\ldots v'_b)\ldots))\)) in} \\
						\prog{+} 
				\end{array}}(v) 
				~\tag{\ref{Dice-and}}
				\\
				=& \dbracket{\begin{array}{l}
						\texttt{if \(\false\) then \false else} \\
						\texttt{if \(\true\) then \true else} \\
						\texttt{let \(\yvar{1s}\) = snd(\((v_1, (v_2, (\ldots v_b)\ldots))\)) in} \\
						\texttt{let \(\yvar{2s}\) = snd(\((v'_1, (v'_2, (\ldots v'_b)\ldots))\)) in} \\
						\prog{+} 
				\end{array}}(v)
				\\
				=& \dbracket{\begin{array}{l}
						\texttt{if \(\true\) then \true else} \\
						\texttt{let \(\yvar{1s}\) = snd(\((v_1, (v_2, (\ldots v_b)\ldots))\)) in} \\
						\texttt{let \(\yvar{2s}\) = snd(\((v'_1, (v'_2, (\ldots v'_b)\ldots))\)) in} \\
						\prog{+} 
				\end{array}}(v)
				= \delta(\true)(v)
				~\tag{\ref{Dice-Ite}}
			\end{flalign*}
			
			\underline{Case 2:} \(r \geq 0.5, r' < 0.5\).
			
			This implies that \(v_1 = \true, v'_1 = \false\). Also, \(r > r'\)
			
			\begin{flalign*}
				&\dbracket{\texttt{p}[\yvar{1} \mapsto \bin{r}{b}, \yvar{2} \mapsto \bin{r'}{b}]}(v) 
				\\	
				=& \sum_{v'} \dbracket{\true \wedge \neg \false}(v') \times
				\sum_{v''} \dbracket{\false \wedge \neg \true}(v'') \times \dbracket{\begin{array}{l}
						\texttt{if \(v'\) then \false else} \\
						\texttt{if \(v''\) then \true else} \\
						\texttt{let \(\yvar{1s}\) = snd(\((v_1, (v_2, (\ldots v_b)\ldots))\)) in} \\
						\texttt{let \(\yvar{2s}\) = snd(\((v'_1, (v'_2, (\ldots v'_b)\ldots))\)) in} \\
						\prog{+} 
				\end{array}}(v)
				\\
				=& \sum_{v'} \delta(\true)(v') \times
				\sum_{v''} \delta(\false)(v'') \times \dbracket{\begin{array}{l}
						\texttt{if \(v'\) then \false else} \\
						\texttt{if \(v''\) then \true else} \\
						\texttt{let \(\yvar{1s}\) = snd(\((v_1, (v_2, (\ldots v_b)\ldots))\)) in} \\
						\texttt{let \(\yvar{2s}\) = snd(\((v'_1, (v'_2, (\ldots v'_b)\ldots))\)) in} \\
						\prog{+} 
				\end{array}}(v) 
				~\tag{\ref{Dice-and}}
				\\
				=& \dbracket{\begin{array}{l}
						\texttt{if \(\true\) then \false else} \\
						\texttt{if \(\false\) then \true else} \\
						\texttt{let \(\yvar{1s}\) = snd(\((v_1, (v_2, (\ldots v_b)\ldots))\)) in} \\
						\texttt{let \(\yvar{2s}\) = snd(\((v'_1, (v'_2, (\ldots v'_b)\ldots))\)) in} \\
						\prog{+} 
				\end{array}}(v)
				= \delta(\false)(v)
				~\tag{\ref{Dice-Ite}}
			\end{flalign*}
			
			\underline{Case 3:} \(r < 0.5, r' < 0.5\).
			
			This implies that \(v_1 = \false, v'_1 = \false\). 
			
			Also, \((v_2, (\ldots v_b)\ldots) = \bin{2r}{b-1}\) and \((v'_2, (\ldots v'_b)\ldots) = \bin{2r'}{b-1}\).
			
			\begin{flalign*}
				&\dbracket{\texttt{p}[\yvar{1} \mapsto \bin{r}{b}, \yvar{2} \mapsto \bin{r'}{b}]}(v) 
				\\	
				=& \sum_{v'} \dbracket{\false \wedge \neg \false}(v') \times
				\sum_{v''} \dbracket{\false \wedge \neg \false}(v'') \times \dbracket{\begin{array}{l}
						\texttt{if \(v'\) then \false else} \\
						\texttt{if \(v''\) then \true else} \\
						\texttt{let \(\yvar{1s}\) = snd(\((v_1, (v_2, (\ldots v_b)\ldots))\)) in} \\
						\texttt{let \(\yvar{2s}\) = snd(\((v'_1, (v'_2, (\ldots v'_b)\ldots))\)) in} \\
						\prog{+}
				\end{array}}(v)
				\\
				=& \sum_{v'} \delta(\false)(v') \times
				\sum_{v''} \delta(\false)(v'') \times \dbracket{\begin{array}{l}
						\texttt{if \(v'\) then \false else} \\
						\texttt{if \(v''\) then \true else} \\
						\texttt{let \(\yvar{1s}\) = snd(\((v_1, (v_2, (\ldots v_b)\ldots))\)) in} \\
						\texttt{let \(\yvar{2s}\) = snd(\((v'_1, (v'_2, (\ldots v'_b)\ldots))\)) in} \\
						\prog{+} 
				\end{array}}(v) 
				~\tag{\ref{Dice-and}}
				\\
				=& \dbracket{\begin{array}{l}
						\texttt{if \(\false\) then \false else} \\
						\texttt{if \(\false\) then \true else} \\
						\texttt{let \(\yvar{1s}\) = snd(\((v_1, (v_2, (\ldots v_b)\ldots))\)) in} \\
						\texttt{let \(\yvar{2s}\) = snd(\((v'_1, (v'_2, (\ldots v'_b)\ldots))\)) in} \\
						\prog{+} 
				\end{array}}(v)
				\\
				=& \dbracket{\begin{array}{l}
						\texttt{let \(\yvar{1s}\) = snd(\((v_1, (v_2, (\ldots v_b)\ldots))\)) in} \\
						\texttt{let \(\yvar{2s}\) = snd(\((v'_1, (v'_2, (\ldots v'_b)\ldots))\)) in} \\
						\prog{+} 
				\end{array}}(v)
				~\tag{\ref{Dice-Ite}}
			\end{flalign*}
			
			\begin{flalign*}
				=& \sum_{v'} \dbracket{\Lsnd{(v_1, (v_2, (\ldots v_b)\ldots))}}(v') \times \sum_{v''} \dbracket{\Lsnd{(v'_1, (v'_2, (\ldots v'_b)\ldots))}}(v'') \times 
				\\
				&\dbracket{\texttt{p'}[\yvar{1s} \mapsto v', \yvar{2s} \mapsto v'']}(v)
				~\tag{\ref{Dice-let}}
				\\
				=& \sum_{v'} \delta((v_2, (\ldots v_b)\ldots))(v') \times \sum_{v''} \delta((v'_2, (\ldots v'_b)\ldots))(v'') \times 
				\dbracket{\texttt{p'}[\yvar{1s} \mapsto v', \yvar{2s} \mapsto v'']}(v)
				~\tag{\ref{Dice-tuple-snd}}
				\\
				=& \dbracket{\texttt{p'}[\yvar{1s} \mapsto (v_2, (\ldots v_b)\ldots), \yvar{2s} \mapsto (v'_2, (\ldots v'_b)\ldots)]}(v)
				~\tag{Considering the only non zero term in accordance to definition of \(\delta\) distribution}
				\\
				=& \dbracket{\texttt{p'}[\yvar{1s} \mapsto \bin{2r}{b-1}, \yvar{2s} \mapsto \bin{2r'}{b-1}]}(v)
				\\
				=& \begin{cases}
					\delta(\true)(v) & 2r < 2r'
					\\	\delta(\false)(v) & \text{otherwise}
				\end{cases}
				= \begin{cases}
					\delta(\true)(v) & r < r'
					\\	\delta(\false)(v) & \text{otherwise}
				\end{cases}
				~\tag{Induction Hypothesis}
			\end{flalign*}
			
			\underline{Case 4:} \(r \geq 0.5, r' \geq 0.5\).
			
			This implies that \(v_1 = \true, v'_1 = \true\). 
			
			Also, \((v_2, (\ldots v_b)\ldots) = \bin{2r - 1}{b-1}\) and \((v'_2, (\ldots v'_b)\ldots) = \bin{2r' - 1}{b-1}\).
			
			\begin{flalign*}
				&\dbracket{\texttt{p}[\yvar{1} \mapsto \bin{r}{b}, \yvar{2} \mapsto \bin{r'}{b}]}(v) 
				\\	
				=& \sum_{v'} \dbracket{\true \wedge \neg \true}(v') \times
				\sum_{v''} \dbracket{\true \wedge \neg \true}(v'') \times \dbracket{\begin{array}{l}
						\texttt{if \(v'\) then \false else} \\
						\texttt{if \(v''\) then \true else} \\
						\texttt{let \(\yvar{1s}\) = snd(\((v_1, (v_2, (\ldots v_b)\ldots))\)) in} \\
						\texttt{let \(\yvar{2s}\) = snd(\((v'_1, (v'_2, (\ldots v'_b)\ldots))\)) in} \\
						\prog{+} 
				\end{array}}(v)
				\\
				=& \sum_{v'} \delta(\false)(v') \times
				\sum_{v''} \delta(\false)(v'') \times \dbracket{\begin{array}{l}
						\texttt{if \(v'\) then \false else} \\
						\texttt{if \(v''\) then \true else} \\
						\texttt{let \(\yvar{1s}\) = snd(\((v_1, (v_2, (\ldots v_b)\ldots))\)) in} \\
						\texttt{let \(\yvar{2s}\) = snd(\((v'_1, (v'_2, (\ldots v'_b)\ldots))\)) in} \\
						\prog{+} 
				\end{array}}(v) 
				~\tag{\ref{Dice-and}}
				\\
				=& \dbracket{\texttt{p'}[\yvar{1s} \mapsto (v_2, (\ldots v_b)\ldots), \yvar{2s} \mapsto (v'_2, (\ldots v'_b)\ldots)]}(v)
				~\tag{As described in Case 3}
				\\
				=& \dbracket{\texttt{p'}[\yvar{1s} \mapsto \bin{2r-1}{b-1}, \yvar{2s} \mapsto \bin{2r'-1}{b-1}]}(v)
				\\
				=& \begin{cases}
					\delta(\true)(v) & 2r - 1 < 2r' - 1
					\\	\delta(\false)(v) & \text{otherwise}
				\end{cases}
				= \begin{cases}
					\delta(\true)(v) & r < r'
					\\	\delta(\false)(v) & \text{otherwise}
				\end{cases}
				~\tag{Induction Hypothesis}
			\end{flalign*}
			
		\end{proof}
		
		\subsubsection{unifObs} helper judgement is of the following form where \(\yvar{}\) is a Dice variable and \(b > 0 \in \mathbb{Z}^+\)
		
		\[\unifobs{\yvar{}}{b} = \prog{}\]
		
		The rules for the judgement are defined as follows:
		
		\begin{align*}
			\inference[]{
				\texttt{fresh \(\yvar{1}\), \(\yvar{bool}\)}
				\\
				\pdf{0}{0} \comp_b \prog{1} \;\;\;\;\;\;\;\;
				\texttt{less\_than(\(\yvar{1}\), \(\yvar{}\), b)} = \prog{2}
			}
			{
				\texttt{unifObs(\(\yvar{}\), b)} = \begin{array}{l}
					\texttt{let \(\yvar{1}\) = \(\prog{1}\) in} \\
					\texttt{let \(\yvar{bool}\) = \(\prog{2}\) in } \\
					\texttt{observe(\(\yvar{bool}\))}
			\end{array}}
			~\tag{unifObs}\label{unifobs}
		\end{align*}
		
		\begin{lemma}\label{lemma:unifobs}
			For any positive integer \(b\) such that \( \unifobs{\yvar{}}{b} = \texttt{p}\), \(r \in [0, 1]_b\) and Dice value \(v'\), the following holds
			\[\dbracket{\texttt{p}[\yvar{} \mapsto \bin{r}{b}]}(v') = \begin{cases}
				r ~& \text{ if } v' = \true
				\\	0 ~& \text{ otherwise}
			\end{cases}\]
		\end{lemma}
		
		\begin{proof}
			By \ref{unifobs}, we have
			
			\(\texttt{p} = \begin{array}{l}
				\texttt{let \(\yvar{1}\) = \(\prog{1}\) in} \\
				\texttt{let \(\yvar{bool}\) = \(\prog{2}\) in } \\
				\texttt{observe(\(\yvar{bool}\))}
			\end{array}\)
			
			where \(\pdf{0}{0} \comp_b \prog{1}\) and \(\lessthan{\(\yvar{1}\)}{\(\yvar{}\)}{b} = \prog{2}\)

			\begin{flalign*}
				&\dbracket{\texttt{p}[\yvar{} \mapsto \bin{r}{b}]}(v') 
				\\
				&= \dbracket{\begin{array}{l}
						\texttt{let \(\yvar{1}\) = \(\prog{1}\) in} \\
						\texttt{let \(\yvar{bool}\) = \(\prog{2}\) in } \\
						\texttt{observe(\(\yvar{bool}\))}
					\end{array}[\yvar{} \mapsto \bin{r}{b}]}(v') \\
				&= \sum_{v_1} \dbracket{\prog{1}[\yvar{} \mapsto \bin{r}{b}]}(v_1) \times \dbracket{\begin{array}{l}
						\texttt{let \(\yvar{bool}\) = \(\prog{2}\) in} \\
						\texttt{observe \(\yvar{bool}\)}
					\end{array}\bigg[\begin{array}{l}\yvar{1} \mapsto v_1 \\ \yvar{} \mapsto \bin{r}{b} \end{array}\bigg]}(v')
				~\tag{\ref{Dice-let}}
				\\
				&= \sum_{r_1 \in [0, 1]_b} \frac{1}{2^b} \times \dbracket{\begin{array}{l}
						\texttt{let \(\yvar{bool}\) = \(\prog{2}\) in} \\
						\texttt{observe \(\yvar{bool}\)}
					\end{array}\bigg[\begin{array}{l}\yvar{1} \mapsto \bin{r_1}{b} \\ \yvar{} \mapsto \bin{r}{b} \end{array}\bigg]}(v')
				~\tag{Lemma \ref{lemma:alpha-0-density}}
				\\
				&= \sum_{r_1 \in [0, 1]_b} \frac{1}{2^b} \times \sum_{v_2} \dbracket{\prog{2}[\yvar{1} \mapsto \bin{r_1}{b}, \yvar{} \mapsto \bin{v}{b}]}(v_2) 
				\times \dbracket{\Lobs{\yvar{bool}}\Bigg[\begin{array}{l}
						\yvar{bool} \mapsto v_2 \\
						\yvar{1} \mapsto \bin{r_1}{b} \\
						\yvar{} \mapsto \bin{r}{b}
					\end{array}
					\Bigg]}(v')
				~\tag{\ref{Dice-let}}
				\\
				&= \frac{1}{2^b} \sum_{r_1 \in [0, 1]_b} \sum_{v_2} \dbracket{\prog{2}[\yvar{1} \mapsto \bin{r_1}{b}, \yvar{} \mapsto \bin{r}{b}]}(v_2) \times \dbracket{\Lobs{v_2}}(v')
				\\
				&= \begin{cases}
					\frac{1}{2^b} \sum_{r_1 \in [0, 1]_b} \dbracket{\prog{2}[\yvar{1} \mapsto \bin{r_1}{b}, \yvar{} \mapsto \bin{r}{b}]}(\true) ~& v' = \true
					\\	0 ~& \text{otherwise}
				\end{cases}
				~\tag{\ref{Dice-observe}}
				\\
				&= \begin{cases}
					\frac{1}{2^b} \sum_{r_1 \in [0, 1]_b, r_1 < r} \delta(\true)(\true) + \frac{1}{2^b} \sum_{r_1 \in [0, 1]_b, r_1 \geq r} \delta(\false)(\true)	~& v' = \true
					\\ 0	~& \text{otherwise}
				\end{cases}
				~\tag{by Lemma~\ref{lemma:lessthan}}
				\\
				&= \begin{cases}
					r	~& v' = \true
					\\ 0	~& \text{otherwise}
				\end{cases}
			\end{flalign*}
		\end{proof}

		\subsubsection{\(\pdf{1}{\beta}\)} Now we describe the judgement \(\comp_b\) for \(\pdf{1}{\beta}\) using the helper judgement \texttt{unifObs} we defined above.

		\begin{lemma}\label{lemma:gp}
			The sum of a geometric progression \([1, r, r^2, r^3 \ldots r^n]\) is computed as follows:
			\[\sum_{0}^{a} r^n = \frac{r^{a + 1} - 1}{r - 1}\]
		\end{lemma}
		
		\begin{lemma}\label{lemma:agp}
			The sum of the arithmetic geometric progression \([0, r, 2r^2, \ldots n r^n]\) is computed as follows:
			\[\sum_{0}^{a} n r^n = \frac{r(ar^{a + 1} - (a + 1)r^a + 1)}{(r-1)^2} \]
		\end{lemma}

		\emph{Proof of Lemma~\ref{lemma:alpha-1-density}}
		
		\begin{proof}
			By ~\ref{Trans-expo1}, if \(\pdf{1}{\beta} \comp_b \prog{}\), then
			
			$\texttt{p} = \begin{array}{l}
				\texttt{let \(\yvar{1}\) = } \prog{1} \texttt{ in }
				\\ \texttt{let \_ = \(\prog{3}\)}
				\\ \texttt{let \(\yvar{2}\) = } \prog{1} \texttt{ in}
				\\ \texttt{let \(\yvar{3}\) = flip\((\theta)\) in}
				\\ \texttt{if \(\yvar{3}\) then \(\yvar{2}\) else \(\yvar{1}\)}
			\end{array}$ 
			
			where  \(\pdf{0}{\beta} \comp_b \prog{1}\) and
			\(\texttt{unifObs(\(\yvar{1}\), b)} = \prog{3}\)
			
			Now, to prove that \(\pdf{1}{\beta}\) and \texttt{p} are \(b\)-equivalent, we need to prove the following by Definition~\ref{def:b-equivalence}.
			
			\[\forall r \in [0, 1]_b, \;\;\;\;\; \int_{r}^{r + \frac{1}{2^b}} \pdf{1}{\beta}(y) \; dy = \dbracket{\texttt{p}}_D(\bin{r}{b})\]
			
			Evaluating the left hand side,
			
			For all \(r \in [0, 1]_b\)
			
			\begin{flalign*}
				\int_{r}^{r + \frac{1}{2^b}} \pdf{1}{\beta}(y) \; dy =& \int_{r}^{r + \frac{1}{2^b}} \frac{y e^{\beta y}}{\int_{0}^{1} z e^{\beta z} \; dz} \; dy
				~\tag{Definition~\ref{def:general-gamma}}
				\\
				=& 
				\frac{\beta(e^{\beta \cdot 2^{-b}}-1)re^{\beta r} + (\beta2^{-b}e^{\beta2^{-b}} - e^{\beta2^{-b}} + 1)e^{\beta r}}{e^{\beta}(\beta - 1) + 1} 
			\end{flalign*}
			
			To evaluate the right hand side, we first evaluate \(\dbracket{\texttt{p}}\)
			
			For any \(r \in [0, 1]_b\), if \(\bin{r}{b} = (v_1, (v_2 (\ldots, v_b))\ldots)\) where \(v_i \in \{0, 1\}\)
			
			\begin{flalign*}
				&\dbracket{\texttt{p}}(\bin{r}{b}) 
				\\
				=& \dbracket{\texttt{p}}(v_1, (v_2 ,(\ldots, v_b)\ldots))
				\\
				=& \sum_{v'} \dbracket{\prog{1}}(v') \times \dbracket{\begin{array}{l}
						\texttt{let \_ = \(\prog{3}\) in} \\
						\texttt{let \(\yvar{2}\) = \(\prog{1}\) in} \\
						\texttt{let \(\yvar{3}\) = flip(\(\theta\)) in} \\
						\texttt{if \(\yvar{3}\) then \(\yvar{2}\) else \(\yvar{1}\)}
					\end{array} [\yvar{1} \mapsto v']}(v_1, (v_2 ,(\ldots, v_b)\ldots))
				~\tag{\ref{Dice-let}}
				\\
				=& \sum_{r' \in [0, 1]_b} \frac{e^{\beta \cdot 2^{-b}} - 1}{e^{\beta} - 1} e^{\beta r'} \times \dbracket{\begin{array}{l}
						\texttt{let \_ = \(\prog{3}\) in} \\
						\texttt{let \(\yvar{2}\) = \(\prog{1}\) in} \\
						\texttt{let \(\yvar{3}\) = flip(\(\theta\)) in} \\
						\texttt{if \(\yvar{3}\) then \(\yvar{2}\) else \(\yvar{1}\)}
					\end{array} [\yvar{1} \mapsto \bin{r'}{b}]}(v_1, (v_2 ,(\ldots, v_b)\ldots))
				~\tag{Lemma~\ref{lemma:alpha-0-density}}
				\\
				=& \sum_{r' \in [0, 1]_b} \frac{e^{\beta \cdot 2^{-b}} - 1}{e^{\beta} - 1} e^{\beta r'} \times
				\sum_{v''} \dbracket{\prog{3} [\yvar{1} \mapsto \bin{r'}{b}]}(v'')
				\\
				&\times
				\dbracket{\begin{array}{l}
						\texttt{let \(\yvar{2}\) = \(\prog{1}\) in} \\
						\texttt{let \(\yvar{3}\) = flip(\(\theta\)) in} \\
						\texttt{if \(\yvar{3}\) then \(\yvar{2}\) else \(\yvar{1}\)}
					\end{array} [\yvar{1} \mapsto \bin{r'}{b}]}(v_1, (v_2 ,(\ldots, v_b)\ldots))
				~\tag{\ref{Dice-let}}
				\\
				=& \sum_{r' \in [0, 1]_b} \frac{e^{\beta \cdot 2^{-b}} - 1}{e^{\beta} - 1} e^{\beta r'} r' \times
				\dbracket{\begin{array}{l}
						\texttt{let \(\yvar{2}\) = \(\prog{1}\) in} \\
						\texttt{let \(\yvar{3}\) = flip(\(\theta\)) in} \\
						\texttt{if \(\yvar{3}\) then \(\yvar{2}\) else \(\bin{r'}{b}\)}
				\end{array}}(v_1, (v_2 ,(\ldots, v_b)\ldots))
				~\tag{Lemma~\ref{lemma:unifobs}}
				\\
				=& 	\sum_{r' \in [0, 1]_b} \frac{e^{\beta \cdot 2^{-b}} - 1}{e^{\beta} - 1} e^{\beta r'} r' \times
				\sum_{v''} \dbracket{\prog{1}}(v'') \times
				\dbracket{\begin{array}{l}
						\texttt{let \(\yvar{3}\) = flip(\(\theta\)) in} \\
						\texttt{if \(\yvar{3}\) then \(v'')\) else \(\bin{r'}{b}\)}
				\end{array}}(v_1, (v_2 ,(\ldots, v_b)\ldots))
				~\tag{\ref{Dice-let}}
				\\
				=& 	\sum_{r' \in [0, 1]_b} \frac{e^{\beta \cdot 2^{-b}} - 1}{e^{\beta} - 1} e^{\beta r'} r' 
				\sum_{r'' \in [0, 1]_b} \frac{e^{\beta \cdot 2^{-b}} - 1}{e^{\beta} - 1} e^{\beta r''} 
				\\
				&\times
				\dbracket{\begin{array}{l}
						\texttt{let \(\yvar{3}\) = flip(\(\theta\)) in} \\
						\texttt{if \(\yvar{3}\) then \(\bin{r''}{b})\) else \(\bin{r'}{b}\)}
				\end{array}}(v_1, (v_2 ,(., v_b).))
				~\tag{Lemma~\ref{lemma:alpha-0-density}}
				\\
				=& 	\sum_{r' \in [0, 1]_b} \frac{e^{\beta \cdot 2^{-b}} - 1}{e^{\beta} - 1} e^{\beta r'} r' 
				\sum_{r'' \in [0, 1]_b} \frac{e^{\beta \cdot 2^{-b}} - 1}{e^{\beta} - 1} e^{\beta r''} \times
				\sum_{v'''} \dbracket{\Lflip{\theta}}(v''') \times
				\\
				&\dbracket{
					\texttt{if \(v'''\) then \(\bin{r''}{b})\) else \(\bin{r'}{b}\)}}(v_1, (v_2 ,(., v_b).))
				~\tag{\ref{Dice-let}}
				\\
				=& 	\sum_{r' \in [0, 1]_b} \frac{e^{\beta \cdot 2^{-b}} - 1}{e^{\beta} - 1} e^{\beta r'} r' 
				\sum_{r'' \in [0, 1]_b} \frac{e^{\beta \cdot 2^{-b}} - 1}{e^{\beta} - 1} e^{\beta r''} \times
				(\theta \dbracket{
					\texttt{if \(\true\) then \(\bin{r''}{b}\) else \(\bin{r'}{b}\)}}(v_1, (v_2 ,(., v_b).)) )
				\\
				&+ (1 - \theta) \dbracket{
					\texttt{if \(\false\) then \(\bin{r''}{b})\) else \(\bin{r'}{b}\)}}(v_1, (v_2 ,(., v_b).)))
				~\tag{\ref{Dice-flip}}
				\\
				=& 	\sum_{r' \in [0, 1]_b} \frac{e^{\beta \cdot 2^{-b}} - 1}{e^{\beta} - 1} e^{\beta r'} r' 
				\sum_{r'' \in [0, 1]_b} \frac{e^{\beta \cdot 2^{-b}} - 1}{e^{\beta} - 1} e^{\beta r''} \times	
				(\theta \delta(\bin{r''}{b})(v_1, (v_2 ,(., v_b).)) )
				\\
				&+ (1 - \theta) \delta(\bin{r'}{b})(v_1, (v_2 ,(., v_b).)))
				~\tag{\ref{Dice-Ite} and \ref{Dice-value}}
			\end{flalign*}
			\begin{flalign*}
				=&  \bigg(\frac{e^{\beta \cdot 2^{-b}} - 1}{e^{\beta} - 1}\bigg)^2	
				(e^{\beta r} \theta \sum_{r' \in [0, 1]_b} e^{\beta r'} r' + e^{\beta r} r (1 - \theta) 	\sum_{r'' \in [0, 1]_b} e^{\beta r''})
				~\tag{Considering only non zero terms resulting from \(\delta\) distribution}
				\\
				=& \bigg(\frac{e^{\beta \cdot 2^{-b}} - 1}{e^{\beta} - 1}\bigg)^2 (e^{\beta r} \theta \frac{2^{-b}((2^b - 1)e^{2^{-b} \beta + \beta} - 2^{b} e^{\beta} + e^{2^{-b} \beta})}{(e^{2^{-b} \beta} - 1)^2} + e^{\beta r}r(1 - \theta) \frac{e^{\beta }-1}{e^{2^{-b} \beta}- 1})
				~\tag{Lemma~\ref{lemma:gp} and ~\ref{lemma:agp}}
				\\
				=& k [\beta(e^{\beta \cdot 2^{-b}}-1)re^{\beta r} + (\beta2^{-b}e^{\beta2^{-b}} - e^{\beta2^{-b}} + 1)e^{\beta r}]
				~\tag{for some constant k which depends on \(\beta, b\)}
			\end{flalign*}

			Evaluating the right hand side now,
			
			\begin{flalign*}
				\dbracket{\texttt{p}}_D(\bin{r}{b}) =& \frac{\dbracket{\texttt{p}}(\bin{r}{b})}{\sum_{r' \in [0, 1]_b} \dbracket{\texttt{p}}(\bin{r'}{b})} = \frac{\beta(e^{\beta \cdot 2^{-b}}-1)re^{\beta r} + (\beta2^{-b}e^{\beta2^{-b}} - e^{\beta2^{-b}} + 1)e^{\beta r}}{e^{\beta}(\beta - 1) + 1}
			\end{flalign*}

			If \(\beta = 0\), the proof proceeds in the same way with every constant computed with the limit (\(\beta \rightarrow 0\))
			
		\end{proof}

		\begin{align*}
			\inference[]{
				\texttt{fresh \(\yvar{1}\), \(\yvar{2}\), \(\yvar{3}\)}
				\;\;\;\;\;\;\;\;
				\beta = 0
				\\
				\pdf{0}{\beta} \comp_b \prog{1}
				\;\;\;\;\;\;\;\;
				\texttt{unifObs(\(\yvar{1}\), b)} = \prog{3}
				\;\;\;\;\;\;\;\;\; \theta = \frac{1}{2^b}
			}
			{\pdf{1}{\beta} \comp_b 
				\begin{array}{l}
					\texttt{let  \(\yvar{1}\) = } \prog{1} \texttt{ in }
					\\ \texttt{let \_ = \(\prog{3}\)}
					\\ \texttt{let \(\yvar{2}\) = } \prog{1} \texttt{ in}
					\\ \texttt{let \(\yvar{3}\) = flip\((\theta)\) in}
					\\ \texttt{if \(\yvar{3}\) then \(\yvar{2}\) else \(\yvar{1}\)}
				\end{array}
			}\label{Trans-expo1zero}~\tag{Trans-expo1zero}
		\end{align*}
		
		\begin{lemma}\label{lemma:alpha-1-beta-0-density}
			\(\forall b \in \mathbb{Z}^+, \beta = 0 \), \texttt{p}, if \(\pdf{1}{\beta} \comp_b \texttt{p}\), then \(\pdf{1}{\beta}\) and \texttt{p} are \(b\)-equivalent.
		\end{lemma}
		
		\begin{proof}
			By ~\ref{Trans-expo1}, if \(\pdf{1}{\beta} \comp_b \prog{}\), then
			
			$\texttt{p} = \begin{array}{l}
				\texttt{let  \(\yvar{1}\) = } \prog{1} \texttt{ in }
				\\ \texttt{let \_ = \(\prog{3}\)}
				\\ \texttt{let \(\yvar{2}\) = } \prog{1} \texttt{ in}
				\\ \texttt{let \(\yvar{3}\) = flip\((\theta)\) in}
				\\ \texttt{if \(\yvar{3}\) then \(\yvar{2}\) else \(\yvar{1}\)}
			\end{array}$ 
			
			where  \(\pdf{0}{\beta} \comp_b \prog{1}\) and
			\(\texttt{unifObs(\(\yvar{1}\), b)} = \prog{3}\)
			
			Now, to prove that \(\pdf{1}{\beta}\) and \texttt{p} are \(b\)-equivalent, we need to prove the following by Definition~\ref{def:b-equivalence}.
			
			\[\forall r \in [0, 1]_b, \;\;\;\;\; \int_{r}^{r + \frac{1}{2^b}} \pdf{1}{\beta}(y) \; dy = \dbracket{\texttt{p}}_D(\bin{r}{b})\]
			
			Evaluating the left hand side,
			
			For all \(r \in [0, 1]_b\)
			
			\begin{flalign*}
				\int_{r}^{r + \frac{1}{2^b}} \pdf{1}{\beta}(y) \; dy =& \int_{r}^{r + \frac{1}{2^b}} \frac{y }{\int_{0}^{1} z \; dz} \; dy = \frac{2r}{2^b} + \frac{1}{4^b}
				~\tag{Definition~\ref{def:general-gamma}}
			\end{flalign*}
			
			To evaluate the right hand side, we first evaluate \(\dbracket{\prog{}}\)
			
			For any \(r \in [0, 1]_b\), if \(\bin{r}{b} = (v_1, (v_2 (\ldots, v_b))\ldots)\) where \(v_i \in \{0, 1\}\)
			
			\begin{flalign*}
				&\dbracket{\prog{}}(\bin{r}{b}) 
				\\
				=& \dbracket{\prog{}}(v_1, (v_2 ,(\ldots, v_b)\ldots))
				\\
				=& \sum_{v'} \dbracket{\prog{1}}(v') \times \dbracket{\begin{array}{l}
						\texttt{let \_ = \(\prog{3}\) in} \\
						\texttt{let \(\yvar{2}\) = \(\prog{1}\) in} \\
						\texttt{let \(\yvar{3}\) = flip(\(\theta\)) in} \\
						\texttt{if \(\yvar{3}\) then \(\yvar{2}\) else \(\yvar{1}\)}
					\end{array} [\yvar{1} \mapsto v']}(v_1, (v_2 ,(\ldots, v_b)\ldots))
				~\tag{\ref{Dice-let}}
				\\
				=& \sum_{r' \in [0, 1]_b} \frac{1}{2^b} \times \dbracket{\begin{array}{l}
						\texttt{let \_ = \(\prog{3}\) in} \\
						\texttt{let \(\yvar{2}\) = \(\prog{1}\) in} \\
						\texttt{let \(\yvar{3}\) = flip(\(\theta\)) in} \\
						\texttt{if \(\yvar{3}\) then \(\yvar{2}\) else \(\yvar{1}\)}
					\end{array} [\yvar{1} \mapsto \bin{r'}{b}]}(v_1, (v_2 ,(\ldots, v_b)\ldots))
				~\tag{Lemma~\ref{lemma:alpha-0-density}}
				\\
				=& \sum_{r' \in [0, 1]_b} \frac{1}{2^b} \times
				\sum_{v''} \dbracket{\prog{3} [\yvar{1} \mapsto \bin{r'}{b}]}(v'')
				\\
				&\times
				\dbracket{\begin{array}{l}
						\texttt{let \(\yvar{2}\) = \(\prog{1}\) in} \\
						\texttt{let \(\yvar{3}\) = flip(\(\theta\)) in} \\
						\texttt{if \(\yvar{3}\) then \(\yvar{2}\) else \(\yvar{1}\)}
					\end{array} [\yvar{1} \mapsto \bin{r'}{b}]}(v_1, (v_2 ,(\ldots, v_b)\ldots))
				~\tag{\ref{Dice-let}}
				\\
				=& \sum_{r' \in [0, 1]_b} \frac{1}{2^b} r' \times
				\dbracket{\begin{array}{l}
						\texttt{let \(\yvar{2}\) = \(\prog{1}\) in} \\
						\texttt{let \(\yvar{3}\) = flip(\(\theta\)) in} \\
						\texttt{if \(\yvar{3}\) then \(\yvar{2}\) else \(\bin{r'}{b}\)}
				\end{array}}(v_1, (v_2 ,(\ldots, v_b)\ldots))
				~\tag{Lemma~\ref{lemma:unifobs}}
				\\
				=& 	\sum_{r' \in [0, 1]_b} \frac{1}{2^b} r' \times
				\sum_{v''} \dbracket{\prog{1}}(v'') \times
				\dbracket{\begin{array}{l}
						\texttt{let \(\yvar{3}\) = flip(\(\theta\)) in} \\
						\texttt{if \(\yvar{3}\) then \(v'')\) else \(\bin{r'}{b}\)}
				\end{array}}(v_1, (v_2 ,(\ldots, v_b)\ldots))
				~\tag{\ref{Dice-let}}
				\\
				=& 	\sum_{r' \in [0, 1]_b} \frac{1}{2^b} r' 
				\sum_{r'' \in [0, 1]_b} \frac{1}{2^b} \times
				\dbracket{\begin{array}{l}
						\texttt{let \(\yvar{3}\) = flip(\(\theta\)) in} \\
						\texttt{if \(\yvar{3}\) then \(\bin{r''}{b})\) else \(\bin{r'}{b}\)}
				\end{array}}(v_1, (v_2 ,(., v_b).))
				~\tag{Lemma~\ref{lemma:alpha-0-density}}
				\\
				=& 	\sum_{r' \in [0, 1]_b} \frac{1}{2^b} r' 
				\sum_{r'' \in [0, 1]_b} \frac{1}{2^b} \times
				\sum_{v'''} \dbracket{\Lflip{\theta}}(v''') \times
				\dbracket{
					\texttt{if \(v'''\) then \(\bin{r''}{b})\) else \(\bin{r'}{b}\)}}(v_1, (v_2 ,(., v_b).))
				~\tag{\ref{Dice-let}}
				\\
				=& 	\sum_{r' \in [0, 1]_b} \frac{1}{2^b} r' 
				\sum_{r'' \in [0, 1]_b} \frac{1}{2^b} \times
				(\theta \dbracket{
					\texttt{if \(\true\) then \(\bin{r''}{b})\) else \(\bin{r'}{b}\)}}(v_1, (v_2 ,(., v_b).)))
				\\
				&+ (1 - \theta) \dbracket{
					\texttt{if \(\false\) then \(\bin{r''}{b})\) else \(\bin{r'}{b}\)}}(v_1, (v_2 ,(., v_b).)))
				~\tag{\ref{Dice-flip}}
				\\
				=& 	\sum_{r' \in [0, 1]_b} \frac{1}{2^b} r' 
				\sum_{r'' \in [0, 1]_b} \frac{1}{2^b} \times	
				(\theta \delta(\bin{r''}{b})(v_1, (v_2 ,(., v_b).)) )
				+ (1 - \theta) \delta(\bin{r'}{b})(v_1, (v_2 ,(., v_b).)))
				~\tag{\ref{Dice-Ite} and \ref{Dice-value}}
				\\
				=&  \bigg(\frac{1}{2^b}\bigg)^2	
				(\theta \sum_{r' \in [0, 1]_b} r' + r (1 - \theta) 	\sum_{r'' \in [0, 1]_b} 1)
				~\tag{Considering only non zero terms resulting from \(\delta\) distribution}
				\\
				=&  \bigg(\frac{1}{2^b}\bigg)^2	
				( \theta \frac{2^b(2^b - 1)}{2 \cdot 2^b} +  r (1 - \theta) 2^b ) 
				= \bigg(\frac{1}{2^b}\bigg)^2	
				( \frac{2^b(2^b - 1)}{2 \cdot 2^b \cdot 2^b} +  r (2^b - 1) )
			\end{flalign*}
			\begin{flalign*}
				=& k (\frac{1}{2} + r \cdot 2^{b})
				~\tag{for some constant k which depends on \(\beta, b\)}
			\end{flalign*}
			
			Evaluating the right hand side now,
			
			\begin{flalign*}
				\dbracket{\prog{}}_D(\bin{r}{b}) = \frac{\dbracket{\prog{}}(\bin{r}{b})}{\sum_{r' \in [0, 1]_b} \dbracket{\prog{}}(\bin{r'}{b})} =& \frac{k(\frac{1}{2} + r2^b)}{\sum_{r' \in [0, 1]_b} k(\frac{1}{2} + r'2^b)}
				\\
				=& \frac{(\frac{1}{2} + r2^b)}{\frac{2^b}{2} + \frac{2^b(2^b - 1)}{2}} = \frac{2r}{2^b} + \frac{1}{4^b}
			\end{flalign*}
			
		\end{proof}

		\subsection{Proof of Lemma~\ref{lemma:alpha>1-density}}
		
		Next, we describe the compilation judgement for \(\pdf{\alpha}{\beta}\) for \(\alpha > 1\). To do so, there is additional helping judgement needed - \pointgamma{}{}{}{}. 
		
		\subsubsection{pointGamma} helper judgement is of the following form where \(\alpha \in \mathbb{Z}^+, \beta \in \mathbb{R}, \epsilon \in \mathbb{R}, b > 0 \in \mathbb{Z}^+\).
		
		\[\pointgamma{\alpha}{\beta}{\epsilon}{b} = \prog{}\]
		
		The rules for the judgement are defined as follows:
		
		\begin{align*}
			\inference[]{\pdf{0}{\beta} \comp_b \prog{1}}{\pointgamma{0}{\beta}{\epsilon}{b} = \prog{1}}~\tag{SPG-0}\label{spg0}
		\end{align*}
		
		\begin{align*}
			\inference[]
			{\alpha > 0
				\;\;\;\;\;\;\;\;
				\texttt{fresh \(\yvar{1}\), \(\yvar{2}\)}
				\\
				\pointgamma{\alpha - 1}{\beta}{\epsilon}{b} = \prog{1}
				\;\;\;\;\;\;\;\;
				\unifobs{\yvar{1}}{b} = \prog{2}
			}
			{ \pointgamma{\alpha}{\beta}{\epsilon}{b} = 
				\begin{array}{l}
					\texttt{let \(\yvar{1}\) = } \prog{1} \texttt{ in } 
					\\ \texttt{let \(\yvar{2}\) = } \Lflip{\frac{1}{1 + \epsilon}} \texttt{ in } 
					\\ \texttt{let } \_ = \Lite{\yvar{2}}{\prog{2}}{\true} \texttt{ in}
					\\  \yvar{1} \end{array}}~\tag{SPG > 0}
			\label{spggt0}
		\end{align*}
		
		\begin{lemma}\label{lemma:pointGamma}
			For any \(b \in \mathbb{Z}^{+}, \alpha \in \mathbb{Z}^{+}, \beta \in \mathbb{R}, \epsilon \in \mathbb{R}\) such that \pointgamma{\alpha}{\beta}{\epsilon}{b} = \(\prog{}\) and \(\forall r \in [0, 1]_b\), the following holds
			
			\[\forall r \in [0, 1]_b \;\;\;\;\;\; \dbracket{\prog{}}_D(\bin{r}{b}) = \frac{(r + \epsilon)^{\alpha}e^{\beta r}}{\sum_{r' \in [0, 1]_b} (r' + \epsilon)^{\alpha}e^{\beta r'}}\]
		\end{lemma}
		
		\begin{proof}
			
			The proof is by induction on \(\alpha\).
			
			\underline{Base Case:} \(\alpha = 0\)
			
			Evaluating the left hand side,
			
			By \ref{spg0}, \(\prog{} = \prog{1}\) where \(\pdf{0}{\beta} \comp_b \prog{1}\).
			
			For all \(r \in [0, 1]_b\) 
			\begin{flalign*}
				\dbracket{\prog{}}_D(\bin{r}{b}) =& \dbracket{\prog{1}}_D(\bin{r}{b})
				= \begin{cases}
					\frac{e^{\beta \cdot 2^{-b}} -1}{e^{\beta} - 1}e^{\beta r} ~& \text{ if }\beta \neq 0
					\\  \frac{1}{2^b} ~& \text{ if }\beta = 0
				\end{cases}
				~\tag{Lemma \ref{lemma:alpha-0-density}}
			\end{flalign*}
			
			Evaluating the right hand side,
			
			\begin{flalign*}
				\frac{(r + \epsilon)^{0}e^{\beta r}}{\sum_{r' \in [0, 1]_b} (r' + \epsilon)^{0}e^{\beta r'}} = \frac{e^{\beta r}}{\sum_{r' \in [0, 1]_b} e^{\beta r'}} = \begin{cases}
					\frac{e^{\beta \cdot 2^{-b}} -1}{e^{\beta} - 1}e^{\beta r} ~& \text{ if }\beta \neq 0
					\\  \frac{1}{2^b} ~& \text{ if }\beta = 0
				\end{cases}
			\end{flalign*}
			
			\underline{Induction Hypothesis:} For \(\alpha' < \alpha\), if \pointgamma{\alpha'}{\beta}{\epsilon}{b} =\(\prog{}\) then 
			\[\forall r \in [0, 1]_b \;\;\;\;\;\; \dbracket{\prog{}}_D(\bin{r}{b}) = \frac{(r + \epsilon)^{\alpha'}e^{\beta r}}{\sum_{r' \in [0, 1]_b} (r' + \epsilon)^{\alpha'}e^{\beta r'}}\]
			
			\underline{Inductive Step:} \(\alpha > 0\)
			
			To evaluate the left hand side, we first evaluate \(\dbracket{\prog{}}\).
			
			By \ref{spggt0}, \(\prog{} = \begin{array}{l}
				\texttt{let \(\yvar{1}\) = } \prog{1} \texttt{ in } 
				\\ \texttt{let \(\yvar{2}\) = } \Lflip{\frac{1}{1 + \epsilon}} \texttt{ in } 
				\\ \texttt{let } \_ = \Lite{\yvar{2}}{\prog{2}}{\true} \texttt{ in}
				\\  \yvar{1} \end{array}\)
			
			where \(\pointgamma{\alpha - 1}{\beta}{\epsilon}{b} = \prog{1}\) and \(\unifobs{\yvar{1}}{b} = \prog{2}\)
			
			For all \(r \in [0, 1]_b\), we have
			
			\begin{flalign*}
				&\dbracket{\prog{}}(\bin{r}{b}) 
				\\
				=& \dbracket{\begin{array}{l}
						\texttt{let \(\yvar{1}\) = } \prog{1} \texttt{ in } 
						\\ \texttt{let \(\yvar{2}\) = } \Lflip{\frac{1}{1 + \epsilon}} \texttt{ in } 
						\\ \texttt{let } \_ = \Lite{\yvar{2}}{\prog{2}}{{\true}} \texttt{ in}
						\\  \yvar{1} \end{array}}(\bin{r}{b})
				\\
				=& \sum_{v'} \dbracket{\prog{1}}(v') \times \dbracket{\begin{array}{l}
						\texttt{let \(\yvar{2}\) = } \Lflip{\frac{1}{1 + \epsilon}} \texttt{ in } 
						\\ \texttt{let } \_ = \Lite{\yvar{2}}{\prog{2}}{{\true}} \texttt{ in}
						\\  \yvar{1} \end{array}
					[\yvar{1} \mapsto v']
				}(\bin{r}{b})
				~\tag{\ref{Dice-let}}
				\\
				=& \sum_{v'} \dbracket{\prog{1}}(v') \times \sum_{v''} \dbracket{\Lflip{\frac{1}{1 + \epsilon}}}(v'') \times
				\\
				&\dbracket{\begin{array}{l} 
						\texttt{let } \_ = \Lite{v''}{\prog{2}}{{\true}} \texttt{ in}
						\\  \yvar{1} \end{array}
					[\yvar{1} \mapsto v']
				}(\bin{r}{b})
				~\tag{\ref{Dice-let}}
				\\
				=& \sum_{v'} \dbracket{\prog{1}}(v') \times \sum_{v''} 
				\dbracket{\Lflip{\frac{1}{1 + \epsilon}}}(v'') \times \dbracket{\Lite{v''}{\prog{2}}{{\true}} [\yvar{1} \mapsto v']}(v'') 
				\\
				&\times \dbracket{v'}(\bin{r}{b})
				~\tag{\ref{Dice-let}}
				\\
				=& \sum_{v'} \dbracket{\prog{1}}(v') \times \sum_{v''}
				(\frac{1}{1 + \epsilon} \dbracket{\prog{2}[\yvar{1} \mapsto v']} + \frac{\epsilon}{1 + \epsilon} \dbracket{{\true}})(v'') \times \dbracket{v'}(\bin{r}{b})
				~\tag{\ref{Dice-flip}, \ref{Dice-Ite}}
				\\
				=& \sum_{r' \in [0, 1]_b} \dbracket{\prog{1}}(\bin{r'}{b}) \times \sum_{v''}
				(\frac{1}{1 + \epsilon} \dbracket{\prog{2}[\yvar{1} \mapsto \bin{r'}{b}]} + \frac{\epsilon}{1 + \epsilon} \dbracket{{\true}})(v'') \times \dbracket{\bin{r'}{b}}(\bin{r}{b})	
				\\
				=& \sum_{r' \in [0, 1]_b} \dbracket{\prog{1}}(\bin{r'}{b}) \times \frac{r' + \epsilon}{1 + \epsilon} \times \dbracket{\bin{r'}{b}}(\bin{r}{b})
				~\tag{Lemma~\ref{lemma:unifobs} and \ref{Dice-value}}
				\\
				=& 	\sum_{r' \in [0, 1]_b} k (r' + \epsilon)^{\alpha - 1}e^{\beta r'} \times \frac{r' + \epsilon}{1 + \epsilon} \times \dbracket{\bin{r'}{b}}(\bin{r}{b})
				~\tag{Induction hypothesis, \(k = \frac{\sum_{v*} \dbracket{\prog{1}}(v*)}{\sum_{r'''} (r + \epsilon)^{\alpha - 1} e^{\beta r}}\)}
				\\
				=& 	\frac{k}{1 + \epsilon} (r + \epsilon)^{\alpha}e^{\beta r}
				~\tag{\ref{Dice-value}}
			\end{flalign*}
			
			Evaluating the right hand side now,
			
			\begin{flalign*}
				\dbracket{\prog{}}_D(\bin{r}{b}) = \frac{\dbracket{\prog{}}(\bin{r}{b})}{\sum_{r' \in [0, 1]_b} \dbracket{\prog{}}(\bin{r'}{b})} = \frac{(r + \epsilon)^{\alpha}e^{\beta r}}{\sum_{r' \in [0, 1]_b} (r' + \epsilon)^{\alpha}e^{\beta r'}}
			\end{flalign*}
		\end{proof}

		\subsubsection{\(\pdf{\alpha}{\beta}\)}
		
		The following rule defines the judgement \(\comp_b\) for \(\pdf{\alpha}{\beta}\)
		
		\begin{align*}
			\inference[]{
				\alpha > 1 \;\;\;\;\;\;
				\texttt{fresh \(\yvar{1}\), \(\yvar{2}\), \(\yvar{3}\), \(\yvar{4}\)}
				\\
				\theta_1 = \frac{\sum_{r' \in [0, 1]_b}r'\int_{r'}^{r' + \frac{1}{2^b}} y^{\alpha - 1}e^{\beta y} \; dy}{\int_{0}^{1} z^{\alpha}e^{\beta z} \; dz}
				\;\;\;\;\;\;\;
				\theta_2 = \frac{-\alpha \int_{0}^{1} y^{\alpha - 1}e^{\beta y} \; dy}{\beta \sum_{r' \in [0, 1]_b} \int_{r'}^{r' + \frac{1}{2^b} } \int_{x = r}^{y} y^{\alpha - 1}e^{\beta y} \; dx \; dy}
				\\
				\theta_3 = \frac{(\alpha - 1) \sum_{r' \in [0, 1]_b} r' \int_{r'}^{r' + \frac{1}{2^b}} y^{\alpha - 2}e^{\beta y} \; dy}{(1 - \theta_2)\beta \sum_{r' \in [0, 1]_b} \int_{r'}^{r' + \frac{1}{2^b} } \int_{x = r}^{y} y^{\alpha - 1}e^{\beta y} \; dx \; dy}
				\\ \pdf{\alpha-1}{\beta} \comp_b \prog{1}
				\\ \pdf{\alpha-2}{\beta} \comp_b \prog{3} 
				\;\;\;\;\;\;\;\; \pointgamma{\alpha-1}{\beta}{\frac{1}{2^b}}{b} = \prog{4}
				\\
				\unifobs{\yvar{1}}{b} = \prog{obs1}
				\;\;\;\;\;\;\;\;
				\unifobs{\yvar{3}}{b} = \prog{obs2}
			}{
				\pdf{\alpha}{\beta} \comp_b 
				\begin{array}{l}
					\texttt{let \(\yvar{1}\) = \(\prog{1}\) in}
					\\ \texttt{let \_ = \(\prog{obs1}\) in}
					\\ \texttt{let \(\yvar{2}\) = \(\prog{1}\) in}
					\\ \texttt{let \(\yvar{3}\) = \(\prog{3}\) in}
					\\ \texttt{let \(\yvar{4}\) = \(\prog{4}\) in}
					\\ \texttt{let \_ = \(\prog{obs2}\) in}
					\\ \texttt{let \(\yvar{f1}\) = flip(\(\theta_1\)) in}
					\\ \texttt{let \(\yvar{f2}\) = flip(\(\theta_2\)) in}
					\\ \texttt{let \(\yvar{f3}\) = flip(\(\theta_3\)) in }
					\\ \texttt{if \(\yvar{f1}\) then \(\yvar{1}\) else} 
					\\ \texttt{if \(\yvar{f2}\) then \(\yvar{2}\) else if \(\yvar{f3}\) then \(\yvar{3}\) else \(\yvar{4}\)}
				\end{array}
			}\label{Trans-expo>1}~\tag{Trans-expo>1}
		\end{align*}

		\emph{Proof for Lemma~\ref{lemma:alpha>1-density}}
		
		\begin{proof}
			The proof is by induction on \(\alpha\).
			
			\underline{Base Case 1:} \(\alpha = 0\)
			
			By Lemma~\ref{lemma:alpha-0-density}, 	\(\forall b, \beta \in \mathbb{R},\) \texttt{p}, if \(\pdf{0}{\beta} \comp_b \texttt{p}\), then \(\pdf{0}{\beta}\) and \texttt{p} are \(b\)-equivalent.
			
			\underline{Base Case 2:} \(\alpha = 1\)
			
			By Lemma~\ref{lemma:alpha-1-density} and ~\ref{lemma:alpha-1-beta-0-density}, \(\forall b, \beta \in \mathbb{R},\) \texttt{p}, if \(\pdf{1}{\beta} \comp_b \texttt{p}\), then \(\pdf{1}{\beta}\) and \texttt{p} are \(b\)-equivalent.
			
			\underline{Induction Hypothesis:} \(\forall \alpha' < \alpha\), let \(\pdf{\alpha'}{\beta} \comp_b \prog{+}\), then \(\pdf{\alpha'}{\beta}\) and \(\prog{+}\) are \(b\)-equivalent.
			
			By Definition~\ref{def:b-equivalence}, we have
			
			\[\forall r \in [0, 1]_b \;\;\;\;\;\; \int_{r}^{r + \frac{1}{2^b}} \pdf{\alpha'}{\beta}(x) \; dx = \dbracket{\prog{+}}_D(\bin{r}{b})\]
			
			\underline{Inductive Step:} \(\forall \alpha > 1\)
			
			By ~\ref{Trans-expo>1}, if \(\pdf{\alpha}{\beta} \comp_b \prog{}\), then
			
			\(\texttt{p} = \begin{array}{l}
				\texttt{let \(\yvar{1}\) = \(\prog{1}\) in}
				\\ \texttt{let \_ = \(\prog{obs1}\) in}
				\\ \texttt{let \(\yvar{2}\) = \(\prog{1}\) in}
				\\ \texttt{let \(\yvar{3}\) = \(\prog{3}\) in}
				\\ \texttt{let \(\yvar{4}\) = \(\prog{4}\) in}
				\\ \texttt{let \_ = \(\prog{obs2}\) in}
				\\ \texttt{let \(\yvar{f1}\) = flip(\(\theta_1\)) in}
				\\ \texttt{let \(\yvar{f2}\) = flip(\(\theta_2\)) in}
				\\ \texttt{let \(\yvar{f3}\) = flip(\(\theta_3\)) in }
				\\ \texttt{if \(\yvar{f1}\) then \(\yvar{1}\) else} 
				\\ \texttt{if \(\yvar{f2}\) then \(\yvar{2}\) else if \(\yvar{f3}\) then \(\yvar{3}\) else \(\yvar{4}\)}
			\end{array}\)
			
			where 
			\[\theta_1 = \frac{\sum_{r' \in [0, 1]_b}r'\int_{r'}^{r' + \frac{1}{2^b}} y^{\alpha - 1}e^{\beta y} \; dy}{\int_{0}^{1} z^{\alpha}e^{\beta z} \; dz}
			\;\;\;\;\;\;\;
			\theta_2 = \frac{-\alpha \int_{0}^{1} y^{\alpha - 1}e^{\beta y} \; dy}{\beta \sum_{r' \in [0, 1]_b} \int_{r'}^{r' + \frac{1}{2^b} } \int_{x = r}^{y} y^{\alpha - 1}e^{\beta y} \; dx \; dy}\]
			
			\[\theta_3 = \frac{(\alpha - 1) \sum_{r' \in [0, 1]_b} r' \int_{r'}^{r' + \frac{1}{2^b}} y^{\alpha - 2}e^{\beta y} \; dy}{(1 - \theta_2)\beta \sum_{r' \in [0, 1]_b} \int_{r'}^{r' + \frac{1}{2^b} } \int_{x = r}^{y} y^{\alpha - 1}e^{\beta y} \; dx \; dy}\]
			
			\[\pdf{\alpha-1}{\beta} \comp_b \prog{1}
			\;\;\;\;\;\;\;\; \pdf{\alpha-2}{\beta} \comp_b \prog{3} \]
			
			\[\pointgamma{\alpha-1}{\beta}{\frac{1}{2^b}}{b} = \prog{4}
			\;\;\;\;\;\;\;\;
			\unifobs{\yvar{1}}{b} = \prog{obs1}
			\;\;\;\;\;\;\;\;
			\unifobs{\yvar{3}}{b} = \prog{obs2}\]

			Now, to prove that \(\pdf{\alpha}{\beta}\) and \texttt{p} are \(b\)-equivalent, we need to prove the following by Definition~\ref{def:b-equivalence}.
			
			\[\forall r \in [0, 1]_b, \;\;\;\;\; \int_{r}^{r + \frac{1}{2^b}} \pdf{\alpha}{\beta}(y) \; dy = \dbracket{\texttt{p}}_D(\bin{r}{b})\]
			
			Evaluating the left hand side,
			
			For all \(r \in [0, 1]_b\)
			
			\begin{flalign*}
				&\int_{r}^{r + \frac{1}{2^b}} \pdf{\alpha}{\beta}(y) \; dy 
				\\
				=& \int_{r}^{r + \frac{1}{2^b}} \frac{y^{\alpha}e^{\beta y} }{\int_{0}^{1} z^{\alpha}e^{\beta z} \; dz} \; dy
				~\tag{Definition~\ref{def:general-gamma}}
				\\
				=& \frac{\int_{r}^{r + \frac{1}{2^b}} (\int_{x = 0}^{y} y^{\alpha - 1}e^{\beta y} \; dx) \; dy}{\int_{0}^{1} z^{\alpha}e^{\beta z} \; dz}
			\end{flalign*}
			\begin{flalign*}
				=& \frac{\int_{r}^{r + \frac{1}{2^b}} (\int_{x = 0}^{r} y^{\alpha - 1}e^{\beta y} \; dx + \int_{x = r}^{y} y^{\alpha - 1}e^{\beta y} \; dx) \; dy}{\int_{0}^{1} z^{\alpha}e^{\beta z} \; dz}
				\\
				=& \frac{\int_{r}^{r + \frac{1}{2^b}} (\int_{x = 0}^{r} y^{\alpha - 1}e^{\beta y} \; dx) \; dy}{\sum_{r' \in [0, 1]_b}\int_{r'}^{r' + \frac{1}{2^b}} (\int_{x = 0}^{r'} y^{\alpha - 1}e^{\beta y} \; dx) \; dy} \frac{\sum_{r' \in [0, 1]_b}\int_{r'}^{r' + \frac{1}{2^b}} (\int_{x = 0}^{r'} y^{\alpha - 1}e^{\beta y} \; dx) \; dy}{\int_{0}^{1} z^{\alpha}e^{\beta z} \; dz}
				\\ 
				&+ 
				\frac{\int_{r}^{r + \frac{1}{2^b}} (\int_{x = r}^{y} y^{\alpha - 1}e^{\beta y} \; dx) \; dy}{\sum_{r' \in [0, 1]_b}\int_{r'}^{r' + \frac{1}{2^b}} (\int_{x = r'}^{y} y^{\alpha - 1}e^{\beta y} \; dx) \; dy} \frac{\sum_{r' \in [0, 1]_b}\int_{r'}^{r' + \frac{1}{2^b}} (\int_{x = r'}^{y} y^{\alpha - 1}e^{\beta y} \; dx) \; dy}{\int_{0}^{1} z^{\alpha}e^{\beta z} \; dz}
				~\tag{Multiplying and dividing by the same expression}
				\\
				=& \frac{\int_{r}^{r + \frac{1}{2^b}} (\int_{x = 0}^{r} y^{\alpha - 1}e^{\beta y} \; dx) \; dy}{\sum_{r' \in [0, 1]_b}\int_{r'}^{r' + \frac{1}{2^b}} (\int_{x = 0}^{r'} y^{\alpha - 1}e^{\beta y} \; dx) \; dy} \frac{\sum_{r' \in [0, 1]_b}\int_{r'}^{r' + \frac{1}{2^b}} (\int_{x = 0}^{r'} y^{\alpha - 1}e^{\beta y} \; dx) \; dy}{\int_{0}^{1} z^{\alpha}e^{\beta z} \; dz}
				\\ 
				&+ 
				\frac{\int_{r}^{r + \frac{1}{2^b}} (\int_{x = r}^{y} y^{\alpha - 1}e^{\beta y} \; dx) \; dy}{\sum_{r' \in [0, 1]_b}\int_{r'}^{r' + \frac{1}{2^b}} (\int_{x = r'}^{y} y^{\alpha - 1}e^{\beta y} \; dx) \; dy} (1 - \frac{\sum_{r' \in [0, 1]_b}\int_{r'}^{r' + \frac{1}{2^b}} (\int_{x = 0}^{r'} y^{\alpha - 1}e^{\beta y} \; dx) \; dy}{\int_{0}^{1} z^{\alpha}e^{\beta z} \; dz})
				\\
				=& \frac{r\int_{r}^{r + \frac{1}{2^b}}y^{\alpha - 1}e^{\beta y} \; dy}{\sum_{r' \in [0, 1]_b} r'\int_{r'}^{r' + \frac{1}{2^b}} y^{\alpha - 1}e^{\beta y} \; dy} \frac{\sum_{r' \in [0, 1]_b}r'\int_{r'}^{r' + \frac{1}{2^b}} y^{\alpha - 1}e^{\beta y} \; dy}{\int_{0}^{1} z^{\alpha}e^{\beta z} \; dz}
				\\ 
				&+ 
				\bigg[\frac{(r + \frac{1}{2^b})^{\alpha - 1} e^{\beta (r + \frac{1}{2^b})}}{2^{b}\beta} + \frac{r(\alpha - 1)}{\beta} \int_{r}^{r + \frac{1}{2^b}} y^{\alpha - 2}e^{\beta y} \; dy - \frac{\alpha}{\beta} \int_{r}^{r + \frac{1}{2^b}} y^{\alpha - 1}e^{\beta y} \; dy \bigg]
				\\
				&\frac{1}{\sum_{r' \in [0, 1]_b}\int_{r'}^{r' + \frac{1}{2^b}} (\int_{x = r'}^{y} y^{\alpha - 1}e^{\beta y} \; dx) \; dy} (1 - \frac{\sum_{r' \in [0, 1]_b}r'\int_{r'}^{r' + \frac{1}{2^b}} y^{\alpha - 1}e^{\beta y} \; dy}{\int_{0}^{1} z^{\alpha}e^{\beta z} \; dz})
				~\tag{Integration by parts}
			\end{flalign*}
			
			To evaluate the right hand side, we first evaluate \(\dbracket{\texttt{p}}\)
			
			For any \(r \in [0, 1]_b\),
			
			\begin{flalign*}
				&\dbracket{\texttt{p}}(\bin{r}{b}) 
				\\
				=& \dbracket{\begin{array}{l}
						\texttt{let \(\yvar{1}\) = \(\prog{1}\) in}
						\\ \texttt{let \_ = \(\prog{obs1}\) in}
						\\ \texttt{let \(\yvar{2}\) = \(\prog{1}\) in}
						\\ \texttt{let \(\yvar{3}\) = \(\prog{3}\) in}
						\\ \texttt{let \(\yvar{4}\) = \(\prog{4}\) in}
						\\ \texttt{let \_ = \(\prog{obs2}\) in}
						\\ \texttt{let \(\yvar{f1}\) = flip(\(\theta_1\)) in}
						\\ \texttt{let \(\yvar{f2}\) = flip(\(\theta_2\)) in}
						\\ \texttt{let \(\yvar{f3}\) = flip(\(\theta_3\)) in }
						\\ \texttt{if \(\yvar{f1}\) then \(\yvar{1}\) else} 
						\\ \texttt{if \(\yvar{f2}\) then \(\yvar{2}\) else if \(\yvar{f3}\) then \(\yvar{3}\) else \(\yvar{4}\)}
				\end{array}}(\bin{r}{b})
			\end{flalign*}
			\begin{flalign*}
				=& \sum_{v'} \dbracket{\prog{1}}(v') \times \dbracket{\begin{array}{l}
						\texttt{let \_ = \(\prog{obs1}\) in}
						\\ \texttt{let \(\yvar{2}\) = \(\prog{1}\) in}
						\\ \texttt{let \(\yvar{3}\) = \(\prog{3}\) in}
						\\ \texttt{let \(\yvar{}\) = \(\prog{4}\) in}
						\\ \texttt{let \_ = \(\prog{obs2}\) in}
						\\ \texttt{let \(\yvar{f1}\) = flip(\(\theta_1\)) in}
						\\ \texttt{let \(\yvar{f2}\) = flip(\(\theta_2\)) in}
						\\ \texttt{let \(\yvar{f3}\) = flip(\(\theta_3\)) in }
						\\ \texttt{if \(\yvar{f1}\) then \(\yvar{1}\) else} 
						\\ \texttt{if \(\yvar{f2}\) then \(\yvar{2}\) else if \(\yvar{f3}\) then \(\yvar{3}\) else \(\yvar{4}\)} 
					\end{array}[\yvar{1} \mapsto v']}(\bin{r}{b})
				~\tag{\ref{Dice-let}}
				\\
				=& \sum_{v'} \dbracket{\prog{1}}(v') \times \sum_{v''} \dbracket{\prog{obs1}[\yvar{1} \mapsto v']}(v'') \times
				\sum_{v'''} \dbracket{\prog{1}}(v''') \times
				\\
				&\dbracket{\begin{array}{l}
						\texttt{let \(\yvar{3}\) = \(\prog{3}\)in}
						\\ \texttt{let \(\yvar{4}\) = \(\prog{4}\) in}
						\\ \texttt{let \_ = \(\prog{obs2}\) in}
						\\ \texttt{let \(\yvar{f1}\) = flip(\(\theta_1\)) in}
						\\ \texttt{let \(\yvar{f2}\) = flip(\(\theta_2\)) in}
						\\ \texttt{let \(\yvar{f3}\) = flip(\(\theta_3\)) in }
						\\ \texttt{if \(\yvar{f1}\) then \(\yvar{1}\) else} 
						\\ \texttt{if \(\yvar{f2}\) then \(\yvar{2}\) else if \(\yvar{f3}\) then \(\yvar{3}\) else \(\yvar{4}\)}
					\end{array}\bigg[\begin{array}{l}
						\yvar{1} \mapsto v'
						\\ \yvar{2} \mapsto v''' \end{array}\bigg]}(\bin{r}{b})
				~\tag{\ref{Dice-let}}
				\\
				=& \sum_{v'} \dbracket{\prog{1}}(v') \times \sum_{v''} \dbracket{\prog{obs1}[\yvar{1} \mapsto v']}(v'') \times
				\sum_{v'''} \dbracket{\prog{1}}(v''') \times
				\sum_{v_4} \dbracket{\prog{3}}(v_4) \times
				\sum_{v_5} \dbracket{\prog{4}}(v_5) \times
				\\
				&\dbracket{\begin{array}{l}
						\\ \texttt{let \_ = \(\prog{obs2}\) in}
						\\ \texttt{let \(\yvar{f1}\) = flip(\(\theta_1\)) in}
						\\ \texttt{let \(\yvar{f2}\) = flip(\(\theta_2\)) in}
						\\ \texttt{let \(\yvar{f3}\) = flip(\(\theta_3\)) in }
						\\ \texttt{if \(\yvar{f1}\) then \(\yvar{1}\) else} 
						\\ \texttt{if \(\yvar{f2}\) then \(\yvar{2}\) else if \(\yvar{f3}\) then \(\yvar{3}\) else \(\yvar{4}\)}
					\end{array}\Bigg[\begin{array}{l}
						\yvar{1} \mapsto v'
						\\ \yvar{2} \mapsto v'''
						\\ \yvar{3} \mapsto v_4
						\\ \yvar{4} \mapsto v_5 \end{array}\Bigg]}(\bin{r}{b})
				~\tag{\ref{Dice-let}}
				\\
				=& \sum_{v'} \dbracket{\prog{1}}(v') \times \sum_{v''} \dbracket{\prog{obs1}[\yvar{1} \mapsto v']}(v'') \times
				\sum_{v'''} \dbracket{\prog{1}}(v''') \times
				\sum_{v_4} \dbracket{\prog{3}}(v_4) \times
				\sum_{v_5} \dbracket{\prog{4}}(v_5) \times
				\\
				&\sum_{v_6} \dbracket{\prog{obs2}[\yvar{3} \mapsto v_4]}(v_6) \times
				\dbracket{\begin{array}{l}
						\\ \texttt{let \(\yvar{f1}\) = flip(\(\theta_1\)) in}
						\\ \texttt{let \(\yvar{f2}\) = flip(\(\theta_2\)) in}
						\\ \texttt{let \(\yvar{f3}\) = flip(\(\theta_3\)) in }
						\\ \texttt{if \(\yvar{f1}\) then \(v'\) else} 
						\\ \texttt{if \(\yvar{f2}\) then \(v'''\) else if \(\yvar{f3}\) then \(v_4\) else \(v_5\)} 
				\end{array}}(\bin{r}{b})
				~\tag{\ref{Dice-let}}
			\end{flalign*}
			\begin{flalign*}
				=& \sum_{v'} \dbracket{\prog{1}}(v') \times \sum_{v''} \dbracket{\prog{obs1}[\yvar{1} \mapsto v']}(v'') \times
				\sum_{v'''} \dbracket{\prog{1}}(v''') \times
				\sum_{v_4} \dbracket{\prog{3}}(v_4) \times
				\sum_{v_5} \dbracket{\prog{4}}(v_5) \times
				\\
				&\sum_{v_6} \dbracket{\prog{obs2}[\yvar{3} \mapsto v_4]}(v_6) \times
				\sum_{v_7} \dbracket{\Lflip{\theta_1}}(v_7) \times
				\sum_{v_8} \dbracket{\Lflip{\theta_2}}(v_8) \times
				\sum_{v_9} \dbracket{\Lflip{\theta_3}}(v_9) \times
				\\
				&\dbracket{\begin{array}{l}
						\texttt{if \(v_7\) then \(v'\) else} 
						\\ \texttt{if \(v_8\) then \(v'''\) else if \(v_9\) then \(v_4\) else \(v_5\)}
				\end{array}}(\bin{r}{b})
				~\tag{\ref{Dice-let}}
				\\
				=& \sum_{v'} \dbracket{\prog{1}}(v') \times \sum_{v''} \dbracket{\prog{obs1}[\yvar{1} \mapsto v']}(v'') \times
				\sum_{v'''} \dbracket{\prog{1}}(v''') \times
				\sum_{v_4} \dbracket{\prog{3}}(v_4) \times
				\sum_{v_5} \dbracket{\prog{4}}(v_5) \times
				\\
				&\sum_{v_6} \dbracket{\prog{obs2}[\yvar{3} \mapsto v_4]}(v_6) \times
				\\
				&(\theta_1 \delta(v') + (1 - \theta_1)\theta_2 \delta(v''') + (1 - \theta_1)(1 - \theta_2)\theta_3 \delta(v_4) + (1 - \theta_1)(1 - \theta_2)(1 - \theta_3) \delta(v_5))(\bin{r}{b})
				~\tag{\ref{Dice-flip}, \ref{Dice-Ite} and \ref{Dice-value}}	
				\\
				=& \sum_{v'} \dbracket{\prog{1}}_D(v') \sum_{v_{s1}} \dbracket{\prog{1}}(v_{s1}) \times \sum_{v''} \dbracket{\prog{obs1}[\yvar{1} \mapsto v']}(v'') \times
				\sum_{v'''} \dbracket{\prog{1}}(v''') \times
				\sum_{v_4} \dbracket{\prog{3}}(v_4) \times
				\\
				&\sum_{v_5} \dbracket{\prog{4}}(v_5) \times
				\sum_{v_6} \dbracket{\prog{obs2}[\yvar{3} \mapsto v_4]}(v_6) \times
				\\
				&(\theta_1 \delta(v') + (1 - \theta_1)\theta_2 \delta(v''') + (1 - \theta_1)(1 - \theta_2)\theta_3 \delta(v_4) + (1 - \theta_1)(1 - \theta_2)(1 - \theta_3) \delta(v_5))(\bin{r}{b})
				~\tag{by ~\ref{normalized-pr}}	
				\\
				=& \sum_{r' \in [0, 1]_b} \frac{\int_{r'}^{r' + \frac{1}{2^b}} x^{\alpha - 1}e^{\beta x} \; dx}{\int_{0}^{1} x^{\alpha - 1}e^{\beta x} \; dx} \sum_{v_{s1}} \dbracket{\prog{1}}(v_{s1}) \times \sum_{v''} \dbracket{\prog{obs1}[\yvar{1} \mapsto r']}(v'') \times
				\sum_{v'''} \dbracket{\prog{1}}(v''') \times
				\\
				&\sum_{v_4} \dbracket{\prog{3}}(v_4) \times
				\sum_{v_5} \dbracket{\prog{4}}(v_5) \times
				\sum_{v_6} \dbracket{\prog{obs2}[\yvar{3} \mapsto v_4]}(v_6) \times
				\\
				&(\theta_1 \delta(\bin{r'}{b}) + (1 - \theta_1)\theta_2 \delta(v''') + (1 - \theta_1)(1 - \theta_2)\theta_3 \delta(v_4) + (1 - \theta_1)(1 - \theta_2)(1 - \theta_3) \delta(v_5))(\bin{r}{b})
				~\tag{Induction Hypothesis}	
				\\
				=& \sum_{r' \in [0, 1]_b} \frac{\int_{r'}^{r' + \frac{1}{2^b}} x^{\alpha - 1}e^{\beta x} \; dx}{\int_{0}^{1} x^{\alpha - 1}e^{\beta x} \; dx} \sum_{v_{s1}} \dbracket{\prog{1}}(v_{s1}) \times r' \times
				\sum_{v'''} \dbracket{\prog{1}}(v''') \times
				\\
				&\sum_{v_4} \dbracket{\prog{3}}(v_4) \times
				\sum_{v_5} \dbracket{\prog{4}}(v_5) \times
				\sum_{v_6} \dbracket{\prog{obs2}[\yvar{3} \mapsto v_4]}(v_6) \times
				\\
				&(\theta_1 \delta(\bin{r'}{b}) + (1 - \theta_1)\theta_2 \delta(v''') + (1 - \theta_1)(1 - \theta_2)\theta_3 \delta(v_4) + (1 - \theta_1)(1 - \theta_2)(1 - \theta_3) \delta(v_5))(\bin{r}{b})
				~\tag{Lemma~\ref{lemma:unifobs}}
				\\
				=& \sum_{r' \in [0, 1]_b} \frac{\int_{r'}^{r' + \frac{1}{2^b}} x^{\alpha - 1}e^{\beta x} \; dx}{\int_{0}^{1} x^{\alpha - 1}e^{\beta x} \; dx} \sum_{v_{s1}} \dbracket{\prog{1}}(v_{s1}) \times r' \times
				\sum_{v'''} \dbracket{\prog{1}}(v''') \sum_{v_{s2}} \dbracket{\prog{1}}(v_{s2}) \times
				\\
				&\sum_{v_4} \dbracket{\prog{3}}(v_4) \times
				\sum_{v_5} \dbracket{\prog{4}}(v_5) \times
				\sum_{v_6} \dbracket{\prog{obs2}[\yvar{3} \mapsto v_4]}(v_6) \times
				\\
				&(\theta_1 \delta(\bin{r'}{b}) + (1 - \theta_1)\theta_2 \delta(v''') + (1 - \theta_1)(1 - \theta_2)\theta_3 \delta(v_4) + (1 - \theta_1)(1 - \theta_2)(1 - \theta_3) \delta(v_5))(\bin{r}{b})
				~\tag{by~\ref{normalized-pr}}	
			\end{flalign*}
			\begin{flalign*}
				=& \sum_{r' \in [0, 1]_b} \frac{\int_{r'}^{r' + \frac{1}{2^b}} x^{\alpha - 1}e^{\beta x} \; dx}{\int_{0}^{1} x^{\alpha - 1}e^{\beta x} \; dx} \sum_{v_{s1}} \dbracket{\prog{1}}(v_{s1}) \times r' \times
				\sum_{r''' \in [0, 1]_b}  \frac{\int_{r'''}^{r''' + \frac{1}{2^b}} x^{\alpha - 1}e^{\beta x} \; dx}{\int_{0}^{1} x^{\alpha - 1}e^{\beta x} \; dx} \sum_{v_{s2}} \dbracket{\prog{1}}(v_{s2}) \times
				\\
				&\sum_{v_4} \dbracket{\prog{3}}(v_4) \times
				\sum_{v_5} \dbracket{\prog{4}}(v_5) \times
				\sum_{v_6} \dbracket{\prog{obs2}[\yvar{3} \mapsto v_4]}(v_6) \times
				\\
				&(\theta_1 \delta(\bin{r'}{b}) + (1 - \theta_1)\theta_2 \delta(\bin{r'''}{b}) + (1 - \theta_1)(1 - \theta_2)\theta_3 \delta(v_4) + (1 - \theta_1)(1 - \theta_2)(1 - \theta_3) \delta(v_5))(\bin{r}{b})
				~\tag{Induction Hypothesis}	
				\\
				=& \sum_{r' \in [0, 1]_b} \frac{\int_{r'}^{r' + \frac{1}{2^b}} x^{\alpha - 1}e^{\beta x} \; dx}{\int_{0}^{1} x^{\alpha - 1}e^{\beta x} \; dx} \sum_{v_{s1}} \dbracket{\prog{1}}(v_{s1}) \times r' \times
				\sum_{r''' \in [0, 1]_b}  \frac{\int_{r'''}^{r''' + \frac{1}{2^b}} x^{\alpha - 1}e^{\beta x} \; dx}{\int_{0}^{1} x^{\alpha - 1}e^{\beta x} \; dx} \sum_{v_{s2}} \dbracket{\prog{1}}(v_{s2}) \times
				\\
				&\sum_{v_4} \dbracket{\prog{3}}_D(v_4) \sum_{v_{s3}} \dbracket{\prog{3}}(v_{s3}) \times
				\sum_{v_5} \dbracket{\prog{4}}(v_5) \times
				\sum_{v_6} \dbracket{\prog{obs2}[\yvar{3} \mapsto v_4]}(v_6) \times
				\\
				&(\theta_1 \delta(\bin{r'}{b}) + (1 - \theta_1)\theta_2 \delta(\bin{r'''}{b}) + (1 - \theta_1)(1 - \theta_2)\theta_3 \delta(v_4) + (1 - \theta_1)(1 - \theta_2)(1 - \theta_3) \delta(v_5))(\bin{r}{b})
				~\tag{by~\ref{normalized-pr}}	
				\\
				=& \sum_{r' \in [0, 1]_b} r'\frac{\int_{r'}^{r' + \frac{1}{2^b}} x^{\alpha - 1}e^{\beta x} \; dx}{\int_{0}^{1} x^{\alpha - 1}e^{\beta x} \; dx} \sum_{v_{s1}} \dbracket{\prog{1}}(v_{s1})
				\sum_{r''' \in [0, 1]_b}  \frac{\int_{r'''}^{r''' + \frac{1}{2^b}} x^{\alpha - 1}e^{\beta x} \; dx}{\int_{0}^{1} x^{\alpha - 1}e^{\beta x} \; dx} \sum_{v_{s2}} \dbracket{\prog{1}}(v_{s2}) \times
				\\
				&\sum_{r_4 \in [0, 1]_b}  \frac{\int_{r_4}^{r_4 + \frac{1}{2^b}} x^{\alpha - 2}e^{\beta x} \; dx}{\int_{0}^{1} x^{\alpha - 2}e^{\beta x} \; dx} \sum_{v_{s3}} \dbracket{\prog{3}}(v_{s3}) \times
				\sum_{v_5} \dbracket{\prog{4}}(v_5) \times
				\sum_{v_6} \dbracket{\prog{obs2}[\yvar{3} \mapsto \bin{r_4}{b}]}(v_6) \times
				\\
				&(\theta_1 \delta(\bin{r'}{b}) + (1 - \theta_1)\theta_2 \delta(\bin{r'''}{b}) + (1 - \theta_1)(1 - \theta_2)\theta_3 \delta(\bin{r_4}{b}) + (1 - \theta_1)(1 - \theta_2)(1 - \theta_3) \delta(v_5))(\bin{r}{b})
				~\tag{Induction Hypothesis}	
				\\
				=& k \sum_{r' \in [0, 1]_b} \frac{r'\int_{r'}^{r' + \frac{1}{2^b}} x^{\alpha - 1}e^{\beta x} \; dx}{\int_{0}^{1} x^{\alpha - 1}e^{\beta x} \; dx} 
				\sum_{r''' \in [0, 1]_b}  \frac{\int_{r'''}^{r''' + \frac{1}{2^b}} x^{\alpha - 1}e^{\beta x} \; dx}{\int_{0}^{1} x^{\alpha - 1}e^{\beta x} \; dx} 
				\sum_{r_4 \in [0, 1]_b}  \frac{\int_{r_4}^{r_4 + \frac{1}{2^b}} x^{\alpha - 2}e^{\beta x} \; dx}{\int_{0}^{1} x^{\alpha - 2}e^{\beta x} \; dx} 
				\sum_{v_5} \dbracket{\prog{4}}(v_5)r_4
				\\
				&(\theta_1 \delta(\bin{r'}{b}) + (1 - \theta_1)\theta_2 \delta(\bin{r'''}{b}) + (1 - \theta_1)(1 - \theta_2)\theta_3 \delta(\bin{r_4}{b}) + (1 - \theta_1)(1 - \theta_2)(1 - \theta_3) \delta(v_5))(\bin{r}{b})
				~\tag{Lemma~\ref{lemma:unifobs} and \(k = \sum_{v_{s1}} \dbracket{\prog{1}}(v_{s1}) \sum_{v_{s2}} \dbracket{\prog{1}}(v_{s2})\sum_{v_{s3}} \dbracket{\prog{3}}(v_{s3})\)}
				\\
				=& k' \sum_{r' \in [0, 1]_b} \frac{r'\int_{r'}^{r' + \frac{1}{2^b}} x^{\alpha - 1}e^{\beta x} \; dx}{\int_{0}^{1} x^{\alpha - 1}e^{\beta x} \; dx} 
				\sum_{r''' \in [0, 1]_b}  \frac{\int_{r'''}^{r''' + \frac{1}{2^b}} x^{\alpha - 1}e^{\beta x} \; dx}{\int_{0}^{1} x^{\alpha - 1}e^{\beta x} \; dx} 
				\sum_{r_4 \in [0, 1]_b}  \frac{r_4\int_{r_4}^{r_4 + \frac{1}{2^b}} x^{\alpha - 2}e^{\beta x} \; dx}{\int_{0}^{1} x^{\alpha - 2}e^{\beta x} \; dx} 
				\\
				&\sum_{r_5} \frac{(r_5 + \frac{1}{2^b})^{\alpha - 1}e^{\beta r_5}}{\sum_{r'_5} (r'_5 + \frac{1}{2^b})^{\alpha - 1}e^{\beta r'_5}} \times
				\\
				&(\theta_1 \delta(\bin{r'}{b}) + (1 - \theta_1)\theta_2 \delta(\bin{r'''}{b}) + (1 - \theta_1)(1 - \theta_2)\theta_3 \delta(\bin{r_4}{b}) + (1 - \theta_1)(1 - \theta_2)(1 - \theta_3) \delta(v_5))(\bin{r}{b})
				~\tag{Lemma~\ref{lemma:pointGamma} and \(k' = k \sum_{v_{s4}} \dbracket{\prog{4}}(v_{s4})\)}
			\end{flalign*}
			\begin{flalign*}
				=& k'' (\theta_1 \frac{r \int_{r}^{r + \frac{1}{2^b}} x^{\alpha - 1}e^{\beta x} \; dx}{\sum_{r_{s1} \in [0, 1]_b} r_{s1} \int_{r_{s1}}^{r_{s1} + \frac{1}{2^b}} x^{\alpha - 1}e^{\beta x} \; dx}
				+(1 - \theta_1)\theta_2  \frac{\int_{r}^{r + \frac{1}{2^b}} x^{\alpha - 1}e^{\beta x} \; dx}{\int_{0}^{1} x^{\alpha - 1}e^{\beta x} \; dx}
				\\
				&+ (1 - \theta_1)(1 - \theta_2) \theta_3
				\frac{r\int_{r}^{r + \frac{1}{2^b}} x^{\alpha - 2}e^{\beta x} \; dx}{\sum_{r_{s2} \in [0, 1]_b} r_{s2}\int_{r_{s2}}^{r_{s2} + \frac{1}{2^b}} x^{\alpha - 2}e^{\beta x} \; dx} 
				\\
				&+ (1 - \theta_1)(1 - \theta_2)(1 - \theta_3) \frac{(r + \frac{1}{2^b})^{\alpha - 1}e^{\beta r}}{\sum_{r_{s3}} (r_{s3} + \frac{1}{2^b})^{\alpha - 1}e^{\beta r_{s3}}})
				~\tag{\(k'' = \frac{k' \sum_{r_{s1} \in [0, 1]_b} r_{s1} \int_{r_{s1}}^{r_{s1} + \frac{1}{2^b} x^{\alpha - 1}e^{\beta x} \; dx}\sum_{r_{s2} \in [0, 1]_b} r_{s2}\int_{r_{s2}}^{r_{s2} + \frac{1}{2^b}} x^{\alpha - 2}e^{\beta x} \; dx}{\int_{0}^{1} x^{\alpha - 1}e^{\beta x} \; dx  \int_{0}^{1} x^{\alpha - 2}e^{\beta x} \; dx}\)}
				\\
				=& k'' \Bigg[\frac{r\int_{r}^{r + \frac{1}{2^b}}y^{\alpha - 1}e^{\beta y} \; dy}{\sum_{r' \in [0, 1]_b} r'\int_{r'}^{r' + \frac{1}{2^b}} y^{\alpha - 1}e^{\beta y} \; dy} \frac{\sum_{r' \in [0, 1]_b}r'\int_{r'}^{r' + \frac{1}{2^b}} y^{\alpha - 1}e^{\beta y} \; dy}{\int_{0}^{1} z^{\alpha}e^{\beta z} \; dz}
				\\ 
				&+ 
				\bigg[\frac{(r + \frac{1}{2^b})^{\alpha - 1} e^{\beta (r + \frac{1}{2^b})}}{2^{b}\beta} + \frac{r(\alpha - 1)}{\beta} \int_{r}^{r + \frac{1}{2^b}} y^{\alpha - 2}e^{\beta y} \; dy - \frac{\alpha}{\beta} \int_{r}^{r + \frac{1}{2^b}} y^{\alpha - 1}e^{\beta y} \; dy \bigg]
				\\
				&\frac{1}{\sum_{r' \in [0, 1]_b}\int_{r'}^{r' + \frac{1}{2^b}} (\int_{x = r'}^{y} y^{\alpha - 1}e^{\beta y} \; dx) \; dy} (1 - \frac{\sum_{r' \in [0, 1]_b}r'\int_{r'}^{r' + \frac{1}{2^b}} y^{\alpha - 1}e^{\beta y} \; dy}{\int_{0}^{1} z^{\alpha}e^{\beta z} \; dz}) \Bigg]
				~\tag{Substituting the values of \(\theta_1, \theta_2, \theta_3\)}
			\end{flalign*}
			
			Evaluating the right hand side now, for all \(r \in [0, 1]_b\)
			
			\begin{flalign*}
				\dbracket{\prog{}}_D(\bin{r}{b}) =& \frac{\dbracket{\prog{}}(\bin{r}{b})}{\sum_{r' \in [0, 1]_b} \dbracket{\prog{}}(\bin{r'}{b})}
				\\
				=& \Bigg[\frac{r\int_{r}^{r + \frac{1}{2^b}}y^{\alpha - 1}e^{\beta y} \; dy}{\sum_{r' \in [0, 1]_b} r'\int_{r'}^{r' + \frac{1}{2^b}} y^{\alpha - 1}e^{\beta y} \; dy} \frac{\sum_{r' \in [0, 1]_b}r'\int_{r'}^{r' + \frac{1}{2^b}} y^{\alpha - 1}e^{\beta y} \; dy}{\int_{0}^{1} z^{\alpha}e^{\beta z} \; dz}
				\\ 
				&+ 
				\bigg[\frac{(r + \frac{1}{2^b})^{\alpha - 1} e^{\beta (r + \frac{1}{2^b})}}{2^{b}\beta} + \frac{r(\alpha - 1)}{\beta} \int_{r}^{r + \frac{1}{2^b}} y^{\alpha - 2}e^{\beta y} \; dy - \frac{\alpha}{\beta} \int_{r}^{r + \frac{1}{2^b}} y^{\alpha - 1}e^{\beta y} \; dy \bigg]
				\\
				&\frac{1}{\sum_{r' \in [0, 1]_b}\int_{r'}^{r' + \frac{1}{2^b}} (\int_{x = r'}^{y} y^{\alpha - 1}e^{\beta y} \; dx) \; dy} (1 - \frac{\sum_{r' \in [0, 1]_b}r'\int_{r'}^{r' + \frac{1}{2^b}} y^{\alpha - 1}e^{\beta y} \; dy}{\int_{0}^{1} z^{\alpha}e^{\beta z} \; dz}) \Bigg]
			\end{flalign*}
		\end{proof}
		
		\subsection{Proof of Theorem~\ref{theorem:semantics-preserving}}
		
		Now that we have described the compilation of generalized-gamma densities to Dice programs, we will be doing the same for mixed-gamma densities.
		
		\emph{Proof for Theorem~\ref{theorem:semantics-preserving}}
		\begin{proof}
			
			By Definition~\ref{def:mixed-gamma}, let \(\Upsilon(x) = \sum_{i = 1}^{N} a_i \pi_{\alpha_i, \beta_i}(x)\) where \(N \in \mathbb{Z}^+; \forall i, \alpha_i \in \mathbb{Z}^+, \beta_i \in \mathbb{R}, a_i \in [0, 1]\) and \(\sum_{i=1}^{N} a_i = 1\). 
			
			The further proof is by induction on N.
			
			\underline{Base Case:} N = 1 
			
			For N = 1, we have \(a_1 = 1\) and \(\Upsilon = \pdf{\alpha}{\beta}\). If \(\Upsilon \comp_b \prog{}\), then \(\Upsilon\) and \(\prog{}\) are \(b\)-equivalent by Lemma~\ref{lemma:alpha>1-density}.
			
			\underline{Inductive Hypothesis:} \(\forall N' < N\), if \(\sum_{i=1}^{N'} a_i \pdf{\alpha_i}{\beta_i} \comp_b \prog{}\), then \(\sum_{i=1}^{N'} a_i \pdf{\alpha_i}{\beta_i}\) and \(\prog{}\) are \(b\)-equivalent. By Definition~\ref{def:binarize}, we have the following:
			
			\[\forall r \in [0, 1]_b \;\;\;\;\;\; \int_{r}^{r + \frac{1}{2^b}} \sum_{i=1}^{N'} a_i \pdf{\alpha_i}{\beta_i}(x) \; dx = \dbracket{\prog{}}_D(\bin{r}{b})\]

			\underline{Inductive Step:} N > 1
			
			By ~\ref{Trans-mix}, \(\prog{} = \begin{array}{l}
				\texttt{let \(\yvar{1}\) = flip(\(a_N\)) in} \\
				\texttt{let \(\yvar{2}\) = \(\prog{1}\) in} \\
				\texttt{let \(\yvar{3}\) = \(\prog{2}\) in} \\
				\texttt{if \(\yvar{1}\) then \(\yvar{2}\) else \(\yvar{3}\)}
			\end{array}\)
			where \(\pdf{\alpha_N}{\beta_N} \comp_b \prog{1}\) and
			\(\sum_{i=2}^{N} \frac{a_i}{1 - a_N} \pdf{\alpha_i}{\beta_i} \comp_b \prog{2}\).

			Now, to prove that \(\Upsilon\) and \texttt{p} are \(b\)-equivalent, we need to prove the following by Definition~\ref{def:b-equivalence}. 
			
			\[\forall r \in [0, 1]_b \;\;\;\; \int_{r}^{r + \frac{1}{2^b}} \sum_{i = 1}^{N} a_i \pi_{\alpha_i, \beta_i}(x) \; dx = \dbracket{\prog{}}_D(\bin{r}{b})\]
			
			To evaluate the right hand side, we first evaluate \(\dbracket{\prog{}}\)
			
			For any \(r \in [0, 1]_b\),
			
			\begin{flalign*}
				&\dbracket{\prog{}}(\bin{r}{b})
				\\
				=& \dbracket{\begin{array}{l}
						\texttt{let \(\yvar{1}\) = flip(\(a_N\)) in} \\
						\texttt{let \(\yvar{2}\) = \(\prog{1}\) in} \\
						\texttt{let \(\yvar{3}\) = \(\prog{2}\) in} \\
						\texttt{if \(\yvar{1}\) then \(\yvar{2}\) else \(\yvar{3}\)}
				\end{array}}(\bin{r}{b})
				\\
				=& \sum_{v'} \dbracket{\Lflip{(a_N)}}(v') \times \dbracket{\begin{array}{l}
						\texttt{let \(\yvar{2}\) = \(\prog{1}\) in} \\
						\texttt{let \(\yvar{3}\) = \(\prog{2}\) in} \\
						\texttt{if \(v'\) then \(\yvar{2}\) else \(\yvar{3}\)}
				\end{array}}(\bin{r}{b})
				~\tag{~\ref{Dice-let}}
				\\
				=& \sum_{v'} \dbracket{\Lflip{(a_N)}}(v') \times
				\sum_{v''} \dbracket{\prog{1}}(v'') \times
				\sum_{v'''} \dbracket{\prog{2}}(v''') \times
				\dbracket{\texttt{if \(v'\) then \(v''\) else \(v'''\)}}(\bin{r}{b})
				~\tag{~\ref{Dice-let}}
				\\
				=& \sum_{v'} \dbracket{\Lflip{(a_N)}}(v') \times
				\sum_{r'' \in [0, 1]_b} \int_{r''}^{r'' + \frac{1}{2^b}} \pdf{\alpha_N}{\beta_N}(x) \; dx \sum_{r_{s1} \in [0, 1]_b} \dbracket{\prog{1}}(\bin{r}{s1}) \times
				\sum_{v'''} \dbracket{\prog{2}}(v''') \times
				\\
				&\dbracket{\texttt{if \(v'\) then \(\bin{r''}{b}\) else \(\bin{r'''}{b}\)}}(\bin{r}{b})
				~\tag{Lemma~\ref{lemma:alpha>1-density}}
				\\
				=& \sum_{v'} \dbracket{\Lflip{(a_N)}}(v') \times
				\sum_{r'' \in [0, 1]_b} \int_{r''}^{r'' + \frac{1}{2^b}} \pdf{\alpha_N}{\beta_N}(x) \; dx \sum_{r_{s1} \in [0, 1]_b} \dbracket{\prog{1}}(\bin{r}{s1}) \times
				\\
				&\sum_{r'''} \int_{r'''}^{r''' + \frac{1}{2^b}} \sum_{i=1}^{N-1} \frac{a_i}{1 - a_N} \pdf{\alpha_i}{\beta_i}(x) \; dx  \sum_{r_{s2} \in [0, 1]_b} \dbracket{\prog{2}}(\bin{r_{s2}}{b}) \times \dbracket{\texttt{if \(v'\) then \(\bin{r''}{b}\) else \(\bin{r'''}{b}\)}}(\bin{r}{b})
				~\tag{Induction Hypothesis}
				\\
				=& k \sum_{v'} \dbracket{\Lflip{(a_N)}}(v') \times
				\sum_{r'' \in [0, 1]_b} \int_{r''}^{r'' + \frac{1}{2^b}} \pdf{\alpha_N}{\beta_N}(x) \; dx \times
				\\
				&\sum_{r'''} \int_{r'''}^{r''' + \frac{1}{2^b}} \sum_{i=1}^{N-1} \frac{a_i}{1 - a_N} \pdf{\alpha_i}{\beta_i}(x) \; dx   \times \dbracket{\texttt{if \(v'\) then \(\bin{r''}{b}\) else \(\bin{r'''}{b}\)}}(\bin{r}{b})
				~\tag{\(k = \sum_{r_{s1} \in [0, 1]_b} \dbracket{\prog{1}}(\bin{r}{s1})\sum_{r_{s2} \in [0, 1]_b} \dbracket{\prog{2}}(\bin{r_{s2}}{b})\)}
				\\
				=& k
				\sum_{r'' \in [0, 1]_b} \int_{r''}^{r'' + \frac{1}{2^b}} \pdf{\alpha_N}{\beta_N}(x) \; dx \sum_{r'''} \int_{r'''}^{r''' + \frac{1}{2^b}} \sum_{i=1}^{N-1} \frac{a_i}{1 - a_N} \pdf{\alpha_i}{\beta_i}(x) \; dx (a_N \delta(\bin{r''}{b})(\bin{r}{b}) \\
				&+ (1 - a_N) \delta(\bin{r'''}{b})(\bin{r}{b}))
				~\tag{~\ref{Dice-flip}, ~\ref{Dice-Ite}, ~\ref{Dice-value}}
				\\
				=& k (a_N \int_{r}^{r + \frac{1}{2^b}} \pdf{\alpha_N}{\beta_N}(x) \; dx + (1 - a_N)\int_{r}^{r + \frac{1}{2^b}} \sum_{i=1}^{N-1} \frac{a_i}{1 - a_N} \pdf{\alpha_i}{\beta_i}(x) \; dx)
				~\tag{Considering only non zero terms}
				\\
				=& k \int_{r}^{r + \frac{1}{2^b}} \sum_{i=1}^{N} a_i \pdf{\alpha_i}{\beta_i}(x) \; dx
			\end{flalign*}
			
			Evaluating the right hand side,
			
			For all \(r \in [0, 1]_b\)
			
			\begin{flalign*}
				\dbracket{\prog{}}_D(\bin{r}{b}) = \frac{\dbracket{\prog{}}(\bin{r}{b})}{\sum_{r' \in [0, 1]_b} \dbracket{\prog{}}(\bin{r'}{b})}
				= \int_{r}^{r + \frac{1}{2^b}} \sum_{i=1}^{N} a_i \pdf{\alpha_i}{\beta_i}(x) \; dx
			\end{flalign*}
		\end{proof}
		
		\section{Proofs of \(b\)-succinctness}
		
		We earlier described the flip counts for a Dice program and now we prove that Dice programs generated by \(\comp_b\) are \(b\)-succinct.
		
		Note that \(\comp_b\) is \(b\)-succinct for a mixed-gamma density \(\Upsilon\) if \(\exists k, \forall \prog{}, b \in \mathbb{Z}^+\), if \(\Upsilon \comp_b \prog{}\), then \(\flipcount{\prog{}} \leq kb\).

		\subsection{Proof of Lemma~\ref{succinct-alpha-0}}
		
		\begin{proof}
			To prove that \(\comp_b\) is \(b\)-succinct for \(\pdf{0}{\beta}\), we need to prove that \(\exists k, \forall \prog{}, b \in \mathbb{Z}^+\), if \(\pdf{0}{\beta} \comp_b \prog{}\), then \(\flipcount{\prog{}} \leq kb\).
			
			For all \(b > 0\), we have by \ref{Trans-expo0}, \(\prog{} = \begin{array}{l}
				\texttt{let \(\yvar{1}\) = flip($\theta_1$) in}
				\\ \texttt{let \(\yvar{2}\) = flip($\theta_2$) in}
				\\ \ldots
				\\ \texttt{let \(\yvar{b}\) = flip($\theta_b$) in}
				\\ \texttt{(\(\yvar{1}\), (\(\yvar{2}\), (\ldots, \(\yvar{b}\)))\ldots))}
			\end{array}\)
			
			Thus,
			\begin{flalign*} 
				\flipcount{\prog{}} =& \flipcount{\begin{array}{l}
						\texttt{let \(\yvar{1}\) = flip($\theta_1$) in}
						\\ \texttt{let \(\yvar{2}\) = flip($\theta_2$) in}
						\\ \ldots
						\\ \texttt{let \(\yvar{b}\) = flip($\theta_b$) in}
						\\ \texttt{(\(\yvar{1}\), (\(\yvar{2}\), (\ldots, \(\yvar{b}\)))\ldots))}
				\end{array}}
				\\
				=& b
				~\tag{Definition~\ref{def:flipcount}}
				\\
				\implies & k = 1
			\end{flalign*}
		\end{proof}
		
		\subsection{Proof of Lemma~\ref{succ:alpha-1-density}}
		
		\begin{lemma}\label{succinct:lessthan}
			For all positive integers \(b\) and Dice variables \(\yvar{1}\), \(\yvar{2}\), if \(\lessthan{\(\yvar{1}\)}{\(\yvar{2}\)}{b} = \prog{}\), then \(\flipcount{\prog{}} = 0\).
		\end{lemma}
		\begin{proof}
			The proof is by induction on \(b\).
			
			\underline{Base Case:} \(b = 1\)
			
			Then by ~\ref{Less-Than-1}, \(\prog{} = (\yvar{2} \wedge \neg \yvar{1})\) and \(\flipcount{\prog{}} = 0\)
			
			\underline{Induction Hypothesis:} \(\forall b' < b\) and dice variables \(\yvar{1}\), \(\yvar{2}\), if \(\lessthan{\(\yvar{1}\)}{\(\yvar{2}\)}{b'} = \prog{+}\), then \(\flipcount{\prog{+}} = 0\)
			
			\underline{Inductive step:} 
			Let \(\lessthan{\(\yvar{1}\)}{\(\yvar{2}\)}{b} = \prog{}\).
			
			Then by \ref{Less-Than>1}
			\(\prog{} = \begin{array}{l}
				\texttt{let \(\yvar{1f}\) = fst(\(\yvar{1}\)) in} \\
				\texttt{let \(\yvar{2f}\) = fst(\(\yvar{2}\)) in} \\
				\texttt{let \(\yvar{1wins}\) = \(\yvar{1f}\) \(\wedge \neg\) \(\yvar{2f}\) in} \\
				\texttt{let \(\yvar{2wins}\) = \(\yvar{2f}\) \(\wedge \neg\) \(\yvar{1f}\) in} \\
				\texttt{if \(\yvar{1wins}\) then \false else} \\
				\texttt{if \(\yvar{2wins}\) then \true else} \\
				\texttt{let \(\yvar{1s}\) = snd(\(\yvar{1}\)) in} \\
				\texttt{let \(\yvar{2s}\) = snd(\(\yvar{2}\)) in} \\
				\prog{+}
			\end{array}\)
			
			where \(\lessthan{\(\yvar{1s}\)}{\(\yvar{2s}\)}{b-1} = \prog{+}\).
			
			\begin{flalign*}
				\flipcount{\prog{}} =& \flipcount{\begin{array}{l}
						\texttt{let \(\yvar{1f}\) = fst(\(\yvar{1}\)) in} \\
						\texttt{let \(\yvar{2f}\) = fst(\(\yvar{2}\)) in} \\
						\texttt{let \(\yvar{1wins}\) = \(\yvar{1f}\) \(\wedge \neg\) \(\yvar{2f}\) in} \\
						\texttt{let \(\yvar{2wins}\) = \(\yvar{2f}\) \(\wedge \neg\) \(\yvar{1f}\) in} \\
						\texttt{if \(\yvar{1wins}\) then \false else} \\
						\texttt{if \(\yvar{2wins}\) then \true else} \\
						\texttt{let \(\yvar{1s}\) = snd(\(\yvar{1}\)) in} \\
						\texttt{let \(\yvar{2s}\) = snd(\(\yvar{2}\)) in} \\
						\prog{+} 
				\end{array}}
				\\
				=& \flipcount{\prog{+}} 
				~\tag{Definition~\ref{def:flipcount}}
				\\
				=& 0
				~\tag{Induction Hypothesis}
			\end{flalign*}
		\end{proof}
		
		\begin{lemma}\label{succinct:unifobs}
			For all positive integers b and Dice variable \(\yvar{}\), if \(\unifobs{\yvar{}}{b} = \prog{}\), then 
			
			\noindent \(\flipcount{\prog{}} = b\)
		\end{lemma}
		
		\begin{proof}
			By \ref{unifobs}, \(\prog{} = \begin{array}{l}
				\texttt{let \(\yvar{1}\) = \(\prog{1}\) in} \\
				\texttt{let \(\yvar{bool}\) = \(\prog{2}\) in } \\
				\texttt{observe(\(\yvar{bool}\))}
			\end{array}\)
			
			where \(\pdf{0}{0} \comp_b \prog{1}\) and
			\(\texttt{less\_than(\(\yvar{1}\), \(\yvar{}\), b)} = \prog{2}\)
			
			\begin{flalign*}
				\flipcount{\prog{}} =& \flipcount{\begin{array}{l}
						\texttt{let \(\yvar{1}\) = \(\prog{1}\) in} \\
						\texttt{let \(\yvar{bool}\) = \(\prog{2}\) in } \\
						\texttt{observe(\(\yvar{bool}\))}
				\end{array}}
				\\
				=& \flipcount{\prog{1}} + \flipcount{\prog{2}} + \flipcount{\Lobs{\yvar{bool}}}
				~\tag{Definition~\ref{def:flipcount}}
				\\
				=& b + 0 + 0 = b
				~\tag{Lemma~\ref{succinct-alpha-0} and Lemma~\ref{succinct:lessthan}}
			\end{flalign*}
		\end{proof}
		
		\emph{Proof of Lemma~\ref{succ:alpha-1-density}}
		
		\begin{proof}
			To prove that \(\forall \beta \in \mathbb{R}, \comp_b\) is \(b\)-succinct for \(\pdf{1}{\beta}\), we need to prove that \(\exists k, \forall \prog{}, b \in \mathbb{Z}^+\), if \(\pdf{1}{\beta} \comp_b \prog{}\), then \(\flipcount{\prog{}} \leq kb\).
			
			\underline{Case 1:} \(\beta \neq 0\)
			
			Then by ~\ref{Trans-expo1}, \(\prog{} = \begin{array}{l}
				\texttt{let \(\yvar{1}\) = } \prog{1} \texttt{ in }
				\\ \texttt{let \_ = \(\prog{3}\)}
				\\ \texttt{let \(\yvar{2}\) = } \prog{1} \texttt{ in}
				\\ \texttt{let \(\yvar{3}\) = flip\((\theta)\) in}
				\\ \texttt{if \(\yvar{3}\) then \(\yvar{2}\) else \(\yvar{1}\)}
			\end{array}\)
			
			where \(\pdf{0}{\beta} \comp_b \prog{1}\) and
			\(\texttt{unifObs(\(\yvar{1}\), b)} = \prog{3}\)
			
			\begin{flalign*}
				\flipcount{\prog{}} =& \flipcount{\begin{array}{l}
						\texttt{let  \(\yvar{1}\) = } \prog{1} \texttt{ in }
						\\ \texttt{let \_ = \(\prog{3}\)}
						\\ \texttt{let \(\yvar{2}\) = } \prog{1} \texttt{ in}
						\\ \texttt{let \(\yvar{3}\) = flip\((\theta)\) in}
						\\ \texttt{if \(\yvar{3}\) then \(\yvar{2}\) else \(\yvar{1}\)}
				\end{array}}
				\\
				=& \flipcount{\prog{1}} + \flipcount{\prog{3}} + \flipcount{\prog{1}} + \flipcount{\Lflip{\theta_1}} 
				~\tag{Definition~\ref{def:flipcount}}
				\\
				=& b + b + b + 1
				~\tag{Lemma~\ref{succinct-alpha-0}, ~\ref{succinct:unifobs}}
				\\
				\leq& 4b \implies k = 4
			\end{flalign*}
			
			\underline{Case 2:} \(\beta = 0\)
			
			Then by ~\ref{Trans-expo1zero}, \(\prog{} = \begin{array}{l}
				\texttt{let \(\yvar{1}\) = } \prog{1} \texttt{ in }
				\\ \texttt{let \_ = \(\prog{3}\)}
				\\ \texttt{let \(\yvar{2}\) = } \prog{1} \texttt{ in}
				\\ \texttt{let \(\yvar{3}\) = flip\((\theta)\) in}
				\\ \texttt{if \(\yvar{3}\) then \(\yvar{2}\) else \(\yvar{1}\)}
			\end{array}\)
			
			where \(\pdf{0}{0} \comp_b \prog{1}\) and
			\(\texttt{unifObs(\(\yvar{1}\), b)} = \prog{3}\)
			
			\begin{flalign*}
				\flipcount{\prog{}} =& \flipcount{\begin{array}{l}
						\texttt{let \(\yvar{1}\) = } \prog{1} \texttt{ in }
						\\ \texttt{let \_ = \(\prog{3}\)}
						\\ \texttt{let \(\yvar{2}\) = } \prog{1} \texttt{ in}
						\\ \texttt{let \(\yvar{3}\) = flip\((\theta)\) in}
						\\ \texttt{if \(\yvar{3}\) then \(\yvar{2}\) else \(\yvar{1}\)}
				\end{array}}
				\\
				=& \flipcount{\prog{1}} + \flipcount{\prog{3}} + \flipcount{\prog{1}} + \flipcount{\Lflip{\theta_1}} 
				~\tag{Definition~\ref{def:flipcount}}
				\\
				=& b + b + b + 1
				~\tag{Lemma~\ref{succinct-alpha-0}, ~\ref{succinct:unifobs}}
				\\
				\leq& 4b \implies k = 4
			\end{flalign*}

		\end{proof}
		
		\subsection{Proof of Lemma~\ref{succ:alpha>1-density}}
		
		\begin{lemma}\label{succ:pointGamma}
			\(\forall \alpha \in \mathbb{Z}^+, \exists k, \forall b \in \mathbb{Z}^{+}, \beta \in \mathbb{R}, \epsilon \in \mathbb{R}, \prog{}\), if \(\pointgamma{\alpha}{\beta}{\epsilon}{b} = \prog{}\), then \(\flipcount{\prog{}} \leq kb\)
		\end{lemma}
		
		\begin{proof}
			The proof is by induction on \(\alpha\).
			
			\underline{Base Case:} \(\alpha = 0\)
			
			By \ref{spg0}, \(\prog{} = \prog{1}\) where \(\pdf{0}{\beta} \comp_b \prog{1}\).
			
			By Lemma~\ref{succinct-alpha-0}, we have \(\flipcount{\prog{}} = \flipcount{\prog{1}} \leq b \implies k = 1\)
			
			\underline{Induction Hypothesis:} \(\forall \alpha' < \alpha, \exists k', \forall b \in \mathbb{Z}^{+}, \beta \in \mathbb{R}, \epsilon \in \mathbb{R}, \prog{}\), if \(\pointgamma{\alpha'}{\beta}{\epsilon}{b} = \prog{}\), then \(\flipcount{\prog{}} \leq k'b\)
			
			\underline{Inductive Step:} \(\alpha > 0\)
			
			By \ref{spggt0}, we have \(\prog{} = \begin{array}{l}
				\texttt{let \(\yvar{1}\) = } \prog{1} \texttt{ in } 
				\\ \texttt{let \(\yvar{2}\) = } \Lflip{\frac{1}{1 + \epsilon}} \texttt{ in } 
				\\ \texttt{let } \_ = \Lite{\yvar{2}}{\prog{2}}{{\true}} \texttt{ in}
				\\  \yvar{1} \end{array}\)
			
			where \(\pointgamma{\alpha - 1}{\beta}{\epsilon}{b} = \prog{1}\) and \(\unifobs{\yvar{1}}{b} = \prog{2}\)
			
			\begin{flalign*}
				\flipcount{\prog{}} =& \flipcount{\begin{array}{l}
						\texttt{let \(\yvar{1}\) = } \prog{1} \texttt{ in } 
						\\ \texttt{let \(\yvar{2}\) = } \Lflip{\frac{1}{1 + \epsilon}} \texttt{ in } 
						\\ \texttt{let } \_ = \Lite{\yvar{2}}{\prog{2}}{{\true}} \texttt{ in}
						\\  \yvar{1} \end{array}}
				\\
				=& \flipcount{\prog{1}} + \flipcount{\Lflip{\frac{1}{1 + \epsilon}}} 
				\\
				\leq& k'b + 1 + \flipcount{\prog{2}} + \flipcount{{\true}}
				~\tag{Induction Hypothesis and Definition~\ref{def:flipcount}}
				\\
				\leq& k'b + 1 + b \leq (k'+2)b 
				~\tag{Lemma \ref{succinct:unifobs}} 
			\end{flalign*}
			
		\end{proof}
		
		\emph{Proof of Lemma~\ref{succ:alpha>1-density}}
		
		\begin{proof}
			The proof is by induction on \(\alpha\).
			
			\underline{Base Case 1:} \(\alpha = 0\)
			By Lemma~\ref{succinct-alpha-0}, \(\forall \beta \in \mathbb{Z}^+, \comp_b\) is \(b\)-succinct for \(\pdf{0}{\beta}\).
			
			\underline{Base Case 2:} \(\alpha = 1\)
			By Lemma~\ref{succ:alpha-1-density}, \(\forall \beta \in \mathbb{Z}^+, \comp_b\) is \(b\)-succinct for \(\pdf{1}{\beta}\).
			
			\underline{Induction Hypothesis:} \(\forall \alpha' < \alpha, \forall \beta \in \mathbb{Z}^+, \comp_b\) is \(b\)-succinct for \(\pdf{\alpha'}{\beta}\).
			
			This implies that:
			\(\forall \alpha' < \alpha, \beta \in \mathbb{Z}^+, \exists k_{\alpha'}, \forall \prog{}, b \in \mathbb{Z}^+, \) if \(\pdf{\alpha}{\beta} \comp_b \prog{},\) then \(\flipcount{\prog{}} \leq k_{\alpha'}b\)
			
			\underline{Inductive Step:} \(\forall \alpha > 1\)
			
			By \ref{Trans-expo>1}, if \(\pdf{\alpha}{\beta} \comp_b \prog{}\), then
			
			\(\texttt{p} = \begin{array}{l}
				\texttt{let \(\yvar{1}\) = \(\prog{1}\) in}
				\\ \texttt{let \_ = \(\prog{obs1}\) in}
				\\ \texttt{let \(\yvar{2}\) = \(\prog{1}\) in}
				\\ \texttt{let \(\yvar{3}\) = \(\prog{3}\) in}
				\\ \texttt{let \(\yvar{4}\) = \(\prog{4}\) in}
				\\ \texttt{let \_ = \(\prog{obs2}\) in}
				\\ \texttt{let \(\yvar{f1}\) = flip(\(\theta_1\)) in}
				\\ \texttt{let \(\yvar{f2}\) = flip(\(\theta_2\)) in}
				\\ \texttt{let \(\yvar{f3}\) = flip(\(\theta_3\)) in }
				\\ \texttt{if \(\yvar{f1}\) then \(\yvar{1}\) else} 
				\\ \texttt{if \(\yvar{f2}\) then \(\yvar{2}\) else if \(\yvar{f3}\) then \(\yvar{3}\) else \(\yvar{4}\)}
			\end{array}\)
			
			where 
			
			\[\pdf{\alpha-1}{\beta} \comp_b \prog{1}
			\;\;\;\;\;\;\;\; \pdf{\alpha-2}{\beta} \comp_b \prog{3} \]
			
			\[\pointgamma{\alpha-1}{\beta}{\frac{1}{2^b}}{b} = \prog{4}
			\;\;\;\;\;\;\;\;
			\unifobs{\yvar{1}}{b} = \prog{obs1}
			\;\;\;\;\;\;\;\;
			\unifobs{\yvar{3}}{b} = \prog{obs2}\]
			
			Then, we have
			
			\begin{flalign*}
				&\flipcount{\prog{}} 
				\\
				=& \flipcount{\begin{array}{l}
						\texttt{let \(\yvar{1}\) = \(\prog{1}\) in}
						\\ \texttt{let \_ = \(\prog{obs1}\) in}
						\\ \texttt{let \(\yvar{2}\) = \(\prog{1}\) in}
						\\ \texttt{let \(\yvar{3}\) = \(\prog{3}\) in}
						\\ \texttt{let \(\yvar{4}\) = \(\prog{4}\) in}
						\\ \texttt{let \_ = \(\prog{obs2}\) in}
						\\ \texttt{let \(\yvar{f1}\) = flip(\(\theta_1\)) in}
						\\ \texttt{let \(\yvar{f2}\) = flip(\(\theta_2\)) in}
						\\ \texttt{let \(\yvar{f3}\) = flip(\(\theta_3\)) in }
						\\ \texttt{if \(\yvar{f1}\) then \(\yvar{1}\) else} 
						\\ \texttt{if \(\yvar{f2}\) then \(\yvar{2}\) else if \(\yvar{f3}\) then \(\yvar{3}\) else \(\yvar{4}\)}
				\end{array}}
				\\
				=& \flipcount{\prog{1}} + \flipcount{\prog{obs1}} + \flipcount{\prog{1}} + \flipcount{\prog{3}}
				\\
				& + \flipcount{\prog{4}} + \flipcount{\prog{obs2}} + \flipcount{\Lflip{\theta_1}} + \flipcount{\Lflip{\theta_2}}
				\\
				& + \flipcount{\Lflip{\theta_3}} + \flipcount{\begin{array}{l}
						\texttt{if \(\yvar{f1}\) then \(\yvar{1}\) else} 
						\\ \texttt{if \(\yvar{f2}\) then \(\yvar{2}\) else if \(\yvar{f3}\) then \(\yvar{3}\) else \(\yvar{4}\)}
				\end{array}}
				\\
				&\leq k_{\alpha-1}b + b + k_{\alpha - 1}b + k_{\alpha - 2}b + k_{\mathit{ptg}}b + b + 3 + 0
				~\tag{Induction Hypothesis, Lemma~\ref{succinct:unifobs} and \ref{succ:pointGamma}, Definition~\ref{def:flipcount}}
				\\
				& \leq (k_{\alpha-1} + 5 + k_{\alpha - 1} + k_{\alpha - 2} + k_{\mathit{ptg}})b
			\end{flalign*}
		\end{proof}
		
		\subsection{Proof for Theorem~\ref{succ:mixed-gamma}}
		
		\begin{proof}
			By Definition~\ref{def:mixed-gamma}, let \(\Upsilon(x) = \sum_{i = 1}^{N} a_i \pi_{\alpha_i, \beta_i}(x)\) where \(N \in \mathbb{Z}^+; \forall i, \alpha_i \in \mathbb{Z}^+, \beta_i \in \mathbb{R}, a_i \in [0, 1]\) and \(\sum_{i=1}^{N} a_i = 1\). 
			
			The further proof is by induction on N.
			
			\underline{Base Case:} N = 1 
			
			For \(N = 1\), we have \(a_1 = 1\) and \(\Upsilon = \pdf{\alpha}{\beta}\), and \(\comp_b\) is \(b\)-succinct for \(\pdf{\alpha}{\beta}\) by Lemma~\ref{succ:alpha>1-density}.
			
			\underline{Induction Hypothesis:} \(\forall N' < N\), \(\comp_b\) is \(b\)-succinct for \(\sum_{i = 1}^{N'} a_i \pdf{\alpha_i}{\beta_i}\).
			
			This implies \(\forall N' < N, \exists k'\) such that if \(\sum_{i = 1}^{N'} a_i\pdf{\alpha_i}{\beta_i} \comp_b \prog{+} \), then \(\flipcount{\prog{+}} \leq k'b\)
			
			\underline{Inductive Step:} \(\forall N > 1\)
			
			By ~\ref{Trans-mix}, \(\prog{} = \begin{array}{l}
				\texttt{let \(\yvar{1}\) = flip(\(a_N\)) in} \\
				\texttt{let \(\yvar{2}\) = \(\prog{1}\) in} \\
				\texttt{let \(\yvar{3}\) = \(\prog{2}\) in} \\
				\texttt{if \(\yvar{}\) then \(\yvar{2}\) else \(\yvar{3}\)}
			\end{array}\)
			
			where \(\pdf{\alpha_N}{\beta_N} \comp_b \prog{1}\) and
			\(\sum_{i=1}^{N-1} \frac{a_i}{1 - a_N} \pdf{\alpha_i}{\beta_i} \comp_b \prog{2}\).
			
			\begin{flalign*}
				\flipcount{\prog{}}
				=& \flipcount{\Lflip{a_N}} + \flipcount{\prog{1}} + \flipcount{\prog{2}} 
				\\
				\leq& 1 + k''b + k'b + 0
				~\tag{Definition~\ref{def:flipcount}, Lemma~\ref{succ:alpha>1-density}, Induction hypothesis, Definition~\ref{def:flipcount}}
				\\
				\leq& (1 + k'' + k')b
			\end{flalign*}
		\end{proof}
		
		\section{Proof of Theorem~\ref{theorem:bitblasting}}
		
		\begin{lemma}
			\(\comp \circ \comp_b\) is a \(b\)-bit blasting function
		\end{lemma}
		
		\begin{proof}
			By Theorem~\ref{succ:mixed-gamma}, \(\comp_b\) is \(b\)-succinct for all mixed-gamma density \(\Upsilon\). This implies \(\forall \Upsilon, \exists k \forall \prog{}, b \in \mathbb{Z}^+\), if \(\Upsilon \comp_b \prog{}\), then \(\flipcount{\prog{}} \leq kb\). This further implies that if \(\prog{} \comp (\varphi, \gamma, w)\), then \(|w|\) is \(\mathcal{O}(b)\). Thus, \(\comp \circ \comp_b\) is a \(b\)-bit blasting function.
		\end{proof}
		
		\begin{lemma}
			\(\comp \circ \comp_b\) is a sound \(b\)-bit discretization function.
		\end{lemma}
		
		\begin{proof}
			By Theorem~\ref{theorem:semantics-preserving}, \(\forall \Upsilon, b \in \mathbb{Z}^+, \prog{}\), if \(\Upsilon \comp_b \prog{}\) then \(\Upsilon\) and \(\prog{}\) are \(b\)-equivalent. By Definition~\ref{def:b-equivalence}, this implies \(\forall r \in [0, 1]_b\), we have the following:
			
			\[\int_{r}^{r + \frac{1}{2^b}} \Upsilon(y) \; dy = \dbracket{\prog{}}_D(\bin{r}{b})\]
			
			Earlier work~\cite{holtzen2020dice} shows that if \(\prog{}\) is a closed Dice program without function calls and suppose \(\{\} \vdash \prog{} : \tau \comp (\varphi, \gamma, w)\). then for any value \(v : \tau\), we have that the following
			
			\[\dbracket{\prog{}}_D(v) = \frac{\mathit{WMC}(((\varphi \iff v) \wedge \gamma), w)}{\mathit{WMC}(\gamma, w)} = \frac{\mathbb{E}_w((\varphi(w) = v) \wedge \gamma)}{\mathbb{E}_w(\gamma)} \text{ if } \mathbb{E}_w(\gamma) \neq 0\]
			
			Both the equations above together imply that \(\comp \circ \comp_b\) is a sound \(b\)-bit discretization function.
		\end{proof}
		
		\newpage
		
		\section{Bounded BDD Size: Proof of Theorem~\ref{theorem:bddsize}}
		
		To prove the upper bound on size of the BDDs compiled from mixed-gamma Dice programs, we make an important observation. That each of the mixed-gamma Dice programs consists of two sets of flips:
		
		\begin{itemize}
			\item The flips that occur as part of a uniform or exponential distribution. 
			\item The flips that occur as guards of \(\Lite{}{}{}\) construct which are independent of the number of bits. 
		\end{itemize}
		
		Further proof entails in two steps. We first prove that every mixed-gamma Dice program satisfies certain invariants and then we prove that programs that satisfy these invariants compile to a BDD of size \(\mathit{poly}(b)\)
		
		We have the following notational assumptions in the following proofs:
		
		\begin{itemize}
			\item We use the notation \((x_1, x_2, \ldots, x_b)\) to denote the b-tuple which would be represented in Dice as b-nested tuple \((x_1, (x_2, (\ldots, x_b))\ldots)\)
			\item Since the mixed-gamma Dice programs do not use the function syntax of Dice, we skip the context \(\Phi\) that maps function names to compiled function bodies in the usage of the judgement \(\comp\).
			\item We use the notation \(\tbool^n\) to denote the type of tuple with \(n\) boolean elements. 
			\item We omit the explicit mention of A-Normal form for tuples for the sake of brevity.
			\item We use the terminology "coin flips" to refer to the \(\Lflip{\theta}\) syntax in the programs.
		\end{itemize}
		
		\subsection{OBDD size of returning Boolean formula}
		
		We show that the Dice programs emitted by \(\comp_b\) have a certain structure and then use that structure to argue about the OBDD size of the returning formula. We first formally define the structural invariant of the program.
		
		\begin{definition}[size of an obdd]
			The size of a reduced ordered binary decision diagram for a Boolean formula \(\phi\) given a variable order \(\mathcal{V}\) or \(\mathit{OBDD}_{\mathcal{V}}(\phi)\) is defined as the number of nodes in the corresponding decision diagram.
		\end{definition}
		
		When the variable order is clear from the context, we use the notation \(\mathit{OBDD}(\phi)\).
		
		\begin{definition}[exponential program and exponential flips]
			Let \(\beta \in \mathbb{R}, b \in \mathbb{Z}^+\) and \(\pdf{0}{\beta} \comp_b\prog{}\), then \(\prog{}\) is an exponential program with \(b\) bits. 
			
			The coin flips that occur in an exponential program are referred to as exponential flips.
		\end{definition}
		
		\begin{definition}[mixture of exponentials]\label{def:mix-expo}
			Let \(\prog{e}\) refer to an exponential program function  with \(b\) bits then \(\prog{m}\) --- a mixture of exponential with \(b\) bits is defined as follows:
			\[\prog{m} ::= \prog{e} \mid \Llet{\yvar{1} = \prog{m}}{\Llet{\yvar{2} = \prog{m}}{\Llet{\yvar{3} = \Lflip{\theta}}{\Lite{\yvar{3}}{\yvar{1}}{\yvar{2}}}}}\]
			
			In the latter rule for \(p_m\), we refer to the \(\Lflip{\theta}\) as an if-guard flip.
		\end{definition}
		
		\begin{definition}[size of a mixture of exponentials]
			The size of a mixture of exponentials (denoted by \(\prog{m}\)) is defined as the number of if-guards and is defined as follows:
			\begin{itemize}
				\item If \(\prog{m}\) is an exponential program, then \(|\prog{m}| = 0\) 
				\item If \(\prog{m} = \Llet{\yvar{1} = \prog{m1}}{\Llet{\yvar{2} = \prog{m2}}{\Llet{\yvar{3} = \Lflip{\theta}}{\Lite{\yvar{3}}{\yvar{1}}{\yvar{2}}}}}\), then \(|\prog{m}| = |\prog{m1}| + |\prog{m2}| + 1\)
			\end{itemize}
		\end{definition}
		\begin{definition}[deconditioned program]\label{def:deconditioning}
			Given a Dice program, \(\prog{}\), replacing the construct \(\Lobs{(.)}\) with \(\true\) and removing dead variables gives a deconditioned program \(\prog{d}\). We use the following notation for deconditioning
			\(\decondition{\prog{}} = \prog{d}\)
		\end{definition}
		
		Note that replacing \(\Lobs{(.)}\) constructs with \(\true\) makes the Dice program free of side effects and unused variables can be removed without changing semantics of the program.
		
		Now, we show that for all programs \(\prog{}\) emitted by \(\comp_b\), \(\decondition{\prog{}}\) is a mixture of exponentials with \(b\) bits.
		
		\begin{lemma}\label{decondition-pointgamma}	
			\(\forall \alpha, \beta, \exists s, \forall \prog{}, \epsilon, b\), if \(\pointgamma{\alpha}{\beta}{\epsilon}{b} = \prog{}\), then \(\mathit{DC}(\prog{})\) is a mixture of exponentials with \(b\) bits of size \(s\).
		\end{lemma}
		
		\begin{proof}
			The proof is by induction on \(\alpha\).
			
			\underline{Base Case:} \(\alpha = 0\)
			
			Then by~\ref{spg0}, \(\prog{} = \prog{1}\) where \(\pdf{0}{\beta} \comp_b \prog{1}\).
			
			Then by Definition~\ref{def:mix-expo}, \(\prog{}\) is a mixture of exponentials with \(b\) bits of size 0.
			
			\underline{Induction Hypothesis:} \(\forall \alpha' < \alpha, \beta, \exists s', \forall \prog{I}, \epsilon, b\), if \(\pointgamma{\alpha'}{\beta}{\epsilon}{b} = \prog{I}\), then \(\mathit{DC}(\prog{I})\) is a mixture of exponentials with \(b\) bits of size \(s'\).
			
			\underline{Inductive Step:} \(\forall \alpha > 0\)
			
			Then by~\ref{spggt0}, \(\prog{} = \begin{array}{l}
				\texttt{let \(\yvar{1}\) = } \prog{1} \texttt{ in } 
				\\ \texttt{let \(\yvar{2}\) = } \Lflip{\frac{1}{1 + \epsilon}} \texttt{ in } 
				\\ \texttt{let } \_ = \Lite{\yvar{2}}{\prog{2}}{\true} \texttt{ in}
				\\  \yvar{1} \end{array}\)
			
			where \(\pointgamma{\alpha - 1}{\beta}{\epsilon}{b} = \prog{1}\) and \(\unifobs{\yvar{1}}{b} = \prog{2}\)
			
			Then by Definition~\ref{def:deconditioning}, 
			
			\begin{flalign*}
				\decondition{\prog{}} = \begin{array}{l}
					\texttt{let \(\yvar{1}\) = } \prog{1} \texttt{ in } 
					\\  \yvar{1} \end{array} = \prog{1}
			\end{flalign*}
			
			Since \(\yvar{1}\) occurs only once in the body of \texttt{let}, \(\Llet{\yvar{1} = \prog{1}}{\prog{1}}\) can be replaced with \(\prog{1}\).
			
			Now, \(\prog{1}\) is a mixture of exponentials with \(b\) bits of size \(s'\) by Induction Hypothesis and thus \(\decondition{\prog{}}\) is a mixture of exponentials with \(b\) bits of size \(s'\) by Definition~\ref{def:mix-expo}.
		\end{proof}
		
		\begin{lemma}\label{lemma:mix-general-gamma}
			\(\forall \alpha, \beta, \exists s, \forall \prog, b\), if \(\pdf{\alpha}{\beta} \comp_b \prog{}\) then \(\mathit{DC}(\prog{})\) is a mixture of exponentials with \(b\) bits of size \(s\) upto variable rearrangements and let in-lining.
		\end{lemma}
		
		\begin{proof}
			The proof is by induction on \(\alpha\).
			
			\underline{Base Case 1:} \(\alpha = 0\)
			
			We have \(\pdf{0}{\beta} \comp_b \prog{}\). By Definition~\ref{def:mix-expo}, \(\prog{}\) is a mixture of exponentials with \(b\) bits of size 0.
			
			\underline{Base Case 2:} \(\alpha = 1\)
			
			By ~\ref{Trans-expo1}, if \(\pdf{1}{\beta} \comp_b \prog{}\), then
			
			$\texttt{p} = \begin{array}{l}
				\texttt{let \(\yvar{1}\) = } \prog{1} \texttt{ in }
				\\ \texttt{let \_ = \(\prog{3}\)}
				\\ \texttt{let \(\yvar{2}\) = } \prog{1} \texttt{ in}
				\\ \texttt{let \(\yvar{3}\) = flip\((\theta)\) in}
				\\ \texttt{if \(\yvar{3}\) then \(\yvar{2}\) else \(\yvar{1}\)}
			\end{array}$ 
			
			where  \(\pdf{0}{\beta} \comp_b \prog{1}\) and
			\(\texttt{unifObs(\(\yvar{1}\), b)} = \prog{3}\)
			
			Then by Definition~\ref{def:deconditioning}, 
			\(\mathit{DC}(\prog{}) = \begin{array}{l}
				\texttt{let \(\yvar{1}\) = } \prog{1} \texttt{ in }
				\\ \texttt{let \(\yvar{2}\) = } \prog{1} \texttt{ in}
				\\ \texttt{let \(\yvar{3}\) = flip\((\theta)\) in}
				\\ \texttt{if \(\yvar{3}\) then \(\yvar{2}\) else \(\yvar{1}\)}
			\end{array} \)
			
			Thus by Definition~\ref{def:mix-expo}, \(\mathit{DC}(\prog{})\) is a mixture of exponentials with \(b\) bits of size 1.
			
			\underline{Induction Hypothesis:} \(\forall \alpha' < \alpha, \beta, \exists s, \forall \prog{I}, b\), if \(\pdf{\alpha'}{\beta} \comp_b \prog{I}\), then \(\mathit{DC}(\prog{I})\) is a mixture of exponentials with \(b\) bits of size \(s\).
			
			\underline{Inductive Step:} \(\forall \alpha > 1\)
			
			By ~\ref{Trans-expo>1}, if \(\pdf{\alpha}{\beta} \comp_b \prog{}\), then
			
			\(\texttt{p} = \begin{array}{l}
				\texttt{let \(\yvar{1}\) = \(\prog{1}\) in}
				\\ \texttt{let \_ = \(\prog{obs1}\) in}
				\\ \texttt{let \(\yvar{2}\) = \(\prog{1}\) in}
				\\ \texttt{let \(\yvar{3}\) = \(\prog{3}\) in}
				\\ \texttt{let \(\yvar{4}\) = \(\prog{4}\) in}
				\\ \texttt{let \_ = \(\prog{obs2}\) in}
				\\ \texttt{let \(\yvar{f1}\) = flip(\(\theta_1\)) in}
				\\ \texttt{let \(\yvar{f2}\) = flip(\(\theta_2\)) in}
				\\ \texttt{let \(\yvar{f3}\) = flip(\(\theta_3\)) in }
				\\ \texttt{if \(\yvar{f1}\) then \(\yvar{1}\) else} 
				\\ \texttt{if \(\yvar{f2}\) then \(\yvar{2}\) else if \(\yvar{f3}\) then \(\yvar{3}\) else \(\yvar{4}\)}
			\end{array}\)
			
			where 
			
			\[\pdf{\alpha-1}{\beta} \comp_b \prog{1}
			\;\;\;\;\;\;\;\; \pdf{\alpha-2}{\beta} \comp_b \prog{3} \]
			
			\[\pointgamma{\alpha-1}{\beta}{\frac{1}{2^b}}{b} = \prog{4}
			\;\;\;\;\;\;\;\;
			\unifobs{\yvar{1}}{b} = \prog{obs1}
			\;\;\;\;\;\;\;\;
			\unifobs{\yvar{3}}{b} = \prog{obs2}\]
			
			\begin{flalign*}
				\mathit{DC}(\prog{}) =& \begin{array}{l}
					\texttt{let \(\yvar{1}\) = \(\mathit{DC}(\prog{1})\) in}
					\\ \texttt{let \(\yvar{2}\) = \(\mathit{DC}(\prog{1})\) in}
					\\ \texttt{let \(\yvar{3}\) = \(\mathit{DC}(\prog{3})\) in}
					\\ \texttt{let \(\yvar{4}\) = \(\mathit{DC}(\prog{4})\) in}
					\\ \texttt{let \(\yvar{f1}\) = flip(\(\theta_1\)) in}
					\\ \texttt{let \(\yvar{f2}\) = flip(\(\theta_2\)) in}
					\\ \texttt{let \(\yvar{f3}\) = flip(\(\theta_3\)) in }
					\\ \texttt{if \(\yvar{f1}\) then \(\yvar{1}\) else} 
					\\ \texttt{if \(\yvar{f2}\) then \(\yvar{2}\) else if \(\yvar{f3}\) then \(\yvar{3}\) else \(\yvar{4}\)}
				\end{array}
			\end{flalign*}
			
			Also by Lemma~\ref{decondition-pointgamma}, \(\mathit{DC}(\prog{4})\) is a mixture of exponentials with \(b\) bits of size \(s_p\) upto variable rearrangement. 
			
			By Induction Hypothesis, \(\exists s_1, s_3\) such that \(\mathit{DC}(\prog{1})\) and \(\mathit{DC}(\prog{3})\) are mixture of exponentials with \(b\) bits of size \(s_1\) and \(s_3\) respectively. Thus, \(\decondition{\prog{}}\) is a mixture of exponentials with \(b\) bits of size \(3 + s_1 + s_3 + s_p\).
		\end{proof}
		
		\begin{lemma}\label{lemma:mix-upsilon}
			\(\forall \Upsilon, \exists s, \forall \prog{}, b\), if \(\Upsilon \comp_b \prog{}\) then \(\mathit{DC}(\prog{})\) is a mixture of exponentials with \(b\) bits of size \(s\).
		\end{lemma}
		
		\begin{proof}
			By Definition~\ref{def:mixed-gamma}, let \(\Upsilon(x) = \sum_{i = 1}^{N} a_i \pi_{\alpha_i, \beta_i}(x)\) where \(N \in \mathbb{Z}^+; \forall i, \alpha_i \in \mathbb{Z}^+, \beta_i \in \mathbb{R}, a_i \in [0, 1]\) and \(\sum_{i=1}^{N} a_i = 1\). 
			
			The further proof is by induction on N.
			
			\underline{Base Case:} N = 1 
			
			For N = 1, we have \(a_1 = 1\) and \(\Upsilon = \pdf{\alpha}{\beta}\). If \(\Upsilon \comp_b \prog{}\), then \(\exists s\) such that \(\mathit{DC}(\prog{})\) is a mixture of exponentials with \(b\) bits of size \(s\) by Lemma~\ref{lemma:mix-general-gamma}.
			
			\underline{Induction Hypothesis:} \(\forall N' < N\), if \(\sum_{i=1}^{N'} a_i \pdf{\alpha_i}{\beta_i} \comp_b \prog{}\), then \(\exists s, \mathit{DC}(\prog{})\) is a mixture of exponentials with \(b\) bits of size \(s\).
			
			\underline{Inductive Step:} \(\forall N > 1\)
			
			By ~\ref{Trans-mix}, \(\prog{} = \begin{array}{l}
				\texttt{let \(\yvar{1}\) = flip(\(a_N\)) in} \\
				\texttt{let \(\yvar{2}\) = \(\prog{1}\) in} \\
				\texttt{let \(\yvar{3}\) = \(\prog{2}\) in} \\
				\texttt{if \(\yvar{1}\) then \(\yvar{2}\) else \(\yvar{3}\)}
			\end{array}\)
			
			where \(\pdf{\alpha_N}{\beta_N} \comp_b \prog{1}\) and
			\(\sum_{i=1}^{N-1} \frac{a_i}{1 - a_N} \pdf{\alpha_i}{\beta_i} \comp_b \prog{2}\).
			
			By Induction Hypothesis, \(\exists s_2\) such that \(\prog{2}\) is a mixture of exponentials with \(b\) bits.
			
			By Lemma~\ref{lemma:mix-general-gamma}, \(\exists s_1\) such that \(\prog{1}\) is a mixture of exponentials with \(b\) bits.
			
			Then by Definition~\ref{def:mix-expo}, \(\decondition{\prog{}}\) is a mixture of exponentials with \(b\) bits of size \(1 + s_1 + s2\).	
		\end{proof}
		
		Now that we have shown that all programs \(\prog{}\) emitted by \(\comp_b\) follow a certain structure, we show that this structure induces a small OBDD size for the returning formula.
		
		\begin{definition}[interleaving variable order]
			Let \(\prog{}\) be a Dice program where coin flips occur either as if-guard flip (denoted by \(\var{g}\)) or exponential flips (denoted by \(\var{f}\)). Let \(\prog{}\) have \(n\) exponential (sub)-programs with \(b\) bits, then \(\var{f}\) has \(nb\) coin flips. In the interleaving variable order, the coin flips occur in the following order:
			\begin{itemize}
				\item \(\var{g}\) in the order they occur in the program (in program order).
				\item Coin flips in \(\var{f}\) used as first element of the \(b\)-tuple in program order.
				\item \(\ldots\)
				\item Coin flips in \(\var{f}\) used as \(j\)-th element of the \(b\)-tuple in program order.
				\item \(\ldots\)
				\item Coin flips in \(\var{f}\) used as \(b\)-th (last) element of the \(b\)-tuple in program order.
			\end{itemize}
		\end{definition}
		
		\begin{example}
			Consider the following program:
			\\
			\texttt{let \(\yvar{11}\) = flip($\theta_1$) in let \(\yvar{21}\) = flip($\theta_2$) in \(\ldots\) let \(\yvar{b1}\) = flip($\theta_b$) in}
			\\ \texttt{let \(\yvar{e1}\) = (\(\yvar{11}\), \(\yvar{21}\), \ldots, \(\yvar{b1}\)) in}
			\\  \texttt{let \(\yvar{12}\) = flip($\theta_1$) in let \(\yvar{22}\) = flip($\theta_2$) in \(\ldots\) let \(\yvar{b2}\) = flip($\theta_b$) in}
			\\ \texttt{let \(\yvar{e2}\) = (\(\yvar{12}\), \(\yvar{22}\), \ldots, \(\yvar{b2}\)) in}
			\\ \texttt{let \(\yvar{f}\) = flip(\(\theta_1\)) in if \(\yvar{1}\) then \(\yvar{e1}\) else \(\yvar{e2}\)}
			
			The flip assigned to \(\yvar{f}\) is an if-guard flip and all other coin flips are exponential flips. Then the interleaving variable order for the coin flips in the above program would be as follows:
			\[[\yvar{f}, \yvar{11}, \yvar{12}, \yvar{21}, \yvar{22}, \ldots \yvar{b1}, \yvar{b2}]\]
		\end{example}
		
		\begin{lemma}~\label{lemma-dcreturn-bdd-soze}
			\(\forall s \in \mathbb{Z}^+, \exists k, \forall \prog{}, b, \Gamma, \varphi, \gamma, w\), if \(\prog{}\) is a mixture of exponentials with \(b\) bits of size \(s\) and \(\Gamma \vdash \prog{} : \bool^b \comp (\varphi, \gamma, w)\), then for the interleaving variable order of \(\prog{}\), \(\obdd{\varphi} \leq kb\)
		\end{lemma}
		
		\begin{proof}
			To prove the above lemma, we first prove the following where \(\varphi_i\) refers to the \(i\)-th element of \(\varphi\):
			
			\(\forall s \in \mathbb{Z}^+, \exists k, \forall \prog{}, b, \Gamma, \varphi, \gamma, w\), if \(\prog{}\) is a mixture of exponentials with \(b\) bits of size \(s\) and \(\Gamma \vdash \prog{} : \bool^b \comp (\varphi, \gamma, w)\), then for the interleaving variable order of \(\prog{}\), \(\forall i, \obdd{\varphi_i} \leq k\)
			
			The proof is by induction on the size of mixture of exponentials, i.e. \(s\).
			
			\underline{Base Case:} \(s = 0\)
			
			If \(s = 0\), then \(\prog{}\) is an exponential program with \(b\) bits, that is \(\exists \beta \in \mathbb{R}, \forall b \in \mathbb{Z}^+\) such that \(\pdf{0}{\beta} \comp_b \prog{}\)
			
			Then by ~\ref{Trans-expo0}, \(\prog{} = \begin{array}{l}
				\texttt{let \(\yvar{1}\) = flip($\theta_1$) in}
				\\ \texttt{let \(\yvar{2}\) = flip($\theta_2$) in}
				\\ \ldots
				\\ \texttt{let \(\yvar{b}\) = flip($\theta_b$) in}
				\\ \texttt{(\(\yvar{1}\), \(\yvar{2}\), \ldots, \(\yvar{b}\))}
			\end{array}\)
			
			Let \(w_i = (\flip{i} \mapsto \theta_i, \overline{\flip{i}} \mapsto 1 - \theta_i)\) and \(D_i = \inference{\texttt{fresh }\flip{i}}{\Gamma \cup \bigcup_{j = 1}^{i-1} \{y_j \mapsto \bool\} \vdash \Lflip{\theta_i}: \bool \comp (\flip{i}, \true, w_i)}[C-Flip]\)
			
			Now consider the following derivation tree.
			
			\begin{flalign*}
				\inference{
					D_1
					&
					\inference{
						\inference{D_b 
							&
							\Gamma \cup \{\yvar{1} \mapsto \bool, ... \yvar{b} \mapsto \bool\} \vdash
							(\yvar{1}, \yvar{2}, \ldots, \yvar{b}) : \bool^b \comp (\yvar{1}, \yvar{2}, \ldots, \yvar{b}) 
						}{\vdots}
					}{\Gamma \cup \{\yvar{1}:\bool\} \vdash \begin{array}{l}
							\texttt{let \(\yvar{2}\) = flip($\theta_2$) in}
							\\ \ldots
							\\ \texttt{let \(\yvar{b}\) = flip($\theta_b$) in}
							\\ \texttt{(\(\yvar{1}\), \(\yvar{2}\), \ldots, \(\yvar{b}\))}
						\end{array}:\bool^b \comp ((\yvar{1}, \flip{2}, \ldots \flip{b}), \true, \bigcup_{i=1}^{b-1} w_i) }[C-let]
				}{\Gamma \vdash \begin{array}{l}
						\texttt{let \(\yvar{1}\) = flip($\theta_1$) in}
						\\ \texttt{let \(\yvar{2}\) = flip($\theta_2$) in}
						\\ \ldots
						\\ \texttt{let \(\yvar{b}\) = flip($\theta_b$) in}
						\\ \texttt{(\(\yvar{1}\), \(\yvar{2}\), \ldots, \(\yvar{b}\))}
					\end{array} : \bool^{b} \comp ((\flip{1}, \flip{2}, \ldots \flip{b}), \true, \bigcup_{i=1}^{b} w_i)}[C-let]
			\end{flalign*}
			
			From the above derivation tree, we have
			\[(\varphi, \gamma, w) = ((\flip{1}, \flip{2}, \ldots \flip{b}), \true, \bigcup_{i=1}^{b} w_i)\]
			We further argue about the size of the OBDD corresponding to the Boolean formula represented by \((\flip{1}, \flip{2}, \ldots \flip{b})\).
			
			The variable order \(\ord{\prog{}}\) is \(\flip{1}, \flip{2}, \ldots \flip{b}\). 
			
			Now, for any \(i \in \{1, 2, \ldots b\}\), \(\varphi_i = \flip{i}\). Since \(\varphi_i\) is independent of the first \(i-1\) variables, there are no nodes corresponding to \(\flip{1}, \flip{2}, \ldots \flip{i-1}\). Now if you fix the value of \(\flip{i}\), \(\varphi_i\) evaluates to \(\true\) or \(\false\) and does not require any more nodes in its BDD. 
			Thus \(\mathit{OBDD}(\varphi_i) = 1 \implies k = 1\)
			
			\underline{Induction Hypothesis:} \(\forall s' < s, \exists k' \forall \prog{}, b, \Gamma, \varphi, \gamma, w\), if \(\prog{}\) is a mixture of exponentials with \(b\) bits of size \(s'\) and \(\Gamma \vdash \prog{'}: \bool^b \comp (\varphi, \gamma, w)\), then for an interleaving variable order of \(\prog{}\), \(\forall i, \mathit{OBDD}(\varphi_i) \leq k'\)
			
			\underline{Inductive Step:} \(\forall s > 0\)
			
			If \(s > 0\), then \(\prog{}\) is of the form 
			
			\(\Llet{\yvar{1} = \prog{1}}{\Llet{\yvar{2} = \prog{2}}{\Llet{\yvar{3} = \Lflip{\theta}}{\Lite{\yvar{3}}{\yvar{1}}{\yvar{2}}}}}\) where \(\prog{1}\) and \(\prog{2}\) are mixture of exponentials with \(b\) bits of size \(s_1 < s\) and \(s_2 < s\) respectively.
			
			Let S1 be \(\Gamma \vdash \prog{1}: \bool^b \comp (\varphi_1, \gamma_1, w_1)\) and S2 be \(\Gamma \vdash \prog{2}: \bool^b \comp (\varphi_2, \gamma_2, w_2)\) where \(w_1\) and \(w_2\) are disjoint, then we have the following derivation tree:
			
			\begin{flalign*}
				\inference{S1 & S2
					&
					\inference{\texttt{fresh }\flip{}}{\Gamma \vdash \Lflip{\theta}: \bool \comp (\flip{}, \true, (\flip{} \mapsto \theta, \overline{\flip{}} \mapsto 1 - \theta))}[C-Flip]
				}{\Gamma \vdash \Lite{\Lflip{\theta}}{\prog{1}}{\prog{2}}
					\comp ((\flip{} \xbroadand{\bool^b} \varphi_1) \xpointor{\bool^b}
					(\overline{\flip{}} \xbroadand{\bool^b} \varphi{2}), \gamma, w_1 \cup w_2 \cup \{\flip{} \mapsto \theta, \overline{\flip{}} \mapsto 1 - \theta \})}[C-Ite]	
			\end{flalign*}
			
			We further argue about the size of the OBDD corresponding to the Boolean formula represented by \(\varphi_i = (\flip{} \xbroadand{\bool^b} \varphi_1) \xpointor{\bool^b}
			(\overline{\flip{}} \xbroadand{\bool^b} \varphi_{2})\).
			
			By the induction hypothesis, we know that for the interleaving variable order of \(\prog{1}\), \(\forall i, \mathit{OBDD}(\varphi_{1i}) \leq k_1\) and for the interleaving variable order of \(\prog{2}\), for which \(\forall i, \mathit{OBDD}(\varphi_{2i}) \leq k_2\).
			
			Now, consider the interleaving variable order for \(\prog{}\) where we interleave the coin flips for the bits of exponentials coming from \(\prog{1}\) and \(\prog{2}\) after all the if-guard coin flips in program order.
			\[\flip{}, \var{g}(\prog{1}), \var{g}(\prog{2}), \ldots\] 
			Since the above variable order maintains the variable order for the subprograms (\(\prog{1}, \prog{2}\)), the induction hypothesis still holds.
			
			For any \(i \in \{1, 2, \ldots b\}\), \(\varphi_i = (\flip{} \wedge \varphi_{1i}) \vee (\overline{\flip{}} \wedge \varphi_{2i})\). 
			
			Conditioning \(\varphi_i\) on \(\flip{}\) gives us two subfunctions as follows:
			\[\varphi_{i} | \flip{} = \varphi_{1i} \;\;\;\; \varphi_{i} | \overline{\flip{}} \ = \varphi_{2i}\]
			\begin{flalign*}
				\mathit{OBDD}(\varphi_i) \leq& 1 + \mathit{OBDD}(\varphi_{1i}) + \mathit{OBDD}(\varphi_{2i}) = 1 + k_1 + k_2~\tag{Induction Hypothesis}
			\end{flalign*}
			
			If for any index \(i\), \(\obdd{\varphi_i} \leq k\), then \(\obdd{\varphi} \leq \sum_{i = 1}^{b} \obdd{\varphi_i} = kb\)
		\end{proof}
		
		\begin{lemma}~\label{lemma:return-bdd-size}
			\(\forall \Upsilon, \prog{}, \varphi, \gamma, w, \exists k, \forall b\), if \(\Upsilon \comp_b \prog{}\) and \(\prog{} \comp (\varphi, \gamma, w)\), then for the interleaving variable order of \(\prog{}\), \(\obdd{\varphi} \leq kb\)
		\end{lemma}
		
		\begin{proof}
			By Lemma~\ref{lemma:mix-upsilon}, \(\forall \Upsilon, \exists s, \forall \prog{}, b\), if \(\Upsilon \comp_b \prog{}\) then \(\mathit{DC}(\prog{})\) is a mixture of exponentials with \(b\) bits of size \(s\). 
			
			By Lemma~\ref{lemma-dcreturn-bdd-soze}, \(\forall s, \exists k, \forall \prog{}, \forall b\), if \(\prog{}\) is a mixture of exponentials with \(b\) bits of size \(s\) and \(\prog{} \comp (\varphi, \gamma, w)\), then for the interleaving variable order of \(\prog{}\), \(\mathit{OBDD}(\varphi) \leq kb\).
			
			Observe that for any arbitrary program \(\prog{}\), if \(\prog{} \comp (\varphi, \gamma, w)\) and \(\decondition{\prog{}} \comp (\varphi', \gamma', w')\), then \(\varphi = \varphi'\) since \(\prog{}\) and \(\mathit{DC}(\prog{})\) only differ in \(\Lobs{}\) construct and dead program variables. Thus, if \(\Upsilon \comp_b \prog{}\) and \(\prog{} \comp (\varphi, \gamma, w)\), then for the interleaving variable order of \(\prog{}\), \(\obdd{\varphi} \leq kb\).
		\end{proof}
		
		\subsection{OBDD size of accepting Boolean formula}
		
		In the following definitions, we use \(\unifobs{\yvar{}}{b}\) to refer to the program \(\prog{obs}\) such that the following judgment holds: \(\unifobs{\yvar{}}{b} = \prog{obs}\)
		
		\begin{definition}[unifobs program]\label{def:unifobs-program}
			A unifobs program \(\prog{}\) with \(b\) bits is defined by the following grammar where \(\prog{e}\) refers to an exponential program with \(b\) bits and \(\prog{obs}\) refers to \(\unifobs{\yvar{}}{b}\).
			\begin{flalign*}
				\texttt{aexp} &:= \yvar{} \mid v
				\\
				\prog{u} &:=  \texttt{aexp} \mid \Lflip{\theta} \mid \prog{e} \mid \Llet{\yvar{} = \prog{u}}{\prog{u}} \mid \prog{obs} \mid \Lite{\texttt{aexp}}{\prog{u}}{\prog{u}}
			\end{flalign*}
		\end{definition}
		
		\begin{definition}[size of a unifobs program]
			The size of a unifobs program \((f, o)\) is a tuple of the number of if-guard flips and the number of \(\unifobs{\yvar{}}{b}\) subprograms.
		\end{definition}
		
		\begin{lemma}\label{lemma:unifobs-pointgamma}
			\(\forall \alpha, \beta, \epsilon, \exists f, o, \forall \prog{}, b\), if \(\pointgamma{\alpha}{\beta}{\epsilon}{b} = \prog{}\) then \(\prog{}\) is a unifobs program with \(b\) bits of size \((f, o)\).
		\end{lemma}
		
		\begin{proof}
			The proof is by induction on \(\alpha\).
			
			\underline{Base Case 1:} \(\alpha = 0\)
			
			Then by~\ref{spg0}, \(\prog{} = \prog{1}\) where \(\pdf{0}{\beta} \comp_b \prog{1}\). \(\prog{}\) is a unifobs program with \(b\) bits of size \((0, 0)\).
			
			\underline{Induction Hypothesis:} \(\forall \alpha' < \alpha, \beta, \epsilon, \exists f', o' \forall \prog{I}, b\), if \(\pointgamma{\alpha'}{\beta}{\epsilon}{b} = \prog{I}\), then \(\prog{I}\) is a unifobs program with \(b\) bits of size \((f', o')\).
			
			\underline{Inductive Step:} \(\forall \alpha > 0\)
			
			Then by~\ref{spggt0}, \(\prog{} = \begin{array}{l}
				\texttt{let \(\yvar{1}\) = } \prog{1} \texttt{ in } 
				\\ \texttt{let \(\yvar{2}\) = } \Lflip{\frac{1}{1 + \epsilon}} \texttt{ in } 
				\\ \texttt{let } \_ = \Lite{\yvar{2}}{\prog{2}}{\true} \texttt{ in}
				\\  \yvar{1} \end{array}\)
			
			where \(\pointgamma{\alpha - 1}{\beta}{\epsilon}{b} = \prog{1}\) and \(\unifobs{\yvar{1}}{b} = \prog{2}\)
			
			By Induction Hypothesis, \(\prog{1}\) is a unifobs program with \(b\) bits of size \((f', o')\). Thus, by Definition~\ref{def:unifobs-program}, \(\prog{}\) is a unifobs program with \(b\) bits of size \((f'+1, o'+1)\).
			
		\end{proof}
		
		\begin{lemma}\label{lemma:unifobs-general-gamma}
			\(\forall \alpha, \beta, \exists f, o, \forall \prog, b\), if \(\pdf{\alpha}{\beta} \comp_b \prog{}\) then \(\prog{}\) is a unifobs program with \(b\) bits of size \((f, o)\).
		\end{lemma}
		
		\begin{proof}
			The proof is by induction on \(\alpha\).
			
			\underline{Base Case 1:} \(\alpha = 0\)
			
			We have \(\pdf{0}{\beta} \comp_b \prog{}\). By Definition~\ref{def:unifobs-program}, \(\prog{}\) is a unifobs program with \(b\) bits of size \((0, 0)\).
			
			\underline{Base Case 2:} \(\alpha = 1\)
			
			By ~\ref{Trans-expo1}, if \(\pdf{1}{\beta} \comp_b \prog{}\), then
			
			$\texttt{p} = \begin{array}{l}
				\texttt{let \(\yvar{1}\) = } \prog{1} \texttt{ in }
				\\ \texttt{let \_ = \(\prog{3}\)}
				\\ \texttt{let \(\yvar{2}\) = } \prog{1} \texttt{ in}
				\\ \texttt{let \(\yvar{3}\) = flip\((\theta)\) in}
				\\ \texttt{if \(\yvar{3}\) then \(\yvar{2}\) else \(\yvar{1}\)}
			\end{array}$ 
			
			where  \(\pdf{0}{\beta} \comp_b \prog{1}\) and
			\(\texttt{unifObs(\(\yvar{1}\), b)} = \prog{3}\)
			
			By Base Case 1, \(\prog{1}\) is a unifobs program with \(b\) bits of size \((0, 0)\). Thus, by Definition~\ref{def:unifobs-program}, \(\prog{}\) is a unifobs program with \(b\) bits of size \((1, 1)\).
			
			\underline{Induction Hypothesis:} \(\forall \alpha' < \alpha, \beta, \exists f', o', \forall \prog{}, b\), if \(\pdf{\alpha'}{\beta} \comp_b \prog{}\), then \(\prog{}\) is a unifobs program with \(b\) bits of size \((f', o')\).
			
			\underline{Inductive Step:} \(\forall \alpha > 1\)
			
			By ~\ref{Trans-expo>1}, if \(\pdf{\alpha}{\beta} \comp_b \prog{}\), then
			
			\(\texttt{p} = \begin{array}{l}
				\texttt{let \(\yvar{1}\) = \(\prog{1}\) in}
				\\ \texttt{let \_ = \(\prog{obs1}\) in}
				\\ \texttt{let \(\yvar{2}\) = \(\prog{1}\) in}
				\\ \texttt{let \(\yvar{3}\) = \(\prog{3}\) in}
				\\ \texttt{let \(\yvar{4}\) = \(\prog{4}\) in}
				\\ \texttt{let \_ = \(\prog{obs2}\) in}
				\\ \texttt{let \(\yvar{f1}\) = flip(\(\theta_1\)) in}
				\\ \texttt{let \(\yvar{f2}\) = flip(\(\theta_2\)) in}
				\\ \texttt{let \(\yvar{f3}\) = flip(\(\theta_3\)) in }
				\\ \texttt{if \(\yvar{f1}\) then \(\yvar{1}\) else} 
				\\ \texttt{if \(\yvar{f2}\) then \(\yvar{2}\) else if \(\yvar{f3}\) then \(\yvar{3}\) else \(\yvar{4}\)}
			\end{array}\)
			
			where 
			
			\[\pdf{\alpha-1}{\beta} \comp_b \prog{1}
			\;\;\;\;\;\;\;\; \pdf{\alpha-2}{\beta} \comp_b \prog{3} \]
			
			\[\pointgamma{\alpha-1}{\beta}{\frac{1}{2^b}}{b} = \prog{4}
			\;\;\;\;\;\;\;\;
			\unifobs{\yvar{1}}{b} = \prog{obs1}
			\;\;\;\;\;\;\;\;
			\unifobs{\yvar{3}}{b} = \prog{obs2}\] 
			
			By Induction Hypothesis, \(\prog{1}\) and \(\prog{3}\) are a unifobs program with \(b\) bits of size \((f_1, o_1)\) and \((f_3, o_3)\) respectively. By Lemma~\ref{lemma:unifobs-pointgamma}, \(\prog{4}\) is a unifobs program wth \(b\) bits of size \((f_4, o_4)\).
			
			Thus, by Definition~\ref{def:unifobs-program}, \(\prog{}\) is a unifobs program with \(b\) bits of size \((2f_1 + f_3 + f_4 + 3, 2o_1 + o_3 + o_4 + 2)\).
		\end{proof}

		\begin{lemma}\label{lemma:all-unifobs-program}
			\(\forall \Upsilon, \exists f, o, \forall \prog{}, b\), if \(\Upsilon \comp_b \prog{}\) then \(\prog{}\) is a unifobs program with \(b\) bits of size \((f, o)\).
		\end{lemma}
		
		\begin{proof}
			By Definition~\ref{def:mixed-gamma}, let \(\Upsilon(x) = \sum_{i = 1}^{N} a_i \pi_{\alpha_i, \beta_i}(x)\) where \(N \in \mathbb{Z}^+; \forall i, \alpha_i \in \mathbb{Z}^+, \beta_i \in \mathbb{R}, a_i \in [0, 1]\) and \(\sum_{i=1}^{N} a_i = 1\). 
			
			The further proof is by induction on N.
			
			\underline{Base Case:} N = 1 
			
			For N = 1, we have \(a_1 = 1\) and \(\Upsilon = \pdf{\alpha}{\beta}\). If \(\Upsilon \comp_b \prog{}\), then \(\exists f, o\) such that \(\prog{}\) is a unifobs program with \(b\) bits of size \((f, o)\) by Lemma~\ref{lemma:unifobs-general-gamma}.
			
			\underline{Induction Hypothesis:} \(\forall N' < N, \exists f', o'\), if \(\sum_{i=1}^{N'} a_i \pdf{\alpha_i}{\beta_i} \comp_b \prog{}\), then \(\prog{}\) is a unifobs program with \(b\) bits of size \((f', o')\).
			
			\underline{Inductive Step:} \(\forall N > 1\)
			
			By ~\ref{Trans-mix}, \(\prog{} = \begin{array}{l}
				\texttt{let \(\yvar{1}\) = flip(\(a_N\)) in} \\
				\texttt{let \(\yvar{2}\) = \(\prog{1}\) in} \\
				\texttt{let \(\yvar{3}\) = \(\prog{2}\) in} \\
				\texttt{if \(\yvar{}\) then \(\yvar{2}\) else \(\yvar{3}\)}
			\end{array}\)
			
			where \(\pdf{\alpha_N}{\beta_N} \comp_b \prog{1}\) and
			\(\sum_{i=1}^{N-1} \frac{a_i}{1 - a_N} \pdf{\alpha_i}{\beta_i} \comp_b \prog{2}\).
			
			By Induction Hypothesis, \(\prog{2}\) is a unifobs program with \(b\) bits of size \((f_2, o_2)\). By Lemma~\ref{lemma:unifobs-general-gamma}, \(\prog{1}\) is a unifobs program with \(b\) bits of size \((f_1, o_1)\).
			
			Thus, \(\prog{}\) is a unifobs program with \(b\) bits of size \((f_1 + f_2 + 1, o_1 + o_2)\).	
		\end{proof}
		
		Recall the definition of the inequality function.
		
		\begin{definition}[inequality function]\label{def:inequality_wbf_restate}
			A \(b\)-bit inequality function, \(\mathit{LT}_b:\{0,1\}^b \times \{0, 1\}^b \rightarrow \{0, 1\}\) takes as input two \(b\)-bit numbers \(x, y\) and outputs $1$ if $ x < y$ and $0$ otherwise. Thus,
			\[\mathit{LT}_b((x_1, \ldots x_b), (y_1, \ldots y_b)) = 
			\begin{cases}
				\neg x_1 y_1 & b = 1
				\\	\neg x_1 y_1 + (\neg x_1 \neg y_1 + x_1y_1)\mathit{LT}_{b-1}((x_2, \ldots, x_b), (y_2, \ldots, y_b)) & b > 1
			\end{cases}\]
		\end{definition}
		
		\begin{lemma}\label{lemma:less_than_wbf}
			\(\forall \prog{}, b, \Gamma\), if \(\lessthan{\(\yvar{1}\)}{\(\yvar{2}\)}{b} = \prog{}\), then
			\(\Gamma \bigcup \{\yvar{1} \mapsto \bool^b, \yvar{2} \mapsto \bool^b\} \vdash \prog{} : \bool \comp (\mathit{LT}_b(\yvar{1}, \yvar{2}), \true, \emptyset)\)
		\end{lemma}
		
		\begin{proof}
			The proof is by induction on the number of bits \(b\).
			
			\underline{Base Case:} \(b = 1\)
			
			Then \texttt{p} = (\(\yvar{2} \wedge \neg \yvar{1}\)) by \ref{Less-Than-1} 
			
			\(\Gamma \bigcup \{\yvar{1} \mapsto \bool, \yvar{2} \mapsto \bool\} \vdash \prog{} : \bool \comp ((\neg \yvar{1} \wedge \yvar{2}), \true, \emptyset)\) and \(\mathit{LT}(\yvar{1}, \yvar{2}) = \neg \yvar{1} \wedge \yvar{2}\)
			
			\underline{Induction Hypothesis:} \(\forall b' < b\)
			
			If \(\lessthan{\(\yvar{1s}\)}{\(\yvar{2s}\)}{b'} = \prog{s}\), 
			
			then
			\(\Gamma \bigcup \{\yvar{1s} \mapsto \bool^{b'}, \yvar{2s} \mapsto \bool^{b'}\} \vdash \prog{} : \bool \comp (\mathit{LT}_{b'}(\yvar{1s}, \yvar{2s}), \true, \emptyset)\)
			
			\underline{Inductive step:} \(\forall b\)
			
			If \(\lessthan{\(\yvar{1}\)}{\(\yvar{2}\)}{b} = \prog{}\), then by~\ref{Less-Than>1}, \(\prog{} =\begin{array}{l}
				\texttt{let \(\yvar{1f}\) = fst(\(\yvar{1}\)) in} \\
				\texttt{let \(\yvar{2f}\) = fst(\(\yvar{2}\)) in} \\
				\texttt{let \(\yvar{1wins}\) = \(\yvar{1f}\) \(\wedge \neg\) \(\yvar{2f}\) in} \\
				\texttt{let \(\yvar{2wins}\) = \(\yvar{2f}\) \(\wedge \neg\) \(\yvar{1f}\) in} \\
				\texttt{if \(\yvar{1wins}\) then \false else} \\
				\texttt{if \(\yvar{2wins}\) then \true else} \\
				\texttt{let \(\yvar{1s}\) = snd(\(\yvar{1}\)) in} \\
				\texttt{let \(\yvar{2s}\) = snd(\(\yvar{2}\)) in} \\
				\prog{+}
			\end{array}\)
			
			where \(\lessthan{\(\yvar{1s}\)}{\(\yvar{2s}\)}{b-1} = \prog{+}\)
			
			We reason through the derivation tree of compiling \(\prog{}\) to a WBF in a top down manner. for the sake of brevity, we only show elements of \(\Gamma\) that occur in \(\prog{}\).
			
			\begin{flalign*}
				S1 :& \Gamma \cup \bigg\{ \begin{array}{l}
					\yvar{1s} \mapsto \bool^{b-1}
					\\ \yvar{2s} \mapsto \bool^{b-1} 
				\end{array}\bigg\} \vdash \prog{+} : \bool \comp (\mathit{LT}_{b-1}(\yvar{1s}, \yvar{2s}), \true, \emptyset)
				~\tag{Induction Hypothesis}
			\end{flalign*}
			
			\begin{flalign*}
				S2 :& \inference{S1 
					&
					\inference{}{\Gamma \cup \bigg\{ \begin{array}{l}
							\yvar{1} \mapsto \bool^b
							\\  \yvar{2} \mapsto \bool^b
						\end{array}\bigg\} \vdash \Lsnd{\yvar{1}} : \bool^{b-1} \comp (\yvar{1r}, \true, \emptyset)}
					\\
					\inference{}{\Gamma \cup \bigg\{ \begin{array}{l}
							\yvar{1} \mapsto \bool^b
							\\  \yvar{2} \mapsto \bool^b
						\end{array}\bigg\} \vdash \Lsnd{\yvar{2}} : \bool^{b-1} \comp (\yvar{2r}, \true, \emptyset)}
				}
				{\Gamma \cup \bigg\{ \begin{array}{l}
						\yvar{1} \mapsto \bool^b
						\\  \yvar{2} \mapsto \bool^b
					\end{array}\bigg\} \vdash \begin{array}{l}
						\texttt{let \(\yvar{1s}\) = snd(\(\yvar{1}\)) in} \\
						\texttt{let \(\yvar{2s}\) = snd(\(\yvar{2}\)) in} \\
						\prog{+}
					\end{array} : \bool \comp (\mathit{LT}_{b-1}(\yvar{1r}, \yvar{2r}), \true, \emptyset)}[C-Let]
			\end{flalign*}
			
			\begin{flalign*}
				&S3 : 
				\\
				&\inference{S2}{\Gamma \cup \Biggl\{\begin{array}{l}
						\yvar{1} \mapsto \bool^b
						\\  \yvar{2} \mapsto \bool^b
						\\  \yvar{1wins} \mapsto \bool
						\\  \yvar{2wins} \mapsto \bool
					\end{array}
					\Bigg\} \vdash \begin{array}{l}
						\texttt{if \(\yvar{1wins}\) then \false else} \\
						\texttt{if \(\yvar{2wins}\) then \true else} \\
						\texttt{let \(\yvar{1s}\) = snd(\(\yvar{1}\)) in} \\
						\texttt{let \(\yvar{2s}\) = snd(\(\yvar{2}\)) in} \\
						\prog{+}
					\end{array} : \bool \comp 
					\begin{array}{l}
						((\yvar{1wins} \wedge \false) \vee (\neg \yvar{1wins} \wedge \\  ((\yvar{2wins} \wedge \true) 
						\\	\vee (\neg \yvar{2wins} \wedge \mathit{LT}_{b-1}(\yvar{1r}, \yvar{2r}))))
						\\ \true, \emptyset)			
				\end{array}}[C-Ite]
			\end{flalign*}
			
			\begin{flalign*}
				S4 :& \inference{S3 
					&
					\inference{}{\Gamma \cup \bigg\{ \begin{array}{l}
							\yvar{1} \mapsto \bool^b
							\\  \yvar{2} \mapsto \bool^b
							\\ \yvar{1f} \mapsto \bool
							\\ \yvar{2f} \mapsto \bool
						\end{array}\bigg\} \vdash \yvar{1f} \wedge \neg \yvar{2f} : \bool \comp (\yvar{1f} \wedge \neg \yvar{2f}, \true, \emptyset)}
					\\
					\inference{}{\Gamma \cup \bigg\{ \begin{array}{l}
							\yvar{1} \mapsto \bool^b
							\\  \yvar{2} \mapsto \bool^b
							\\ \yvar{1f} \mapsto \bool
							\\ \yvar{2f} \mapsto \bool
						\end{array}\bigg\} \vdash \yvar{2f} \wedge \neg \yvar{1f} : \bool \comp (\yvar{2f} \wedge \neg \yvar{1f}, \true, \emptyset)}
				}
				{\Gamma \cup \bigg\{ \begin{array}{l}
						\yvar{1} \mapsto \bool^b
						\\  \yvar{2} \mapsto \bool^b
						\\  \yvar{1f} \mapsto \bool
						\\  \yvar{2f} \mapsto \bool 
					\end{array}\bigg\} \vdash \begin{array}{l}
						\texttt{let \(\yvar{1wins}\) = \(\yvar{1f}\) \(\wedge \neg\) \(\yvar{2f}\) in} \\
						\texttt{let \(\yvar{2wins}\) = \(\yvar{2f}\) \(\wedge \neg\) \(\yvar{1f}\) in} \\
						\texttt{if \(\yvar{1wins}\) then \false else} \\
						\texttt{if \(\yvar{2wins}\) then \true else} \\
						\texttt{let \(\yvar{1s}\) = snd(\(\yvar{1}\)) in} \\
						\texttt{let \(\yvar{2s}\) = snd(\(\yvar{2}\)) in} \\
						\prog{+}
					\end{array} : \bool \comp \begin{array}{l}
						(\neg \yvar{1f} \wedge \yvar{2f}) \vee 
						\\	(((\neg \yvar{1f} \wedge \neg \yvar{2f})\vee
						\\	(\yvar{1f} \wedge \yvar{2f})) \wedge 
						\\	\mathit{LT}_{b-1}(\yvar{1r}, \yvar{2r})), \true, \emptyset)
				\end{array} }[C-Let]
			\end{flalign*}
			
			\begin{flalign*}
				S5 :& \inference{S4 
					&
					\inference{}{\Gamma \cup \bigg\{ \begin{array}{l}
							\yvar{1} \mapsto \bool^b
							\\  \yvar{2} \mapsto \bool^b
						\end{array}\bigg\} \vdash \Lfst{\yvar{1}} : \bool^{b-1} \comp (\yvar{1l}, \true, \emptyset)}
					\\
					\inference{}{\Gamma \cup \bigg\{ \begin{array}{l}
							\yvar{1} \mapsto \bool^b
							\\  \yvar{2} \mapsto \bool^b
						\end{array}\bigg\} \vdash \Lfst{\yvar{2}} : \bool^{b-1} \comp (\yvar{2l}, \true, \emptyset)}
				}
				{\Gamma \cup \bigg\{ \begin{array}{l}
						\yvar{1} \mapsto \bool^b
						\\  \yvar{2} \mapsto \bool^b
					\end{array}\bigg\} \vdash \begin{array}{l}
						\texttt{let \(\yvar{1f}\) = fst(\(\yvar{1}\)) in} \\
						\texttt{let \(\yvar{2f}\) = fst(\(\yvar{2}\)) in} \\
						\texttt{let \(\yvar{1wins}\) = \(\yvar{1f}\) \(\wedge \neg\) \(\yvar{2f}\) in} \\
						\texttt{let \(\yvar{2wins}\) = \(\yvar{2f}\) \(\wedge \neg\) \(\yvar{1f}\) in} \\
						\texttt{if \(\yvar{1wins}\) then \false else} \\
						\texttt{if \(\yvar{2wins}\) then \true else} \\
						\texttt{let \(\yvar{1s}\) = snd(\(\yvar{1}\)) in} \\
						\texttt{let \(\yvar{2s}\) = snd(\(\yvar{2}\)) in} \\
						\prog{+}
					\end{array} : \bool \comp \begin{array}{l}
						(\neg \yvar{1l} \wedge \yvar{2l}) \vee 
						\\	(((\neg \yvar{1l} \wedge \neg \yvar{2l})\vee
						\\	(\yvar{1l} \wedge \yvar{2l})) \wedge 
						\\	\mathit{LT}_{b-1}(\yvar{1r}, \yvar{2r})), \true, \emptyset)
				\end{array}}[C-Let]
			\end{flalign*}
			
			And \((\neg \yvar{1l} \wedge \yvar{2l}) \vee (((\neg \yvar{1l} \wedge \neg \yvar{2l})\vee
			(\yvar{1l} \wedge \yvar{2l})) \wedge 
			\mathit{LT}_{b-1}(\yvar{1r}, \yvar{2r})) = \mathit{LT}_b(\yvar{1}, \yvar{2})\)
			
		\end{proof}
		
		We restate Lemma~\ref{lemma:less-than-bdd-size} here.
		\begin{lemma}\label{lemma:less-than-bdd-size-restate}
			\(\exists k, \forall b,\) for the variable order \(x_1, y_1, x_2, y_2, \ldots, x_b, y_b\), the size of the OBDD, that is  \(\mathit{OBDD}(\mathit{LT}((x_1, x_2, \ldots, x_b), (y_1, y_2, \ldots, y_b))) \leq kb\).
		\end{lemma}
		
		\begin{proof}
			
			Let \(f\) be a function over \(v_1, v_2, \ldots v_n\) and \(m\) be the number of distinct subfunctions of \(f\) obtained by conditioning on \(v_1, \ldots v_{i-1}\)
			that depend on \(v_i\). A reduced OBDD for \(f\) using variable
			ordering \(v_1, v_2, \ldots v_n\) contains exactly \(m\) nodes labeled with \(v_i\)~\cite{sieling_wegener}.
			
			By Definition~\ref{def:inequality_wbf}, we have the following:
			
			\[\mathit{LT}_b((x_1, \ldots x_b), (y_1, \ldots y_b)) = 
			\begin{cases}
				\neg x_1 y_1 & b = 1
				\\	\neg x_1 y_1 \vee (\neg x_1 \neg y_1 \vee x_1y_1)\mathit{LT}_{b-1}((x_2, \ldots, x_b), (y_2, \ldots, y_b)) & b > 1
			\end{cases}\]
			
			Conditioning on variables till \(x_i\), we have the following:
			
			\(\forall i < b, \mathit{LT}_b((x_1, x_2, \ldots x_b), (y_1, y_2, \ldots y_b)) | (x_1, y_1, \ldots x_i) = \begin{cases}
				\neg y_i + y_i \mathit{LT}_{b - i}(x_{i+1}, \ldots y_b)
				\\ \neg y_i \mathit{LT}_{b - i}(x_{i+1}, \ldots y_b)
			\end{cases}\)
			
			Thus, for every \(y_i\), we can only have two nodes.
			
			Conditioning on variables till \(y_i\), we have the following:
			
			\(\forall i < b, \mathit{LT}_b((x_1, x_2, \ldots x_b), (y_1, y_2, \ldots y_b)) | (x_1, y_1, \ldots x_i) = \begin{cases}
				\true
				\\ \false
				\\ \mathit{LT}_{b - i}(x_{i+1}, \ldots y_b)
			\end{cases}\)
			
			Thus, for every \(x_i\), we can only have a single node. So, the total number of nodes in the OBDD would be \(3b\). That is, k = 3.
		\end{proof}
		
		\begin{lemma}\label{lemma:unifobs-compile}
			If \(\unifobs{\yvar{}}{b} = \prog{obs}\) then \(\Gamma \cup \{\yvar{} : \bool^b\} \vdash \prog{obs}:\bool \comp (\true, \mathit{LT}_b((f_1, f_2, \ldots, f_b), \yvar{}), w)\) where \(w\) assigns weights to the Boolean random variables \(f_1, f_2, \ldots, f_b\)
		\end{lemma}
		
		\begin{proof}
			By \ref{unifobs}, \(\prog{obs} = \begin{array}{l}
				\texttt{let \(\yvar{1}\) = \(\prog{1}\) in} \\
				\texttt{let \(\yvar{bool}\) = \(\prog{2}\) in } \\
				\texttt{observe(\(\yvar{bool}\))}
			\end{array}\)
			
			where \(\pdf{0}{0} \comp_b \prog{1}\) and
			\(\texttt{less\_than(\(\yvar{1}\), \(\yvar{}\), b)} = \prog{2}\)
			
			We first compile subprograms to weighted Boolean formulas.
			
			Now, \(\pdf{0}{0} \comp_b \prog{1}\)
			
			Then by ~\ref{Trans-expo0}, \(\prog{1} = \begin{array}{l}
				\texttt{let \(\yvar{1}\) = flip($\theta_1$) in}
				\\ \texttt{let \(\yvar{2}\) = flip($\theta_2$) in}
				\\ \ldots
				\\ \texttt{let \(\yvar{b}\) = flip($\theta_b$) in}
				\\ \texttt{(\(\yvar{1}\), \(\yvar{2}\), \ldots, \(\yvar{b}\))}
			\end{array}\)
			
			Let \(w_i = (\flip{i} \mapsto \theta_i, \overline{\flip{i}} \mapsto 1 - \theta_i)\) and \(D_i = \inference{\texttt{fresh }\flip{i}}{\Gamma \cup \bigcup_{j = 1}^{i-1} \{y_j \mapsto \bool\} \vdash \Lflip{\theta_i}: \bool \comp (\flip{i}, \true, w_i)}[C-Flip]\)
			
			Now consider the following derivation tree.
			
			\begin{flalign*}
				\inference{
					D_1
					&
					\inference{
						\inference{D_b 
							&
							\Gamma \cup \{\yvar{1} \mapsto \bool, ... \yvar{b} \mapsto \bool\} \vdash
							(\yvar{1}, \yvar{2}, \ldots, \yvar{b}) : \bool^b \comp (\yvar{1}, \yvar{2}, \ldots, \yvar{b}) 
						}{\vdots}
					}{\Gamma \cup \{\yvar{1}:\bool\} \vdash \begin{array}{l}
							\texttt{let \(\yvar{2}\) = flip($\theta_2$) in}
							\\ \ldots
							\\ \texttt{let \(\yvar{b}\) = flip($\theta_b$) in}
							\\ \texttt{(\(\yvar{1}\), \(\yvar{2}\), \ldots, \(\yvar{b}\))}
						\end{array}:\bool^b \comp ((\yvar{1}, \flip{2}, \ldots \flip{b}), \true, \bigcup_{i=1}^{b-1} w_i) }[C-let]
				}{S1 :: \Gamma \vdash \begin{array}{l}
						\texttt{let \(\yvar{1}\) = flip($\theta_1$) in}
						\\ \texttt{let \(\yvar{2}\) = flip($\theta_2$) in}
						\\ \ldots
						\\ \texttt{let \(\yvar{b}\) = flip($\theta_b$) in}
						\\ \texttt{(\(\yvar{1}\), \(\yvar{2}\), \ldots, \(\yvar{b}\))}
					\end{array} : \bool^{b} \comp ((\flip{1}, \flip{2}, \ldots \flip{b}), \true, \bigcup_{i=1}^{b} w_i)}[C-let]
			\end{flalign*}
			
			Now, we have \(\texttt{less\_than(\(\yvar{1}\), \(\yvar{}\), b)} = \prog{2}\). By Lemma~\ref{lemma:less_than_wbf}, we have the following: 
			
			\(S2 :: \Gamma \bigcup \{\yvar{1} \mapsto \bool^b, \yvar{} \mapsto \bool^b\} \vdash \prog{2} : \bool \comp (\mathit{LT}_b(\yvar{1}, \yvar{}), \true, \emptyset)\)
			
			\begin{flalign*}
				\inference{
					S1
					&
					\inference{
						S2
						&
						\Gamma \cup \{\yvar{}:\bool^b, \yvar{1}: \bool^b, \yvar{bool}:\bool\} \vdash \Lobs{(\yvar{bool})} : \bool \comp (\true, \yvar{bool}, \emptyset)
					}
					{
						\Gamma \cup \{\yvar{}:\bool^b, \yvar{1}:\bool^b\} \vdash \Llet{\yvar{bool} = \prog{2}}{\Lobs{(\yvar{bool})}} : \bool \comp (\true, \mathit{LT}_b(\yvar{1}, \yvar{}), \emptyset)
					}[C-let]
				}
				{
					\Gamma \cup \{\yvar{} : \bool^b\} \vdash \begin{array}{l}
						\texttt{let \(\yvar{1}\) = \(\prog{1}\) in} \\
						\texttt{let \(\yvar{bool}\) = \(\prog{2}\) in } \\
						\texttt{observe(\(\yvar{bool}\))}
					\end{array} : \bool \comp (\true, \mathit{LT}_b((\flip{1}, \flip{2}, \ldots \flip{b}), \yvar{}), \bigcup_{i=1}^{b} w_i)
				}
			\end{flalign*}
		\end{proof}
		
		\begin{lemma}\label{lemma:unifobs-bdd-size}
			\(\forall f, o \exists k, \forall \prog{}, b, \varphi, \gamma, w\), if \(\prog{}\) is a unifobs program with \(b\) bits of size \((f, o)\) and \(\prog{} \comp (\varphi, \gamma, w)\), then for the interleaving variable order \(\obdd{\gamma} \leq kb\).
		\end{lemma}
		
		\begin{proof}
			Consider \(\prog{}\), a unifobs program with \(b\) bits of size \((f, o)\). This implies that \(\var{g}\) (if-guard flips) has \(f\) flips. Let \(\var{f}(ij)\) refer to the \(i\)-th most significant bit of \(j\)-th exponential. Now consider the interleaving variable order  
			\[ \var{g}, \var{f}(11), \var{f}(21), \ldots, \var{f}(n1),
			\var{f}(12), \ldots, \var{f}(n2),
			\ldots,
			\var{f}(1b), \ldots, \var{f}(nb)\] 
			
			Conditioning \(\gamma \mid \var{g}\), there are at most \(2^f\) subfunctions. Let's refer to them as \(\gamma_c\). We further have \(\gamma_c = \bigwedge_{i = 1}^{i < o} \gamma_i\) where each \(\gamma_i = \mathit{LT}_b((x_1, \ldots x_b), (y_1, \ldots y_b))\) (Lemma~\ref{lemma:unifobs-compile})
			
			Now, conditioning \(\gamma_c\) on the first \(l\) bits of all exponential programs involved gives at most \(2^o\) subfunctions (Lemma~\ref{lemma:less-than-bdd-size}) and there are \(b\) bits which imply that each \(\gamma_c\) have \(b2^o\) nodes in its OBDD~\cite{wegener2000branching}.
			
			Thus, we have \(\obdd{\gamma} \leq 2^{(f+o)}b\).
		\end{proof}

		\begin{lemma}\label{lemma:accept-bdd-size}
			\(\forall \Upsilon, \exists k, \forall \prog{}, \varphi, \gamma, w, \forall b\), if \(\Upsilon \comp_b \prog{}\) and \(\prog{} \comp (\varphi, \gamma, w)\), then for the interleaving variable order of \(\prog{}\), \(\obdd{\gamma} \leq kb\)
		\end{lemma}
		
		\begin{proof}
			By Lemma~\ref{lemma:all-unifobs-program}, we have that \(\forall \Upsilon, \exists f, o, \forall \prog{}, b\), if \(\Upsilon \comp_b \prog{}\), then \(\prog{}\) is a unifobs program with \(b\) bits of size \((f, o)\).
			
			By Lemma~\ref{lemma:unifobs-bdd-size}, we have that \(\forall f, o, \exists k, \forall \prog{}, b, \varphi, \gamma, w\), if \(\Gamma \vdash \prog{} : \bool^b \comp (\varphi, \gamma, w)\), then \(\obdd{\gamma} \leq kb\) for the interleaving variable order.
			
			Combining them we have the above lemma. 
		\end{proof}
		
		\begin{theorem}
			\(\forall \Upsilon, \exists k, \forall \prog{}, \varphi, \gamma, w, \forall b\), if \(\Upsilon \comp_b \prog{}\) and \(\prog{} \comp (\varphi, \gamma, w)\), then there exists a variable order of Boolean random variables in \(w\) such that \(\obdd{\varphi} + \obdd{\gamma} \leq kb\)
		\end{theorem}
		
		\begin{proof}
			The theorem follows from lemma~\ref{lemma:return-bdd-size} and ~\ref{lemma:accept-bdd-size}.
		\end{proof}

		\section{Proof of Theorem~\ref{expectation} and ~\ref{variance}}
		
		\begin{theorem}
			Let D be a discrete probability distribution over the interval $[0, 2^n)$ represented as a distribution over $n$ bits as $(b_n, b_{n-1}, \ldots, b_1)$, then the expectation of D can be computed using linearity of expectation as follows:
			\[\mathbb{E}[D] = \sum_{i = 0}^{2^n-1} i \cdot \mathit{pr}(i) = \mathbb{E}(\sum_{j = 1}^{n} 2^{j-1}b_j) = \sum_{j=1}^{n}2^{j-1} \cdot \mathit{pr}(b_j) \]
		\end{theorem}
		
		\begin{proof}
			Let X be a random variable that has distribution D. Let \((b_n, b_{n-1}, \ldots, b_1)\) be the binary representation of X. then, we have the following:
			\begin{flalign*}
				X &= \sum_{j = 1}^{n} 2^{j-1} b_j
				\\
				\mathbb{E}(X) &= \mathbb{E}(\sum_{j = 1}^{n} 2^{j-1} b_j) = \sum_{j = 1}^{n} 2^{j-1} \mathbb{E}(b_j)~\tag{by linearity of expectation}
				\\	
				\mathbb{E}(D) &= \sum_{j = 1}^{n} 2^{j-1} \mathit{pr}(b_j)~\tag{For Boolean variables \(\mathbb{E}(b) = \mathit{pr}(b)\)}
			\end{flalign*}
		\end{proof}
		
		\begin{theorem}
			Let D be a discrete probability distribution over the interval $[0, 2^n)$ represented as a distribution over $n$ bits as $(b_n, b_{n-1}, \ldots, b_1)$, then the variance of D can be computed as follows:
			\[\mathit{Var}[D] = \sum_{i=0}^{2^n - 1} i^2 \mathit{pr}(i) - (\mathbb{E}(D))^2 = \mathit{Var}(\sum_{j = 1}^{n} 2^{j-1}b_j) = \sum_{k = 1}^{n} \sum_{l = 1}^{n} 2^{l+k-2} [\mathit{pr}(b_l \wedge b_k) - \mathit{pr}(b_l)\mathit{pr}(b_k)] \]
		\end{theorem}
		
		\begin{proof}
			Let X be a random variable that has distribution D. Let \((b_n, b_{n-1}, \ldots, b_1)\) be the binary representation of X. then, we have the following:
			\begin{flalign*}
				X &= \sum_{j = 1}^{n} 2^{j-1} b_j
				\\
				\mathit{Var}(X) &= \mathit{Var}(\sum_{j = 1}^{n} 2^{j-1} b_j) = \sum_{l, k = 1}^{n} 2^{l-1}2^{k-1}\mathit{Cov}(b_l, b_k)~\tag{by Bienaym\'e's identity}
				\\	
				&= \sum_{l=1}^{n} \sum_{k=1}^{n} 2^{l + k - 2} (\mathbb{E}(b_l \wedge b_k) - \mathbb{E}(b_l)\mathbb{E}(b_k)) = \sum_{l=1}^{n} \sum_{k=1}^{n} 2^{l + k - 2} (\mathit{pr}(b_l \wedge b_k) - \mathit{pr}(b_l)\mathit{pr}(b_k))~\tag{For Boolean variables \(\mathbb{E}(b) = \mathit{pr}(b)\)}
			\end{flalign*}
		\end{proof}
		
		\section{OBDD for uniform distribution}

\begin{wrapfigure}{l}{0.25\textwidth}
			\begin{tikzpicture}
				\def\lvl{20pt}

				\node (f1) at (0, 0) [bddnode] {$f_1$};
				\node (f2) at ($(f1) + (20bp, -\lvl)$) [bddnode] {$f_2$};
				\node (f3) at ($(f2) + (20bp, -\lvl)$) [bddnode] {$f_3$};
				
				\node[draw, rounded corners] at ($(f1) + (-30bp, 0bp)$)  {0.5};
				\node[draw, rounded corners] at ($(f1) + (-30bp, -20bp)$)  {0.5};
				\node[draw, rounded corners] at ($(f1) + (-30bp, -40bp)$)  {0.5};

				\node (false) at ($(f3) + (-30bp, -1*\lvl)$) [bddterminal] {$\false$};
				
				\node (true) at ($(f3) + (-15bp, -1*\lvl)$) [bddterminal] {$\true$};

				\node (b2) at ($(f1) + (0bp, 20bp)$) [bddroot] {$b_1$};
				\node (b1) at ($(b2) + (20bp, 0bp)$) [bddroot] {$b_2$};
				\node (b0) at ($(b1) + (20bp, 0bp)$) [bddroot] {$b_3$};

				\begin{scope}[on background layer]
					\draw [highedge] (f1) -- (true);
					\draw [lowedge] (f1) -- (false);
					\draw [highedge] (f2) -- (true);
					\draw [lowedge] (f2) -- (false);
					\draw [highedge] (f3) -- (true);
					\draw [lowedge] (f3) -- (false);

					\draw[-stealth] (b2) -- (f1);
					\draw[-stealth] (b1) -- (f2);
					\draw[-stealth] (b0) -- (f3);

				\end{scope}
			\end{tikzpicture}
			\caption{
				Compiled BDD for 3-bit blasted uniform distribution.
			}\label{fig:unif_bdd}
		\end{wrapfigure}
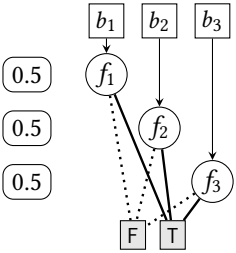

		In Figure~\ref{fig:unif_bdd}, we 3-bit blast a uniform distribution, i.e.\ we have a discrete uniform distribution over 8 values (0, 0.125, 0.25, $\ldots$, 0.875). The figure shows a 3-rooted BDD where each root labeled as $b_1, b_2$ and $b_3$ represents a bit in the returned value of 3 bits. Consider that we fix the value of the flip corresponding to the node $f_1$ to be true, then for $b_1$, we would reach the terminal 1 and its value would be assigned 1. The operation of weighted model counting (WMC) would calculate the probability of $b_1$ to be 1 as \(0.5\) as node $f_1$ has probability \(0.5\) to be true. Similarly WMC can be used for other roots of this BDD.\@ Since each bit needs one BDD node, the overall BDD size grows linearly with the number of bits. Since WMC runs in time linear in the BDD size, probabilistic inference for an exponential distribution would run linear in the number of bits $\mathcal{O}(b)$. 
		
		\section{Details on Experiments in Section~\ref{sec:motivation} and Section~\ref{sec:experiments}}
		
		\subsection{Gene Expression: Scaling with Logical Constraints}
		
		We showed example of scaling with logical constraints in Section~\ref{sec:gene}. We use this subsection to discuss empirical results in detail. We compare against various various probabilistic inference algorithms on this example $-$ WebPPL rejection sampling, WebPPL MCMC, WebPPL SMC, Stan and Psi. To run Stan on this example, we made use of SlicStan~\cite{Gorinova_2021} to get the Stan program with marginalized discrete variables. Figure~\ref{fig:logical_plot} reports the absolute error achieved by different probabilistic programming systems as the number of discrete variables (T) increases with the timeout of 20 minutes. The absolute error reported for \hybit{} is obtained by running it with 20 bits and 1024 pieces.

		As it was expected, \hybit{} scales well with increasing number of discrete variables and consistently reports low absolute error ($4.7 \times 10^{-7}$). This is due to \hybit{}'s ability to exploit discrete structure. Psi, when it does not timeout report the exact probability but is not able to handle more than 10 discrete variables. The error exhibited by Stan increases as number of discrete variables increase but it isn't able to draw a single sample when the number of discrete variables go beyond 15. The decreasing number of samples can be attributed to the fact that the marginalizing out discrete random variables does not take into account the context-specific independences exhibited by the discrete structure in this program. WebPPL rejection sampling times out for this program and AQUA does not support this program. 
		
		\subsection{Comparison of piece-wise approximation with central-limit theorem}
		
\begin{figure}[h]
			\includegraphics[width = 0.5\textwidth]{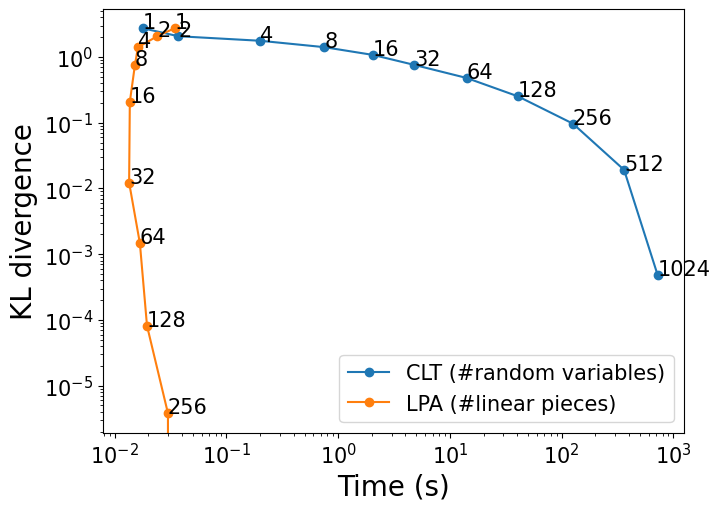}
			\caption{Log-log plot comparing Central limit
				theorem and Linear piece-wise approximation}\label{clt-lpa}
		\end{figure}

		A normal distribution can be approximated by summing independent and identically distributed random variables in accordance to the central limit theorem. We sum up uniformly distributed random variables to achieve a gaussian distribution; we refer to this approach as CLT.\@ We compare the linear piece-wise approximation of a normal distribution (LPA) with CLT in terms of accuracy in Figure~\ref{clt-lpa}. It can be clearly seen that the piece-wise approximation achieves higher accuracy in much less time. This is due to the construction of piece-wise distribution being more efficient than summing the uniformly distributed random variables in CLT.\@ It is also worth noting here that piece-wise approximation is generalizable to other continuous distributions as well whereas that is not the case with CLT.\@
		
	\end{document}